\documentclass[10pt]{iopart}

\usepackage{graphicx}
\usepackage{amssymb}
 \usepackage{amsthm}
\usepackage{enumitem}
\expandafter\let\csname equation*\endcsname\relax

\expandafter\let\csname endequation*\endcsname\relax
\expandafter\let\csname eqalign*\endcsname\relax

\expandafter\let\csname endeqalign*\endcsname\relax
\usepackage{amsmath}

\usepackage{mathtools}
\usepackage{float}
\usepackage{microtype}
\usepackage{tabularx}

\begin{document}

%\begin{frontmatter}


\title[Spectroscopy inferences of plasma-molecule interactions]{A novel hydrogenic spectroscopic technique for inferring the role of plasma-molecule interaction on power and particle balance during detached conditions}



\author{K. Verhaegh$^{1,2,3}$, B. Lipschultz$^2$, C. Bowman$^2$, B.P. Duval$^3$, U. Fantz$^{4,5}$, A. Fil$^{2,1}$, J.R. Harrison$^1$, D. Moulton$^1$, O. Myatra$^2$, D. W\"{u}nderlich$^4$, F. Federici$^2$, D.S. Gahle$^{6,1}$, A. Perek$^7$, M. Wensing$^3$, the TCV Team$^{*}$ and the EuroFusion MST1 team$^{**}$}
\address{$^1$ Culham Centre for Fusion Energy, Culham, United Kingdom} 
\address{$^2$ York Plasma Institute, University of York, United Kingdom}
\address{$^3$ \'{E}cole Polytechnique F\'{e}d\'{e}rale de Lausanne (EPFL), Swiss Plasma Center (SPC), CH-1015 Lausanne, Switzerland}
\address{$^4$ Max Planck Institute for Plasma Physics, Garching bei M\"{u}nchen, Germany}
\address{$^5$ Augsburg University, Augsburg, Germany}
\address{$^6$ SUPA, University of Strathclyde, Glasgow, United Kingdom}
\address{$^7$ DIFFER, Eindhoven, The Netherlands}
\address{$^*$ See author list of "S. Coda et al 2019 Nucl. Fusion 59 112023"}
\address{$^**$ See author list of "B. Labit et al 2019 Nucl. Fusion 59 086020"}
\ead{kevin.verhaegh@ukaea.uk}

\begin{abstract}
Detachment, an important mechanism for reducing target heat deposition, is achieved through reductions in power, particle and momentum; which are induced through plasma-atom and plasma-molecule interactions. Experimental research in how those reactions precisely contribute to detachment is limited. 

Both plasma-atom as well as plasma-molecule interactions can result in excited hydrogen atoms which emit atomic line emission. In this work, we investigate a new Balmer Spectroscopy technique for Plasma-Molecule Interaction - BaSPMI. This first disentangles the Balmer line emission from the various plasma-atom and plasma-molecule interactions and secondly quantifies their contributions to particle (ionisation and recombination) and power balance (radiative power losses). Its performance is verified using synthetic diagnostic techniques of both attached and detached TCV and MAST-U SOLPS-ITER simulations. 

%During detachment, the observed $H\alpha$ emission often strongly increases, which could be an indicator for plasma-molecule interactions involving $H_2^+$ and/or $H^-$. Our analysis technique quantifies the $H\alpha$ emission due to plasma-molecule interactions and uses this to 1) quantify the Balmer line emission contribution due to $H_2^+$ and/or $H^-$; 2) subsequently estimate its resulting particle sinks/sources and radiative power losses. 

We find that $H_2$ plasma chemistry involving $H_2^+$ and/or $H^-$ can substantially elevate the $H\alpha$ emission during detachment, which we show is an important precursor for Molecular Activated Recombination (MAR). An example illustration analysis of the full BaSPMI technique shows that the hydrogenic line series, even $Ly\alpha$ as well as the medium-n Balmer lines, can be significantly influenced by plasma-molecule interactions by tens of percent. That has important implications for using atomic hydrogen spectroscopy for diagnosing divertor plasmas. 
\end{abstract}

%\vspace{1pc}
\noindent{\it Keywords}: Tokamak divertor; Molecules; plasma;  SOLPS-ITER; Plasma spectroscopy; Power/particle balances; Detachment

\section{Introduction}
\label{ch:introduction}

Divertor detachment is predicted to be crucial for handling the power exhaust of future fusion devices, such as ITER \cite{Pitts2013,Loarte2007,Pitcher1997}. Divertor detachment implies a simultaneous reduction of the target plasma temperature, target ion flux and the target pressure. This is achieved through atomic and molecular processes driving power losses, momentum losses and particle losses (through a reduction of ion sources and/or increases of ion sinks). All three losses play an important role in the detached state and require detailed characterisation \cite{Stangeby2018,Verhaegh2019,Krasheninnikov2017}. 

The hydrogenic line series (such as the Balmer line series) has been routinely monitored in tokamaks using both line of sight spectroscopy as well as filtered camera imaging systems. Those measurements can be used to study some of the plasma-atom interactions involved in detachment. First, this involved studying the increase of the Electron-Ion Recombination (EIR) ion sink during detachment \cite{Terry1999,Lipschultz1998,Verhaegh2017}. Later studies involved estimating ion sources \cite{Verhaegh2019,Verhaegh2019a,Lomanowski2020}, as well as the power lost due to ionisation \cite{Verhaegh2019a}.  

Plasma-molecule interactions involve both \emph{collisions} and \emph{reactions} which impact power, particle and momentum balance. $H_2$ becomes rovibronically (e.g. rotationally, vibrationally and electronically) excited through \emph{collisions} between the electrons and $H_2$ \cite{Krasheninnikov1997,Fantz2002,Fantz2001,Sakamoto2017,Hollmann2006,Groth2019,Kukushkin2017,Wischmeier2004}. De-excitation of electronically excited molecules can result in $H_2$ Fulcher band emission. Vibrationally excited molecules strongly promote the creation of $H^-$ and $H_2^+$ (for $T_e$ between 1-4 eV). $H_2^+$ and $H^-$ can undergo \emph{reactions} with the plasma resulting in Molecular Activated Recombination (MAR) and Molecular Activated Ionisation (MAI) ion sinks/sources \cite{Sakamoto2017,Kukushkin2017,Krasheninnikov2017,Terry1998,Terakado2019}.

Experimental investigations on plasma-molecule interactions in tokamak divertors are, in general, few and are typically based on measuring the  $H_2$ Fulcher band spectra \cite{Fantz2002,Fantz2001,Hollmann2006,Groth2019}. Such measurements provide useful information on the rovibrational structure of $H_2$ \cite{Fantz2002} and thus provide \emph{direct} evidence of plasma-molecule \emph{collisions} as the molecules get rovibronically excited by the plasma. It can also provide $H_2$ dissociation estimates \cite{Hollmann2006}. Models can be used to \emph{extrapolate} those Fulcher band measurements (from plasma-molecule \emph{collisions} resulting in excited molecules) to MAR/MAI plasma-molecule \emph{reaction} estimates \cite{Fantz2002}, on which the Fulcher band provides no \emph{direct} information. The $H_2$ Fulcher band is however complicated to diagnose given its limited brightness and that measuring it fully requires a relative wide wavelength (590-640 nm) range as well as a relatively high spectral resolution \cite{Fantz2002} to resolve its band structure.

$H^-$ and $H_2^+$ undergo reactions with the plasma leading to MAR or MAI, which can also lead to \emph{excited atoms} modifying the hydrogenic line series emission \cite{Fantz2002,Fantz2001,Wuenderlich2016,Sakamoto2017,Terakado2019}, particularly $H\alpha$ and $H\beta$. Such molecule-derived modifications to the Balmer line series and their associated radiative losses have not yet been studied experimentally before in tokamak divertors and provide an alternative way of estimating MAR/MAI as well as atomic line radiation related to $H_2$ plasma chemistry.

\subsection{This work and its outline}

In this work, we describe an analysis technique which can quantify the contributions of plasma-molecule interactions to the Balmer line emission and use that to estimate the role plasma-molecule interactions play on particle and power balance during detachment. Our technique - Balmer Spectroscopy Plasma-Molecule Interaction (BaSPMI) is explained in section \ref{ch:analysis}. BaSPMI first executes the technique previously developed by the authors \cite{Verhaegh2019a,Verhaegh2017} to separate the \emph{atomic} process contributions (electron-impact excitation and electron-ion recombination) from the analysis of medium-n ($n = 5,6,7$) Balmer lines (section \ref{ch:analysis_atomic}). $H\alpha$ and $H\beta$ emission is generally more sensitive to plasma-molecule interactions than the medium-n Balmer lines. We extrapolate the \emph{atomic process information} from the \emph{medium-n Balmer lines} to $H\alpha$ and $H\beta$. We compare this to the measured $H\alpha$ and $H\beta$ to estimate the contribution of excited atoms related to $H_2$ plasma chemistry to $H\alpha$ and $H\beta$ (section \ref{ch:analysis_DaMol}). Using collisional-radiative model results from Yacora (on the Web) \cite{Wuenderlich2016,Wunderlich2020}, Balmer line emission attributed to $H_2$ plasma chemistry involving $H_2^+, H^-$ is quantitatively separated using the ratio of the sum of the molecular process contributions of $H\alpha$ and $H\beta$ (section \ref{ch:analysis_DaMolSepa}). Those contributions are then used individually to: %Throughout this, a certain range of possible $H_2$ concentrations in the plasma is assumed as function of $T_e$, similar to \cite{Stangeby2018} (based on SOLPS-ITER simulations of TCV), which is accounted for in all Balmer lines. 
\begin{itemize}
    \item Estimate Molecular Activated ion sinks (Recombination) /sources (Ionisation) - MAR/MAI for each emission channel (section \ref{ch:analysis_molMARMAI}).
    \item Estimate the contribution of plasma-molecule interactions to:
    \begin{itemize}
        \item the entire hydrogenic spectra providing radiative loss estimates for excited atoms arising from plasma interactions with $H_2, H_2^+$ and $H^-$ (section \ref{ch:analysis_molrad}).
        \item the medium-n Balmer lines, which is accounted for self-consistently (section \ref{ch:analysis_ContaminIterate}).
    \end{itemize}
\end{itemize} 

The applicability of this technique is verified using synthetic diagnostic data from TCV and MAST-U SOLPS simulations in section \ref{ch:analysis_solps}. Here the analysis estimates, based on a synthetic spectrometer signal analysed through BaSPMI, are compared against the values directly obtained from the simulation. The performance of the technique is further tested by artificially removing emission process contributions from the synthetic spectrometer signals and checking the analysis response (section \ref{ch:analysis_indepth_syndiag}). We find that the analysis behaves as expected: the synthetic diagnostic analysis estimates are in quantitative agreement (within uncertainty) with the direct values obtained from the simulation.
 
BaSPMI has been applied to a set of TCV experimental data and appears in \cite{Verhaegh2020a}. For a complete picture of BaSPMI we have also included a brief example of the TCV experimental data analysis in this paper in section \ref{ch:results}. This shows the capabilities of BaSPMI to separate various hydrogen emission lines in its different emission pathways (e.g. electron-ion recombination (of $H^+$), electron-impact excitation (of $H$) and related to $H_2$ plasma chemistry (involving $H_2$, $H_2^+$ and $H^-$). %this model is applied to experimental measurements on TCV. There, it is also shown that once we detach during a density ramp on TCV, the measured $H\alpha$ emission bifurcates from its atomic process estimate. We provide a brief illustration of utlizing BaSPMI for separating Balmer line emission contributions sinks using the TCV experimental data analysed in \cite{Verhaegh2020a} for two different plasma conditions in section \ref{ch:results}.

We further discuss the Balmer line emission associated with $H_2$ (as opposed to $H_2^+$ and $H^-$) in section \ref{ch:fH2_Treatment}, which we show is expected to be negligible in the discussed detached divertor conditions. As the analysis relies on the fact that $H_2$ plasma chemistry results in additional $H\alpha$ emission, other processes which could result in additional $H\alpha$ emission may interfere with this analysis. These other contributions are discussed and estimated in section \ref{ch:OtherDa_Emiss}. We discuss how using this analysis for MAR/MAI estimates compares against using model extrapolations from the $H_2$ Fulcher band measurements in section \ref{ch:discus_fulcher}. The dependence of this analysis chain on molecular data is further discussed in section \ref{ch:discussion_isotope}. The application of BaSPMI to more reactor-like tokamak environments is discussed in section \ref{ch:discus_applicability}, together with analysis enhancement suggestions.

The development of BaSPMI was motivated by observing that the 'atomic extrapolated $H\alpha$' bifurcates from the measured $H\alpha$ at the detachment onset \cite{Verhaegh2018}. Our theoretical analysis indicates that this bifurcation is a particularly powerful indicator for plasma-molecule interactions. This is in agreement with the application of the \emph{full BaSPMI analysis} on experimental TCV data in \cite{Verhaegh2020a}, where this bifurcation is shown to correspond to the onset of MAR as well as Balmer line emission related to $H_2^+$ chemistry. In this work we show that this bifurcation can be used for \emph{quantitative} MAR estimates, which are in agreement with those from the full BaSPMI analysis (section \ref{ch:Dalpha_mon}). The full self-consistent analysis chain BaSPMI is, however, required for estimating the impact of $H_2$ plasma chemistry on the total hydrogenic spectra which can be important for ionisation source estimates as highlighted in \cite{Verhaegh2020a}.

$H_2$ plasma chemistry involves reactions which can result in excited atoms and thus atomic line emission. It is important to account for this when analysing the hydrogenic Balmer line series. This work provides an analysis technique - BaSPMI to dissect the hydrogen Balmer line emission into its various components and use this to perform a power and particle balance analysis accounting for both plasma-atom interactions as well as plasma-molecule interactions.

\section{Spectral analysis techniques of inferring information on plasma-molecule interaction from the Balmer spectra}
\label{ch:analysis}

\begin{figure}[htb]
    \centering
    \includegraphics[width=0.2\linewidth]{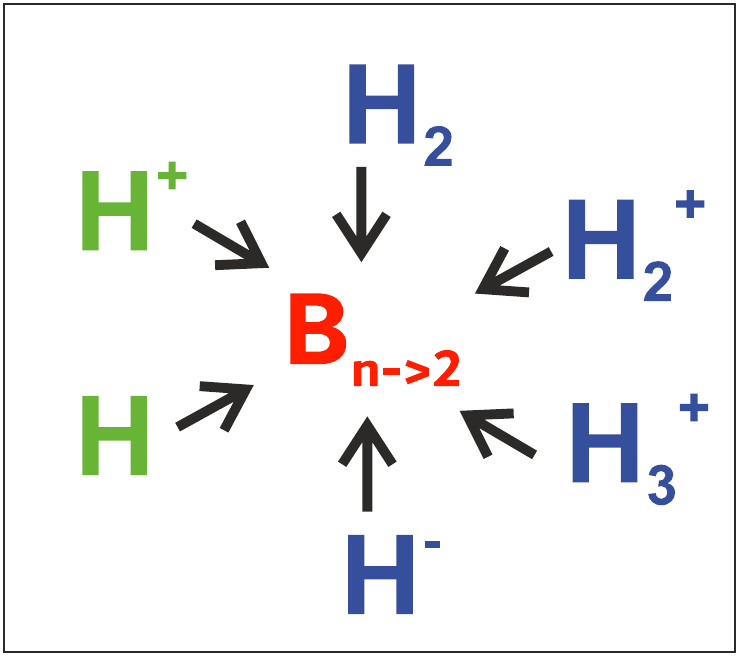}
    \caption{Schematic overview of the various reaction channels resulting in hydrogenic atomic Balmer line emission, adopted from \cite{Wuenderlich2016}. Blue indicates contributions related to $H_2$ plasma chemistry, green indicates purely "atomic" interactions and red indicates the total emission.}
    \label{fig:emiss_pathways}
\end{figure}

The goal of our analysis technique BaSPMI is to quantify the contribution of plasma-molecule interactions to $H\alpha$ \footnote{In this work we denote $H$ as hydrogen since most of the atomic/molecular data is available for hydrogen. In experiments, however, generally deuterium is used - more information can be found in section \ref{ch:discussion_isotope}} and use this to provide quantitative estimates on the influence of molecules on power losses; particle (ion) sources/sinks and Balmer line emission. A schematic overview of the contribution of the various plasma-atom and plasma-molecule interactions to excited hydrogen neutrals (which emit hydrogenic line emission) are shown in \ref{fig:emiss_pathways}. 

%The problem with this, however, is that the \emph{atomic} contribution to the $H\alpha, H\beta$ emissivities is estimated using medium-n Balmer lines (n=5,6) \emph{assuming that all medium-n Balmer line emission is due to atomic processes only}. That, however, may not be true. Using the 'molecular' estimate of the $H\alpha$ emission, we propagate that molecular contribution to the medium-n Balmer lines in an iterative fashion (section \ref{ch:analysis_ContaminIterate}), making the analysis self-consistent. 

\begin{figure}[H]
    \centering
    \includegraphics[width=\linewidth]{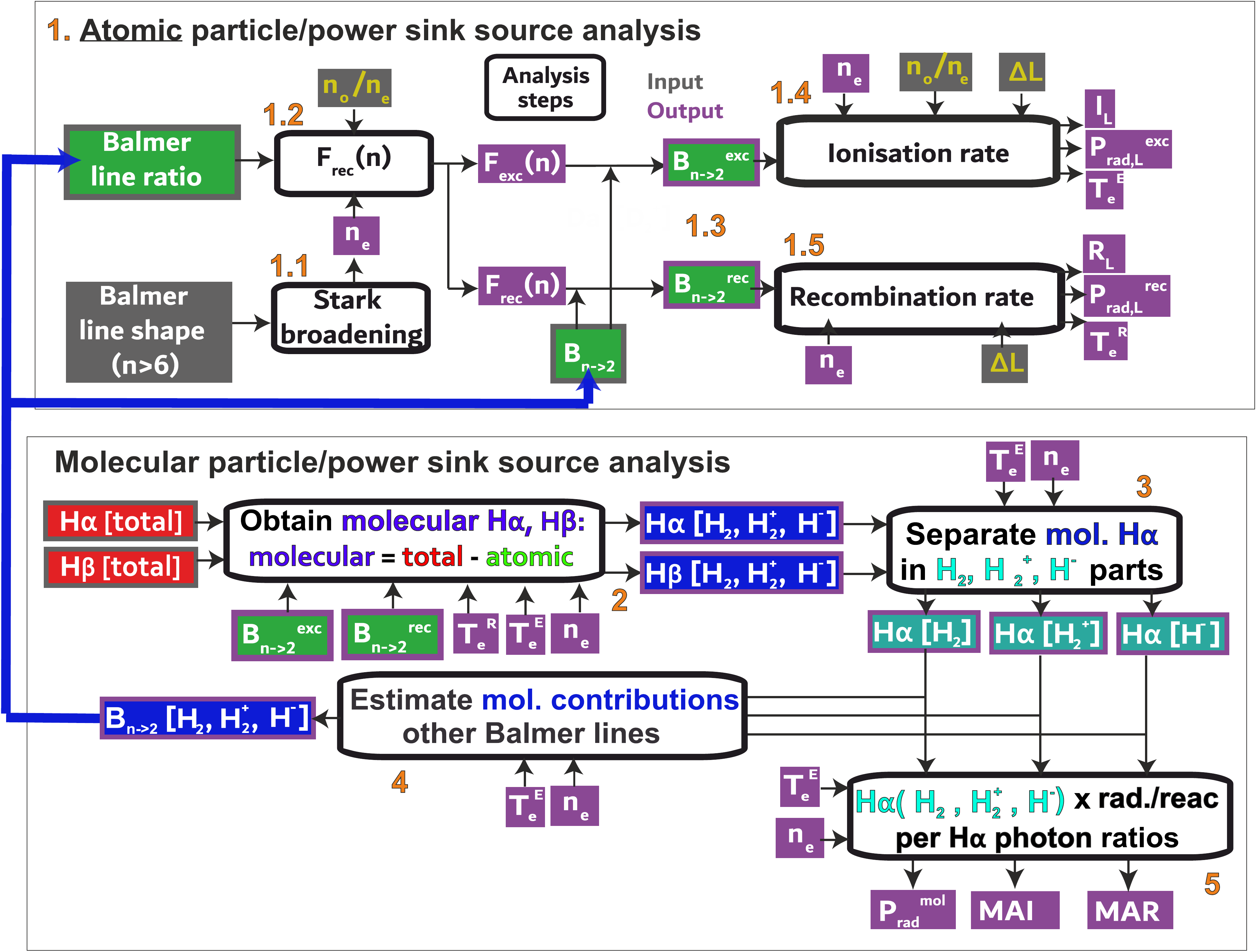}
    \caption{Schematic overview of the full analysis routine. The atomic analysis part has been adopted from \cite{Verhaegh2019a}. The orange numbers indicate the sequence of the various steps. The nomenclature used is adopted from \cite{Verhaegh2019a}: $n_o/n_e$ - neutral fraction; $\Delta L$ - emission pathlength; $n_e$ - electron density (from Stark broadening \cite{Verhaegh2019a}); $T_e^E$ - estimated electron temperature in the excitation region; $T_e^R$ - estimated electron temperature in the recombination region; $B_{n\rightarrow2}^{exc}$ - atomic Balmer line emission due to excitation; $B_{n\rightarrow2}^{rec}$ - atomic Balmer line emission due to recombination. The steps within the blocks 'Obtain molecular $H\alpha$'; 'Separate mol. $H\alpha$'; 'Estimate mol. contributions'; '$H\alpha (H_2, H_2^+, H^-)$ x rad./reac per $H\alpha$ photon ratios' are shown in more detailed in figures \ref{fig:dalpha_sepa}, \ref{fig:dalpha_sepa_mol}, \ref{fig:dalpha_extrap}, \ref{fig:dalpha_rates} respectively.}
    \label{fig:schem_analysis}
\end{figure}

The analysis developed in this work builds upon the Balmer line analysis techniques developed previously by the authors in \cite{Verhaegh2019a}, of which we provide a summary in section \ref{ch:analysis_atomic}. For the analysis we utilise the measurements of $H\alpha$, $H\beta$ in addition to two medium-n Balmer lines (n=5,6,7) \cite{Verhaegh2019a}. The analysis works on the basis of assigning all \emph{measured} Balmer line emission to the sum of 1) the \emph{expected} Balmer line emission from plasma-atom interactions (involving $H, H^+$) and 2) $H_2$ plasma chemistry related contributions (involving $H_2, H_2^+, H^-$). Contributions from $H_3^+$ are ignored since our post-processing (in section \ref{ch:analysis_solps}) indicates that its contribution to the Balmer line emission is negligible ($\ll0.1 \%$). A flowchart of the analysis scheme is provided in figure \ref{fig:schem_analysis} and consists of several steps.

\begin{enumerate}[label*=\arabic*.]
    \item We apply the analysis technique from \cite{Verhaegh2019a} on the medium-n Balmer lines, \emph{which considers only atomic processes} (e.g. electron-impact excitation of $H$ and electron-ion recombination of $H^+$). Initially we attribute all medium-n Balmer line emission to only atomic processes. The analysis from \cite{Verhaegh2019a} consists of several sub-steps and more information can be found in section \ref{ch:analysis_atomic}:
    \begin{enumerate}[label*=\arabic*.]
        \item We infer the electron density from the Balmer line shape through Stark broadening \cite{Verhaegh2019a,Verhaegh2018}.
        \item The fraction of the medium-n Balmer line ratio due to electron-impact excitation $F_{exc} (n)$ and electron-ion recombination $F_{rec} (n) = 1 - F_{exc} (n)$ is determined from the ratio of two medium-n Balmer lines. This uses an assumed possible range of neutral fractions $n_o/n_e$. 
        \item These fractions are multiplied with the measured medium-n Balmer line brightness to obtain the Balmer line brightnesses due to electron-impact excitation ($B_{n\rightarrow2}^{exc}$) and electron-ion recombination ($B_{n\rightarrow2}^{rec}$).
        \item The (line-integrated) ionisation rate $I_L$, radiative power loss due to electron-impact excitation $P_{rad,L}^{exc}$ and respective excitation region temperature $T_e^E$ is estimated from $B_{n\rightarrow2}^{exc}$ using an assumed range of possible neutral fractions $n_o/n_e$ and pathlengths $\Delta L$.
        \item The (line-integrated) recombination rate $R_L$, radiative power loss due to electron-ion recombination $P_{rad,L}^{rec}$ and respective recombination region temperature $T_e^R$ is estimated from $B_{n\rightarrow2}^{rec}$ using an assumed range of possible pathlengths $\Delta L$.
    \end{enumerate}
    \item The \emph{sum of the contributions} of $H_2$ plasma chemistry (involving $H_2, H_2^+$ and $H^-$) to $H\alpha$ and $H\beta$ are estimated using the measured $H\alpha, H\beta$ brightnesses and outputs from the "atomic particle/power sink/source analysis" as will be explained in section \ref{ch:analysis_DaMol}.
    \item The \emph{individual contributions} ($H_2, H_2^+$ and $H^-$) of plasma-chemistry to $H\alpha$ are \emph{separated} using the ratio between the sum of those contributions to $H\alpha$ and $H\beta$ as will be explained in section \ref{ch:analysis_DaMolSepa}.
    \item The individual contributions of $H_2$ plasma-chemistry to $H\alpha$ are used to estimate the individual contributions of $H_2$ plasma-chemistry to the medium-n Balmer line as will be explained in section \ref{ch:analysis_ContaminIterate}. This information is used to modify the atomic process contributions to the medium-n Balmer line brightnesses in step 1, which is then iterated up until step 4 until a converged result is obtained.
    \item After a converged result is obtained, the individual contributions of $H\alpha$ associated with $H_2$ plasma-chemistry are used to estimate (line-integrated) MAI ion sources, MAR ion sinks as well as the (line-integrated) radiated power due to excited atoms after plasma-molecule reactions involving $H_2$, $H_2^+$ and $H^-$ - $P_{rad,L}^{mol}$. This will be explained in sections \ref{ch:analysis_molMARMAI} and \ref{ch:analysis_molrad}. 
\end{enumerate}

There are two different versions of the analysis we can apply, ranging in complexity: 1) a 'simple' version: include only 'atomic' emission channels for the medium-n Balmer lines (e.g. no iteration applied) and optionally estimate the molecular component of $H\alpha$ and assume this is purely due to $H_2^+$ to obtain MAR/MAI/radiative loss rates (see section \ref{ch:Dalpha_mon}); 2) the complex 'full' version, which does apply the iterative technique and separates $H\alpha$ into its $H_2, H_2^+, H^-$ contributions. We have applied the 'full' version to the results unless otherwise specified.

%The analysis techniques are verified utilising a synthetic diagnostic approach, similar to \cite{Verhaegh2019a,Verhaegh2018}, in section \ref{ch:analysis_syndiagver}. This is performed for both TCV and MAST-U simulations.

We summarise the reactions on which BaSPMI provides estimates, in terms of radiative loss and particle sinks/sources in table \ref{tab:reaction_outputs}. Note that this table is not an overview of all the important plasma-molecule interactions. Most notably, the table does not contain the reactions where $H_2^+$ \& $H^-$ are being 'created' as these do not directly lead to Balmer line emission (but the destruction of these species, as shown in table \ref{tab:reaction_outputs}, does).

\begin{table}[]
\begin{tabularx}{\linewidth}{lllX}
\footnotesize{Reactions}          & \footnotesize{Ion bal.} & \footnotesize{Emission} & \footnotesize{Comment}                                 \\
$e^- + H \rightarrow e^- + H$               & N/A              & \checkmark             & \footnotesize{Electron impact excitation (of $H$)}                     \\
$e^- + H \rightarrow 2 e^- + H^+$           & \checkmark                & N/A            & \footnotesize{Ionisation}                              \\
$e^- + H^+ \rightarrow H$                   & \checkmark                & \checkmark             & \footnotesize{(Radiative) Electron-Ion Recombination (of $H^+$) - EIR}  \\
$2 e^- + H^+ \rightarrow e^- + H$           & \checkmark                & \checkmark             & \footnotesize{(Three body) Electron-Ion Recombination (of $H^+$) - EIR} \\
$e^- + H_2 \rightarrow e^- + H + H$         & N/A              & \checkmark             & \footnotesize{Dissociation}                            \\
$e^- + H_2 \rightarrow 2 e^- + H + H^+$     & \checkmark                & \checkmark             & \footnotesize{Electron impact dissociative ionisation (part of MAI chain)}    \\
$e^- + H_2^+ \rightarrow 2 e^- + H^+ + H^+$ & \checkmark                & N/A           & \footnotesize{Electron impact dissociative ionisation (MAI)}  \\
$e^- + H_2^+ \rightarrow H + H^+ + e^-$           & N/A              & \checkmark             & \footnotesize{Dissociation (part of MAD (or MAI\footnotemark) chain)}  \\
$e^- + H_2^+ \rightarrow H + H$             & \checkmark                & \checkmark             & \footnotesize{Dissociative recombination (part of MAR (or MAD\footnotemark[\value{footnote}]) chain)} \\
$H^+ + H^- \rightarrow H + H^+ + e^-$         & N/A                & \checkmark             & \footnotesize{Proton impact ionisation (part of MAD chain)}  \\
$H^+ + H^- \rightarrow H + H$               & \checkmark                & \checkmark             & \footnotesize{Mutual neutralisation (part of MAR chain)}
\end{tabularx}
    \caption{Overview of the various reactions on which the analysis provides information in terms of particle ($H^+$ ion) balance (bal.) and radiative power loss (radiation). If the analysis provides information on it, it is denoted with a '\checkmark' (whereas N/A implies not applicable). MAR/MAI/MAD mean Molecular Activated Recombination/Ionisation/Dissociation}
\label{tab:reaction_outputs}
\end{table}

\footnotetext{Whether interactions with $H_2^+$ are part of a MAR, MAD or MAI chain depends on the reaction process which \emph{created} $H_2^+$ (e.g. whether it is molecular charge exchange $H^+ + H_2 \rightarrow H_2^+ + H$ or $H_2$ ionisation $e^- + H_2 \rightarrow 2 e^- + H_2^+$). This is explained in section \ref{ch:analysis_molMARMAI}}

%Important processes which play a role in the creation of $H_2^$ is molecular 'charge exchange' or 'ion conversion': $H_2 + H^+ \rightarrow H_2^+ + H$ and for $H^-$ is $H_2 + e^- \rightarrow H^- + H$ (both reactions depend on the distribution of the $H_2$ vibrational states). 

\subsection{Atomic Balmer line analysis and analysis framework}
\label{ch:analysis_atomic}

The basic steps of the atomic Balmer line analysis technique of the upper block of figure \ref{fig:schem_analysis} were discussed above and some important additional details are discussed below here. More information can be found in \cite{Verhaegh2019a}. %This analysis technique, presented schematically in the first block of figure \ref{fig:schem_analysis}, first starts with the measurement of a Balmer line shape, from which the Stark density $n_e$ is extracted using techniques from \cite{Verhaegh2017,Verhaegh2019a,Verhaegh2018}. Secondly, a Balmer line ratio, typically n=6/n=5, together with $n_e$ and a neutral fraction ($n_o/n_e$) estimate, is utilised to infer the fraction of emission of a particular Balmer line ($B_{n\rightarrow 2}$) due to recombination ($F_{rec} (n)$) and excitation ($F_{exc} = 1 - F_{rec} (n)$). Those fractions are then used to split the Balmer line brightness into an excitation  ($B_{n\rightarrow2}^{exc} = (1 - F_{rec} (n)) \times B_{n\rightarrow2}$) and recombination ($B_{n\rightarrow2}^{rec} = F_{rec} (n) \times B_{n\rightarrow2}$) component. These excitation/recombination brightnesses are analysed separately, together with assumptions on $n_o/n_e$ and the characteristic pathlength of the emission $\Delta L$, to provide ionisation/recombination estimates, as well as estimates of the characteristic temperatures of the excitation and recombination regions along the line of sight ($T_e^E, T_e^R$ respectively). 

In this analysis the emission is modelled using a collisional-radiative model by a 0D 'semi slab-like' plasma model. Here the emission region has a pathlength (e.g. width) $\Delta L$, and an electron density $n_e$, while a different temperature is ascribed to the electron-impact excitation (of $H$) - $T_e^E$ and electron-ion recombination (of $H^+$) - $T_e^R$ regions (essentially a 'dual slab' model). For simplicity, this model assumes that the $H^+$ density equals the electron density ($n_{H^+} = n_e$, ignoring impurities); which is expected to have a negligible impact \cite{Verhaegh2017,Verhaegh2018,Verhaegh2019a} on this analysis. The emission for the excitation/recombination region is determined using results from collisional radiative modelling from ADAS \cite{Summers2006,OMullane} in the form of Photon Emission Coefficients (PECs - photons $m^3 s^{-1}$). The PEC is defined as the population coefficient ($\frac{n_p}{n_e n_{0}}$ where $n_p$ is the population density of the $p$ state and $n_{0} = \sum_p n_p$ is the total density of the emitter (sum of the population densities)) multiplied with the respective Einstein coefficient $A_{pq}$ for a $p\rightarrow q$ transition: $PEC (p,q) = A_{pq} \frac{n_p}{n_e n_{ground}}$ \cite{Summers2006,OMullane}.

%Not all of the required parameters of the analysis are well known or can be directly measured; instead the range for them has to be estimated. Estimates of the neutral fraction range are obtained from divertor modelling, whereas the path-length is estimated from the width of the ion target flux profile measured by Langmuir probes. Further information on these estimates can be found in \cite{Verhaegh2019a,Verhaegh2018}.

All the analysis shown in this work is done in a 'probabilistic' manner, which is also employed for all plasma-molecule interaction related estimates \cite{Verhaegh2019a}. For \emph{each} input parameter in figure \ref{fig:schem_analysis}, depending on their uncertainty, a 'Probability Density Function' (PDF) is ascribed. The peak of this parameter corresponds to the measured input parameter, whereas its width and shape corresponds to the expected uncertainty of this parameter. According to those PDFs, samples of input values for each parameter in figure \ref{fig:schem_analysis} are obtained through Monte Carlo sampling. These are then propagated to the output parameters, yielding a PDF for the output parameters from which the estimates and their uncertainties are obtained. %For each set of samples, the output parameters in figure \ref{fig:schem_analysis} are determined which, using non-parametric Kernel Density Estimates \cite{Botev2010}, can be mapped to a PDF for each \emph{output} parameter, providing us with information on the most likely estimate (peak of the PDF) as well as its uncertainty. More information on this technique can be found in \cite{Verhaegh2019a,Verhaegh2018}.

The full atomic \& molecular analysis requires implementing $H\alpha$ \& $H\beta$ brightnesses in addition to the two medium-n Balmer lines used in the atomic analysis in \cite{Verhaegh2019a}. This required modification to the PDF description of the relative brightnesses with respect to \cite{Verhaegh2019a}, which has to be similar for all possible line ratios. This was achieved using multivariate normal distributions with a set correlation strength according to \cite{Hinkley1969}; which leads to normal distributions for all the various line ratios ($\sigma = 0.075$) as well as the line intensities ($\sigma = 0.15$) \cite{Hinkley1969}. In addition to \cite{Verhaegh2019a}, we have also included random, uncorrelated, uncertainties in both the atomic and molecular collisional radiative model coefficients (e.g.PECs and reaction rates from ADAS \cite{Summers2006,OMullane}, Yacora \cite{Wuenderlich2016,Wunderlich2020} and AMJUEL \cite{Kotov2007,Reiter2008,Reiter2005}); which are parameterised by uniform probability density functions. For the atomic rates/emission coefficients an uncertainty of 12.5\% is assumed; while this is assumed to be 25\% for the molecular related coefficients.

\subsection{Inferring molecular contributions to  $H\alpha$ emission}
\label{ch:analysis_DaMol}

\begin{figure}[htb]
    \centering
    \includegraphics[width=0.8\linewidth]{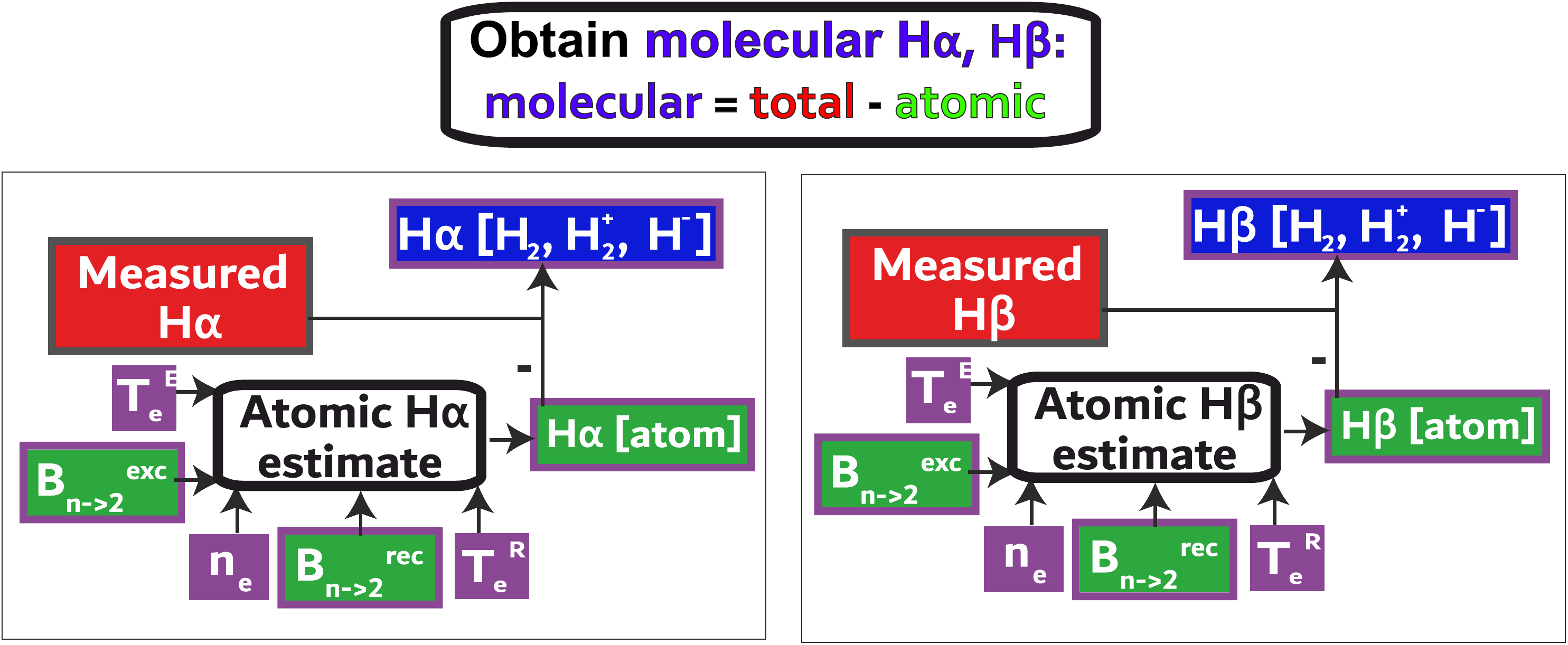}
    \caption{Schematic analysis flow chart for separating the atomic and molecular contributions from the $H\alpha$ and $H\beta$ emission. This represents step 2 'Obtain molecular $H\alpha, H\beta$' in figure \ref{fig:schem_analysis}.}
    \label{fig:dalpha_sepa}
\end{figure}

After the medium-n Balmer lines are analysed from the viewpoint of "atomic" interactions, those results are used with measured $H\alpha, H\beta$ brightnesses to estimate the contribution of $H_2$ plasma chemistry to $H\alpha$ and $H\beta$, which is illustrated in figure \ref{fig:dalpha_sepa}, which is step 2 in figure \ref{fig:schem_analysis}.

This is achieved by assuming that the \emph{total measured $H\alpha$, $H\beta$} ($B_{3,4\rightarrow2}^{total}$) equals its \emph{atomic} part ($B_{3,4\rightarrow2}^{atom}$) plus its \emph{molecular} part ($B_{3,4\rightarrow2}^{H_2, H_2^+, H^-}$) - as shown in equation \ref{eq:DaMol}. That assumption is further discussed for TCV in section \ref{ch:OtherDa_Emiss}. 

\begin{equation}
    B_{3,4\rightarrow2}^{H_2, H_2^+, H^-} = B_{3,4\rightarrow2}^{tot, measured} - B_{3,4\rightarrow2}^{atom}
    \label{eq:DaMol}
\end{equation}

The output information from the \emph{atomic} analysis of the medium-n Balmer lines (figure \ref{fig:schem_analysis}) is utilised to \emph{extrapolate} the \emph{atomic parts} of the medium-n Balmer line brightnesses of a Balmer line (typically $n=5,6,7$) to $H\alpha$ and $H\beta$, yielding the \emph{atomic parts} of the $H\alpha$ and $H\beta$ brightnesses. Utilising the recombination/excitation inferred temperatures ($T_e^E$, $T_e^R$) and the Stark inferred density ($n_e$), the individual excitation ($B_{n\rightarrow2}^{exc}$) and recombination ($B_{n\rightarrow2}^{rec}$) medium-n Balmer line brightnesses are extrapolated to $H\alpha$ and $H\beta$ ($B_{3,4\rightarrow2}^{atom,extrapolated}$) as shown in equation \ref{eq:DaUp} and schematically in figure \ref{fig:dalpha_sepa}, \cite{Verhaegh2019a}. %To reduce uncertainties, $n_1$ corresponds to the lower-n and $n_2$ corresponds to the upper-n Balmer line used in the atomic analysis.

\begin{equation}
    B_{3,4\rightarrow2}^{atom, extrapolated} = B_{n\rightarrow2}^{exc} \frac{PEC_{3\rightarrow2}^{exc} (n_e, T_e^E)}{PEC_{n\rightarrow2}^{exc} (n_e, T_e^E)} + B_{n\rightarrow2}^{rec} \frac{PEC_{3\rightarrow2}^{rec} (n_e, T_e^R)}{PEC_{n\rightarrow2}^{rec} (n_e, T_e^R)} 
    \label{eq:DaUp}
\end{equation}

% will be multiplied with a 'reactions per $H\alpha$ (molecular) photon' (or a 'radiation per $H\alpha$ (molecular) photon') ratio and to obtain estimates of radiative power loss due to plasma-molecule collisions.

The extrapolated 'atomic' contribution to  $H\alpha$ and $H\beta$ is then subtracted from the total measured $H\alpha$ and $H\beta$ brightnesses (equation \ref{eq:DaMol}, figure \ref{fig:dalpha_sepa}) to estimate the $H\alpha$ and $H\beta$ brightnesses associated with $H_2$ plasma. 

\subsection{Separating multiple molecular contributions to $H\alpha$ emission}
\label{ch:analysis_DaMolSepa}

\begin{figure}[htb]
    \centering
    \includegraphics[width=0.5\linewidth]{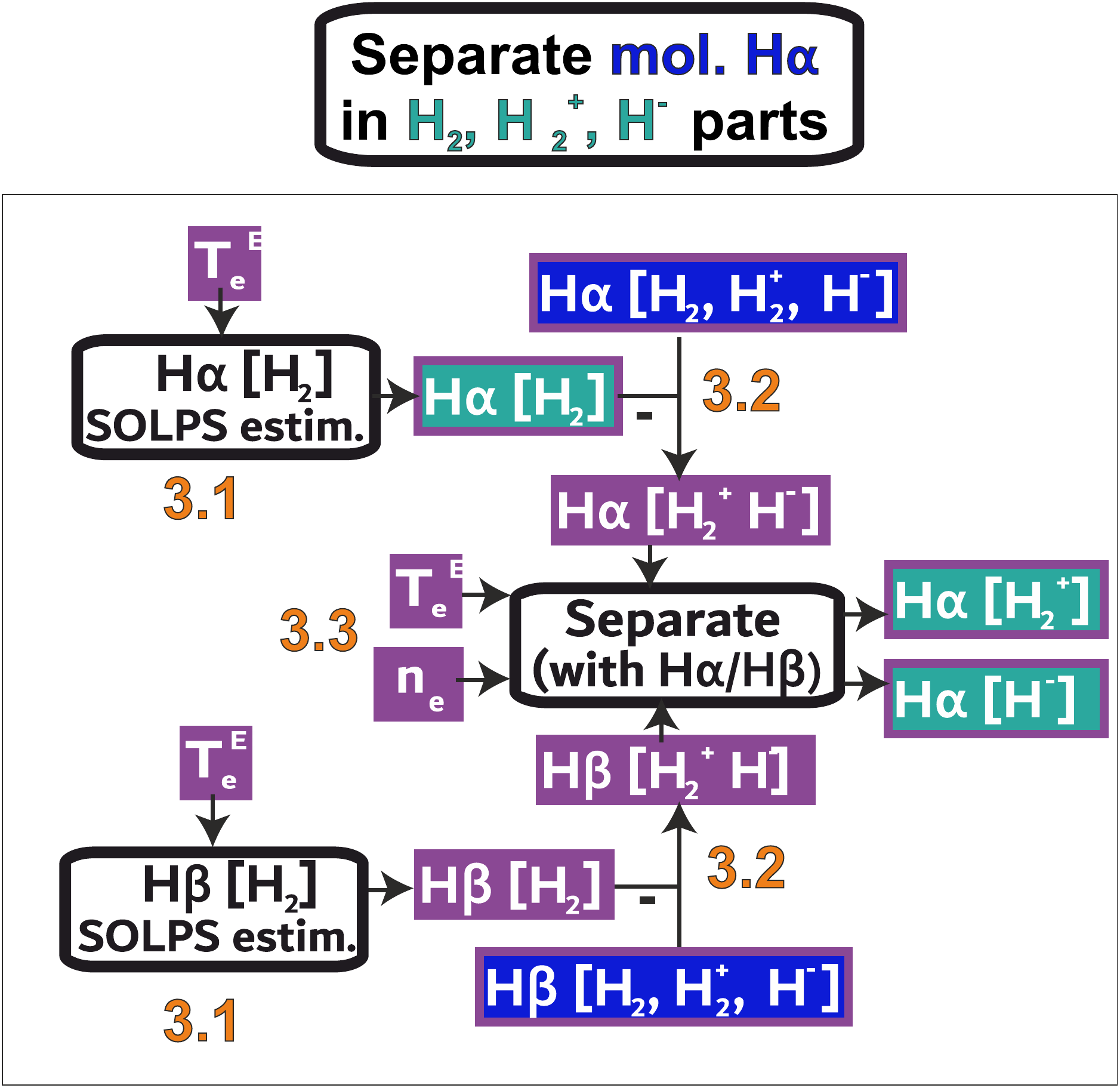}
    \caption{Schematic analysis flow chart for separating the various pathways of the molecular $H\alpha$ emission. This represents step 3 'Separate mol.$H\alpha$ in $H_2$, $H_2^+$, $H^-$ parts' in figure \ref{fig:schem_analysis}.}
    \label{fig:dalpha_sepa_mol}
\end{figure}

Now that we obtained an estimate for $B_{3,4\rightarrow2}^{H_2, H_2^+, H^-}$, we will separate the various contributions using the Balmer line emission model for $B_{3,4\rightarrow2}^{H_2, H_2^+, H^-}$ highlighted in section \ref{ch:analysis_BalmerEmissModel}. The steps for this are highlighted in figure \ref{fig:dalpha_sepa_mol} (which is step 3 in figure \ref{fig:schem_analysis}) and use the electron density and electron temperature obtained from Stark broadening and the atomic analysis:

\begin{enumerate}[label*=3.\arabic*.]
    \item The $H_2$ contribution of $H\alpha$ and $H\beta$ are estimated using an assumed relation between the $H_2$ density and the electron temperature, which is obtained from SOLPS simulations (more information is provided below and in section \ref{ch:fH2_Treatment}).
    \item This $H_2$ contribution is subtracted from the total $H\alpha$ and $H\beta$ brightnesses attributed to plasma-molecule interactions to obtain the $H\alpha$ and $H\beta$ brightnesses attributed to $H_2^+$ and $H^-$.
    \item The ratio of those $H\alpha$ and $H\beta$ brightnesses are used to separate the $H\alpha$ emission attributed to $H_2^+$ and $H^-$. 
\end{enumerate}

We have used SOLPS-ITER simulation results from TCV and MAST-U to establish a relation between the expected $H_2$ density times the pathlength $\Delta L$ and the (excitation) electron temperature \cite{Stangeby2017} - $g_{H_2} (T_e^E) \approx \Delta L n_{H_2}$, which is used to estimate the Balmer line brightnesses attributed to $H_2$ - $B_{n\rightarrow2}^{H_2}$ as shown in equation \ref{eq:DaH2}. After having estimated $B_{3,4 \rightarrow 2}^{H_2}$, this is used to estimate the $H\alpha$ and $H\beta$ emission attributed to $H_2^+$ and $H^-$: $B_{3,4 \rightarrow2}^{H_2^+} + B_{3,4 \rightarrow 2}^{H^-} = B_{3,4 \rightarrow 2}^{H_2, H_2^+, H^-} - B_{3,4 \rightarrow2}^{H_2}$.

\begin{equation}
    B_{n\rightarrow2}^{H_2} = g_{H_2} (T_e^E) n_e PEC_{n\rightarrow2}^{H_2} (n_e, T_e^E)
    \label{eq:DaH2}
\end{equation}

\begin{figure}[htb]
    \centering
    \includegraphics[width=0.8\linewidth]{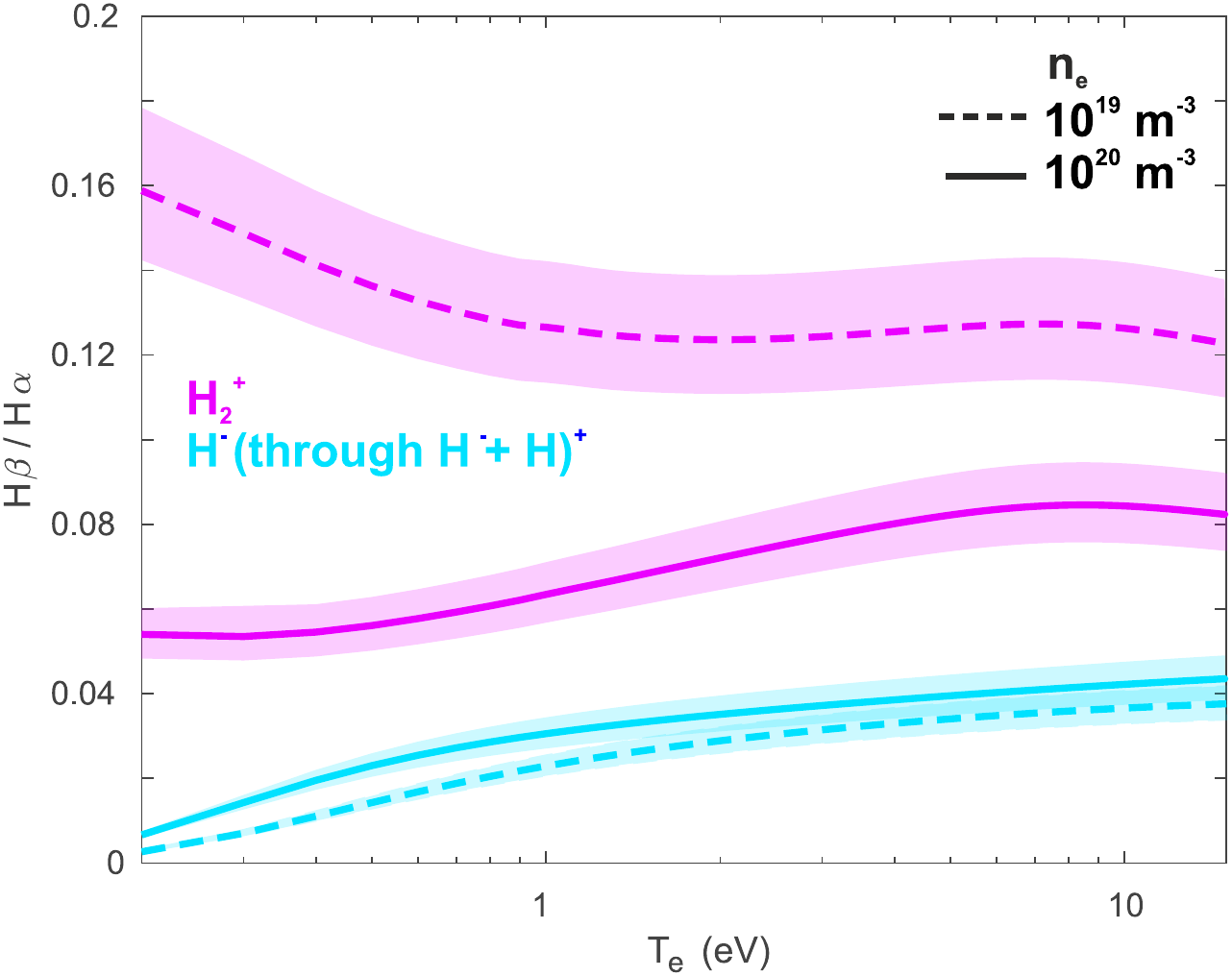}
    \caption{$H\beta/H\alpha$ ratio using YACORA PEC coefficients for $H_2^+$ and $H^- + H^+$ for two different electron densities as function of $T_e$ with shaded uncertainties (based on the assumed 25 \% uncertainty for molecular coefficients).}
    \label{fig:RatHmHp_RatH2p}
\end{figure}

 Plasma-molecule interactions involving $H_2^+$ and $H^-$ lead to different $H\beta/H\alpha$ ratios as shown in figure \ref{fig:RatHmHp_RatH2p}, which is calculated using data from Yacora (on the Web) \cite{Wuenderlich2016,Wunderlich2020}. This distinction can be used to quantitatively separate emission contributions from $H^-$ and $H_2^+$ using equation \ref{eq:DaSplit}, which can be readily obtained when the $B_{3,4\rightarrow2}^{H_2^+,H^-}$ brightnesses are expressed using a plasma-slab model (equation \ref{eq:BalMolSimpl}). We use the Stark broadening derived electron density $n_e$ and the electron impact excitation emission derived temperature ($T_e^E$) to interrogate the required PECs as this is a more reliable overall temperature (with uncertainty) indicator of the plasma - especially for a hotter plasma \cite{Verhaegh2019a} - which ultimately is important for MAI estimates (section \ref{ch:analysis_molMARMAI}). However, using the electron-ion recombination derived temperature instead for any of the molecular estimates would not change any of the obtained conclusions in the tested conditions, apart from reducing MAI rates.  %Before we can do this, however, we must first obtain the molecular contributions of $H\alpha$ and $H\beta$ (see section \ref{ch:analysis_DaMol}). Afterwards, we must subtract the contribution of $H_2$ to $H\alpha$ and $H\beta$, see figure \ref{fig:dalpha_sepa_mol}. The contribution of $H_2$ to the 'molecular' part of a Balmer line ($\frac{B_{n\rightarrow2}^{H_2}}{B_{n\rightarrow2}^{H_2, H_2^+, H^-}}$) is found to be minor/negligible. It is included in the calculation for all Balmer lines, however, and our treatment for estimating a possible range of $B_{n\rightarrow2}^{H_2}$ is described in appendix \ref{ch:fH2_Treatment}. 

%Schematically, our approach for this, highlighted previously in figure \ref{fig:schem_analysis}, is further expanded schematically in figure \ref{fig:dalpha_sepa_mol}, which we will discuss throughout this section.

%After subtracting this $H_2$ contribution from both $H\alpha$ and $H\beta$, we are left with estimates of the $H\alpha$ and $H\beta$ emission due to $H_2^+$ and $H^-$: $B_{3,4 \rightarrow2} + B_{3,4 \rightarrow 2} = B_{3,4 \rightarrow 2}^{H_2, H_2^+, H^-} - B_{3,4 \rightarrow2}^{H_2}$. Using the ratio of those separated contributions $\left.(H\beta/H\alpha)\right|_{H^-, H_2^+} \equiv \frac{B_{4\rightarrow2}^{H_2^+} + B_{4\rightarrow2}^{H^-}}{B_{3\rightarrow2}^{H_2^+} + B_{3\rightarrow2}^{H^-}}$, we can quantitatively separate the emission due to $H_2^+$ and $H^-$ as illustrated in equation \ref{eq:DaSplit}, in which we have used the Stark density and the excitation temperature to evaluate the required PECs. We have chosen the excitation temperature for this as this is a more reliable overal temperature (with uncertainty) indicator of the plasma - especially for a hotter plasma \cite{Verhaegh2019a} - which ultimately is important for MAI estimates (section \ref{ch:analysis_molMARMAI}). However, using the recombination temperature instead for any of the molecular estimates would not change any of the obtained conclusions from the analysis.

\begin{equation}
\begin{split}
    f_{H_2^+, mol. H\alpha} &\equiv \frac{B_{3\rightarrow2}^{H_2^+}}{B_{3\rightarrow2}^{H_2^+} + B_{3\rightarrow2}^{H^-}} = \frac{1}{1 + C} \\
    C &=\frac{PEC_{3\rightarrow2}^{H^-} (n_e, T_e^E) \bigg[ PEC_{3\rightarrow2}^{H_2^+} (n_e, T_e^E)  \left.(H\beta/H\alpha)\right|_{H^-, H_2^+} - PEC_{4\rightarrow2}^{H_2^+}(n_e, T_e^E) \bigg]}{PEC_{4\rightarrow2}^{H^-}(n_e, T_e^E)  - PEC_{3\rightarrow2}^{H^-}(n_e, T_e^E)  \left.(H\beta/H\alpha)\right|_{H^-, H_2^+}}
    \end{split}
    \label{eq:DaSplit}
\end{equation}

Now we have all the information required to determine all the emission contributions to $H\alpha$, which are summarised in equation \ref{eq:DaSplitFull}.

\begin{equation}
\begin{split}
    B_{3\rightarrow2}^{atom, extrapolated} &= B_{n\rightarrow2}^{exc} \frac{PEC_{3\rightarrow2}^{exc} (n_e, T_e^E)}{PEC_{n\rightarrow2}^{exc} (n_e, T_e^E)} + B_{n\rightarrow2}^{rec} \frac{PEC_{3\rightarrow2}^{rec} (n_e, T_e^R)}{PEC_{n\rightarrow2}^{rec} (n_e, T_e^R)} \\
    {B_{3\rightarrow2}^{H_2, H_2^+, H^-}} &= B_{3\rightarrow2}^{tot, measured} - B_{3\rightarrow2}^{atom, extrapolated} \\
    {B_{3\rightarrow2}^{H_2}} &= g_{H_2} n_e (T_e^E) PEC_{n\rightarrow2}^{H_2} (n_e, T_e^E) \\
    {B_{3\rightarrow2}^{H_2^+}} &= (B_{3\rightarrow2}^{H_2, H_2^+, H^-} - B_{3\rightarrow2}^{H_2}) \times f_{H_2^+, mol. H\alpha} \\
     {B_{3\rightarrow2}^{H^-}} &= (B_{3\rightarrow2}^{H_2, H_2^+, H^-} - B_{3\rightarrow2}^{H_2}) \times (1-f_{H_2^+, mol. H\alpha}) 
    \label{eq:DaSplitFull}
\end{split}
\end{equation}

\subsection{Molecular contributions to $n>4$ Balmer line emission}
\label{ch:analysis_ContaminIterate}

\begin{figure}[htb]
    \centering
    \includegraphics[width=0.5\linewidth]{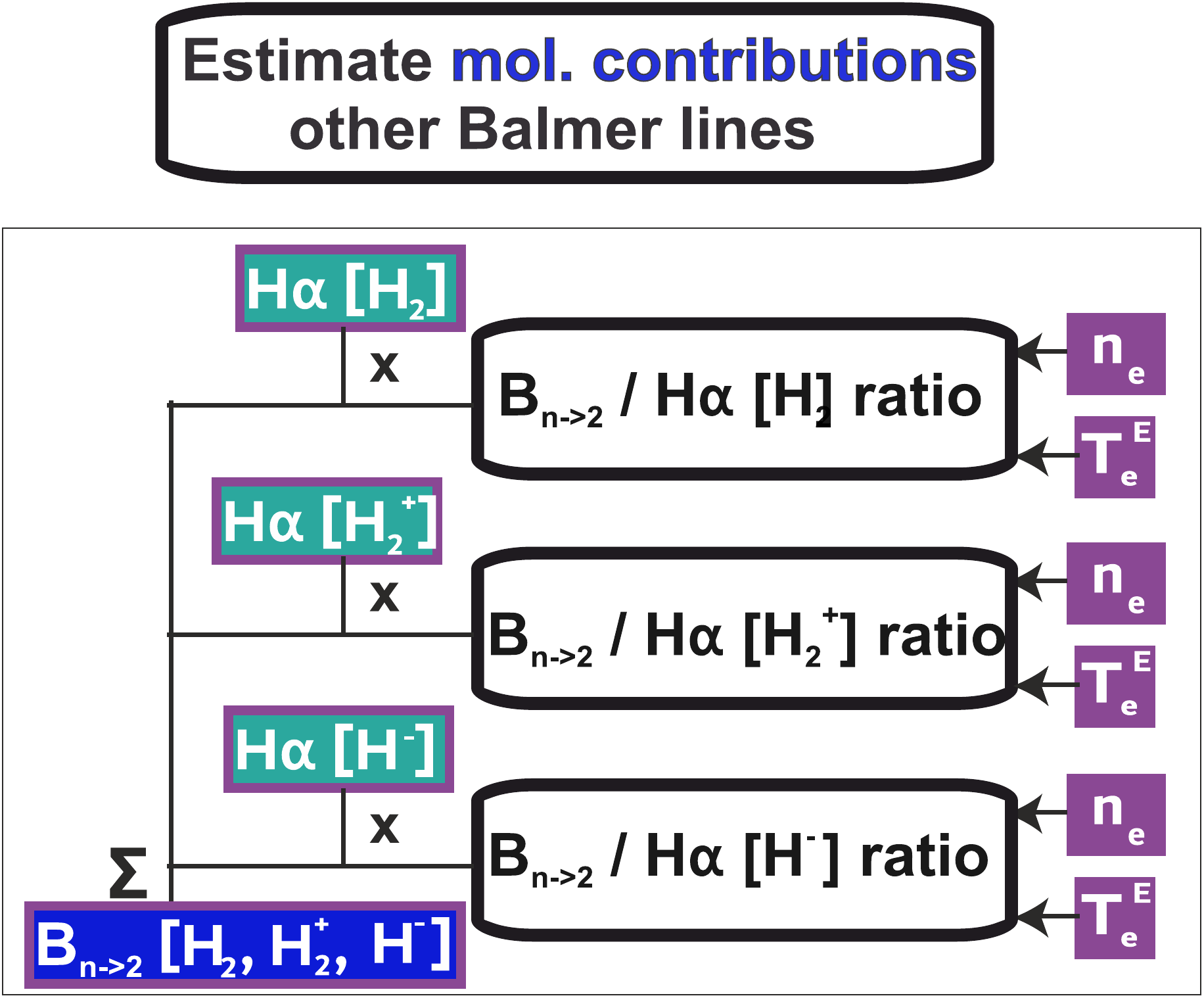}
    \caption{Schematic analysis flow chart for estimating the molecular contributions to the other Balmer lines based on the various $H\alpha$ 'molecular' emission channels. This represents step 4 'Separate mol. contributions other Balmer lines' in figure \ref{fig:schem_analysis}.}
    \label{fig:dalpha_extrap}
\end{figure}

Up until this point in the analysis, we have not taken into account that the medium-n Balmer lines can also be influenced by plasma-molecule interactions. Although plasma-molecule interactions predominantly impact $H\alpha$ and $H\beta$, the impact on the medium-n Balmer lines may not be fully negligible. We can account for this by enforcing consistency between the molecular contributions to $H\alpha$ \& $H\beta$ and the medium-n Balmer lines used in the atomic part of the analysis. 

This is achieved by first extrapolating $B_{3\rightarrow2}^{H_2, H_2^+, H^-}$ to the medium-n Balmer lines - $B_{n\rightarrow2}^{H_2, H_2^+, H^-}$ - figure \ref{fig:dalpha_extrap}, which represents step 4 in figure \ref{fig:schem_analysis}. This extrapolation is achieved by estimating the ratio between the medium-n Balmer lines and $H\alpha$ for $H_2, H_2^+, H^-$ separately and multiplying those ratios with the respective $B_{n\rightarrow2}^{H_2}, B_{n\rightarrow2}^{H_2^+}, B_{n\rightarrow2}^{H^-}$ brightnesses, which are summed to provide $B_{n\rightarrow2}^{H_2, H_2^+, H^-}$. 

Secondly, using $B_{n\rightarrow2}^{H_2, H_2^+, H^-}$ for the medium-n Balmer lines, the atomic contribution of the medium-n Balmer lines is estimated - $B_{n\rightarrow2}^{atom}$ (equation \ref{eq:MolContamin}). Here, it is assumed that the total measured medium-n Balmer line brightness is its atomic part plus its molecular part - which was used for $H\alpha$ in equation \ref{eq:DaMol}. 

With those updated $B_{n\rightarrow2}^{atom}$ estimates for the medium-n Balmer lines, the entire analysis is re-executed, yielding modified values for all estimates - including a new extrapolated $B_{n\rightarrow2}{H_2, H_2^+, H^-}$ for the medium-n Balmer lines. This is repeated iteratively (see appendix \ref{ch:analysis_atomic}) until these extrapolated brightnesses reach a converged value which has a fully self-consistent solution between $B_{3\rightarrow2}^{H_2, H_2^+, H^-}$ and $B_{n\rightarrow2}^{H_2, H_2^+, H^-}$.

\begin{equation}
\begin{split}
    B_{n\rightarrow2}^{tot, measured} &= B_{n\rightarrow2}^{H_2, H_2^+, H^-} + B_{n\rightarrow2}^{atom} \\
    B_{n\rightarrow2}^{atom} &= B_{n\rightarrow2}^{tot, measured} - {B_{n\rightarrow2}^{H_2}} - {B_{n\rightarrow2}^{H_2^+}} - {B_{n\rightarrow2}^{H^- + H^{+}}} \\
    B_{n\rightarrow2}^{atom} &= B_{n\rightarrow2}^{tot, measured} - B_{3\rightarrow2}^{H_2} \frac{PEC_{n\rightarrow2}^{H_2} (n_e, T_e^E)}{PEC_{3\rightarrow2}^{H_2} (n_e, T_e^E)} - B_{3\rightarrow2}^{H_2^+} \frac{PEC_{n\rightarrow2}^{H_2^+} (n_e, T_e^E)}{PEC_{3\rightarrow2}^{H_2^+} (n_e, T_e^E)} - \\ &B_{3\rightarrow2}^{H^-} \frac{PEC_{n\rightarrow2}^{H^- + H^+} (n_e, T_e^E)}{PEC_{3\rightarrow2}^{H^- + H^+} (n_e, T_e^E)} 
    \label{eq:MolContamin}
    \end{split}
\end{equation}

When we compare results with and without this iterative approach (section \ref{ch:Dalpha_mon}) we find that the $B_{3\rightarrow2}^{H_2, H_2^+, H^-}$ estimate is insensitive to this iteration considering its uncertainties. The iterative approach, however, is important for obtaining accurate estimates of the excitation emission component of the medium-n Balmer lines, which are important for ionisation estimates.  

%Using this approach in equation \ref{eq:MolContamin}, the atomic part of the Balmer line emission $n$ is estimated. Cases where the molecular estimation is larger than the total measured Balmer line emission are removed from the probabilistic data set. 
%The results of this equation are then fed back   %During this iteration, the "setting" of the Monte Carlo probabilistic iteration the same (e.g. if, for instance, if the emission according to the Monte Carlo iteration is underestimated by 10\% of Balmer line $n_1$ - this is kept the same throughout the iteration). 

\subsection{Inferring radiative losses and MAI/MAR from plasma-molecule interactions}

\begin{figure}[htb]
    \centering
    \includegraphics[width=\linewidth]{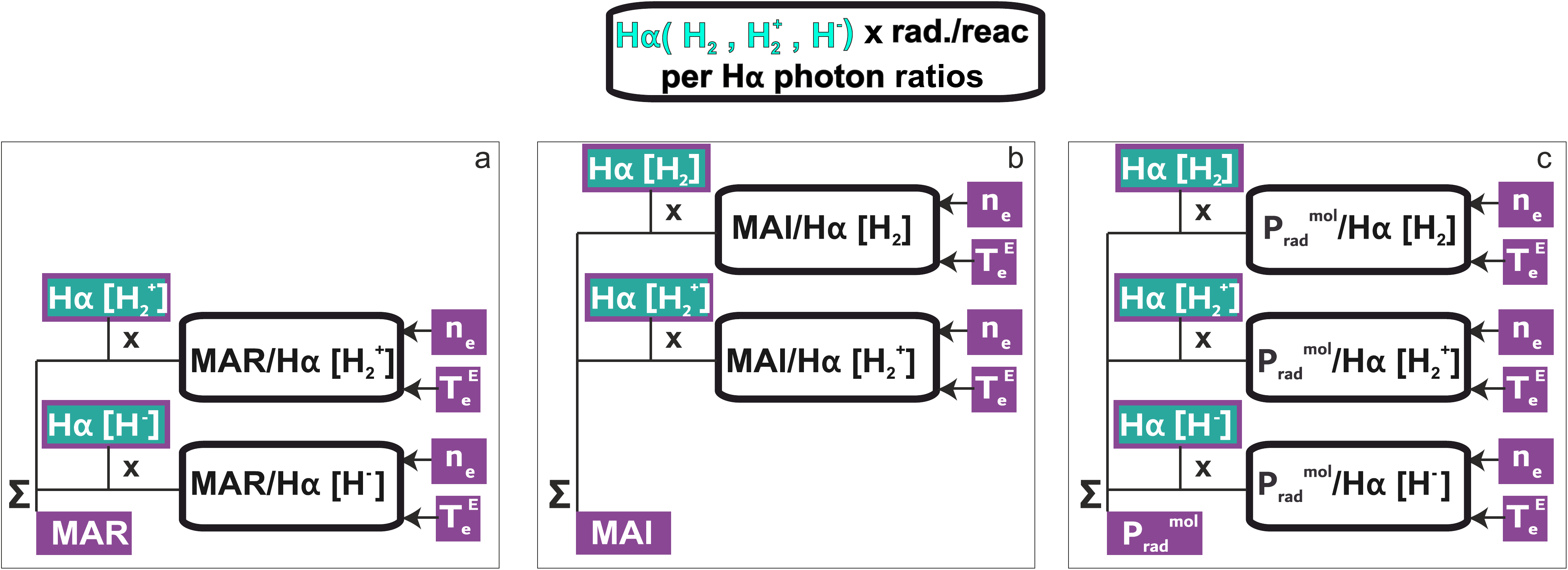}
    \caption{Schematic analysis flow chart for estimating MAR (a), MAI (b) and radiative loss rates (c) from the separated $H\alpha$ 'molecular' emission pathways.  This represents step 5 '$H\alpha (H_2, H_2^+, H^-)$ x rad./reac per $H\alpha$ photon ratios' in figure \ref{fig:schem_analysis}.}
    \label{fig:dalpha_rates}
\end{figure}

The separated brightnesses of $H\alpha$ are used to determine the various atomic reaction rates/power losses (as is done in \cite{Verhaegh2019a}), as well as the various MAR/MAI ion sinks/sources and hydrogenic radiative power losses (table \ref{tab:reaction_outputs}) related to excited atoms after plasma-molecule interaction. The analysis steps of this approach are shown schematically in figure \ref{fig:dalpha_rates}, which is step 5 in figure \ref{fig:schem_analysis}. For all the cases in figure \ref{fig:dalpha_rates}, the separated $H\alpha$ brightnesses are multiplied with the "effective radiative loss (figure \ref{fig:dalpha_rates} c) (or MAI/MAR reaction rate - figure \ref{fig:dalpha_rates} a/b) per emitted $H\alpha$ photon" using the Stark broadening inferred $n_e$ and electron-impact excitation derived $T_e^E$. This provides radiative losses (or MAI/MAR reaction rates) for each process, which are summed to provide the total hydrogenic radiative power loss estimates and MAR/MAI rates.

\subsubsection{Inferring hydrogenic line radiative losses from plasma-molecule interactions}
\label{ch:analysis_molrad}

Although $H\alpha$ emission does not lead to significant radiative losses directly, considering most plasma radiation is in the VUV \cite{McLean2019}, it can be an indicator for significant radiative losses. $H\alpha$ emission, corresponding to the $3\rightarrow2$ transition, directly implies also the presence of $Ly\beta$ ($3\rightarrow1$) emission. Utilising the associated Einstein coefficients and photon energies, 6.5 times more radiative loss arises due to $Ly\beta$ than $H\alpha$ (e.g. $\frac{E_{Ly\beta} \times {A_{31}/A_{32}}}{E_{H\alpha}} \approx 6.5$). Since $H\alpha$ indicates a transition to the $n=2$ excited state, the enhancement of $H\alpha$ should also lead to some enhancement of the $n=2$ excited state, which subsequently results in $Ly\alpha$ emission - which carries 5.8 times more energy than a $H\alpha$ photon. 

It is thus clear that, at a minimum, a power loss of the order of ten more than the power loss of $H\alpha$ itself is associated with related (V)UV emission. Since this only covers the influence of plasma-molecule interactions on the $n=3$ populated state, this is a conservative estimate of the radiative losses due to plasma-molecule interactions. For example, plasma-molecule interactions could potentially directly lead to an enhancement of the $n=2$ populated state, and thus directly enhance $Ly\alpha$ radiation losses.

It is important to repeat that the power loss estimated here is \emph{radiation from hydrogenic (atomic) emission lines} arising from \emph{excited atoms} after \emph{plasma-molecule interactions}. \emph{This is different} from radiative losses associated with molecular band emission which has been the subject of previous research \cite{Groth2019,McLean2019}, where the brightness of several molecular (Fulcher, Werner (VUV)) bands were measured and its associated radiative power loss was estimated to be negligible. Therefore, the atomic radiative losses from plasma-molecule interactions likely plays a dominant role in the radiative losses attributed to plasma-molecule interactions in detached plasmas.

To estimate radiative power losses due to plasma-molecule interactions, we utilise Yacora (on the Web) \cite{Wuenderlich2016,Wunderlich2020} to model the most dominant lines ($n<7$) of the atomic Balmer and Lyman spectra associated with plasma-molecule interactions. These are multiplied with their respective photon energies and summed to estimate the radiated hydrogenic (atomic) power loss due to excited atoms after plasma-molecule interaction. This power is then divided by the $H\alpha$ emission attributed to those channels, obtaining a ratio representing 'total radiated energy (eV) per $H\alpha$ photon' for each individual emission channel (equation \ref{eq:PradDaTOT}). We represent this as ($\frac{P_{rad, L}^{H_2, H_2^+, H^-}}{B_{3\rightarrow2}^{H_2, H_2^+, H^-}}$) where $P_{rad, L}^{H_2, H_2^+, H^-}$ is a line-integrated radiation rate in $W/m^2$, which can be determined by multiplying the respective brightness with the respective 'total radiation per $H\alpha$ photon' coefficient: $P_{rad, L}^{H_2, H_2^+, H^-} = B_{3\rightarrow2}^{H_2, H_2^+, H^-} \times \frac{P_{rad, L}^{H_2, H_2^+, H^-}}{B_{3\rightarrow2}^{H_2, H_2^+, H^-}}$. 

\begin{equation}
\begin{aligned}
    \frac{P_{rad, L}^{H_2}}{B_{3\rightarrow2}^{H_2}} &=\sum_{i=2,3,4,5,6} \sum_{j=1,2; i>j} \frac{PEC_{i\rightarrow j}^{H_2} (n_e, T_e)}{PEC_{3\rightarrow2}^{H_2} (n_e, T_e)} \\
    \frac{P_{rad, L}^{H_2^+}}{B_{3\rightarrow2}^{H_2^+}}&=\sum_{i=2,3,4,5,6} \sum_{j=1,2; i>j} \frac{PEC_{i\rightarrow j}^{H_2^+} (n_e, T_e)}{PEC_{3\rightarrow2}^{H_2^+} (n_e, T_e)} \\
    \frac{P_{rad, L}^{H^-}}{B_{3\rightarrow2}^{H^-}}&=\sum_{i=2,3,4,5} \sum_{j=1,2; i>j} \frac{PEC_{i\rightarrow j}^{H^- + H^+} (n_e, T_e)}{PEC_{3\rightarrow2}^{H^- + H^+} (n_e, T_e)}  \end{aligned}
    \label{eq:PradDaTOT}
\end{equation}

\begin{figure}[htb]
    \centering
    \includegraphics[width=0.7\linewidth]{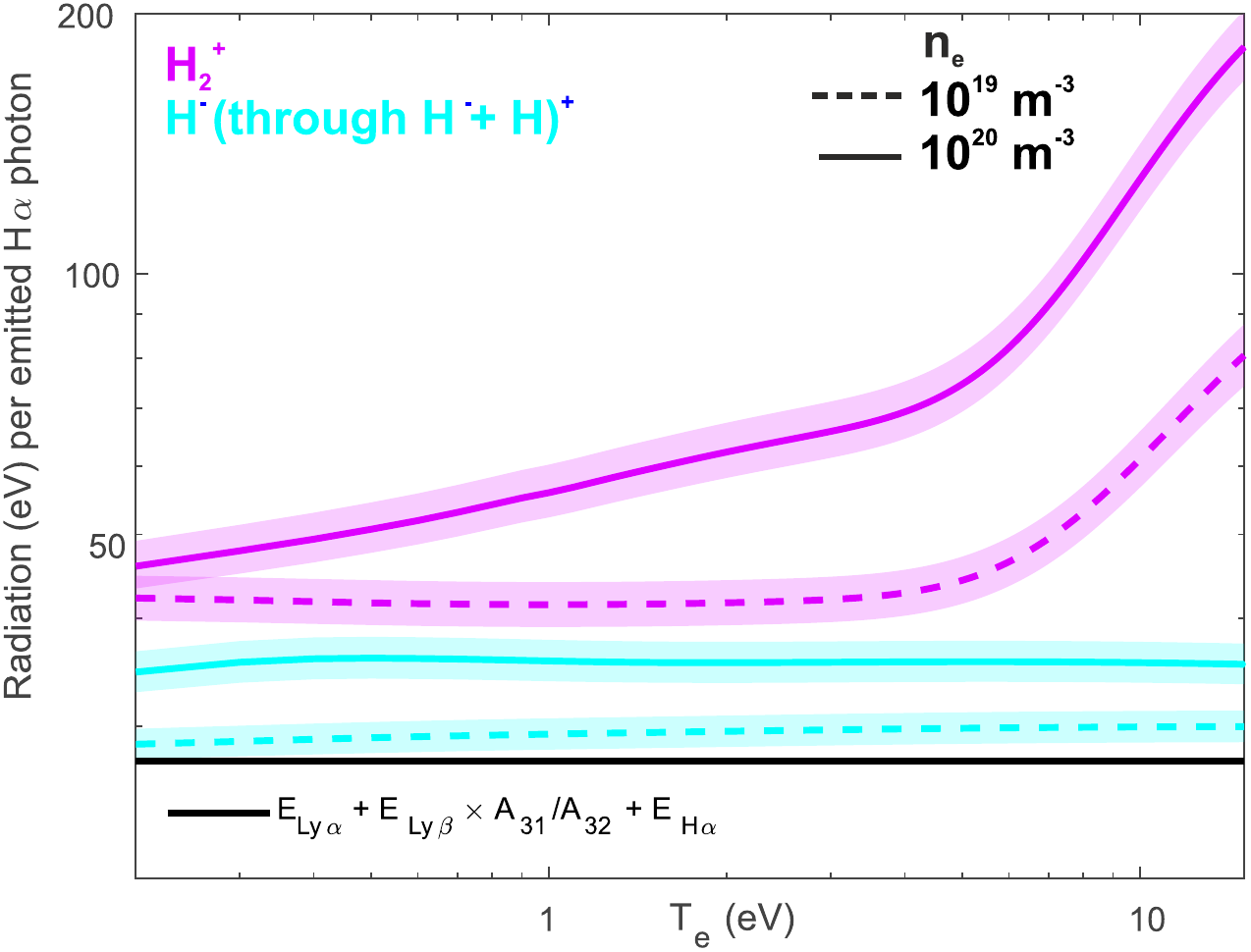}
    \caption{Radiative loss (eV) per emitted $H\alpha$ photon derived from molecular contributions associated with $H_2^+$ and $H^-$ (assuming excited $H$ atoms arising from plasma-molecule interactions involving $H^-$ comes from the $H^- + H^+$ reaction) at different electron densities. The uncertainty margins arise from the assumed 25 \% uncertainty in all PECs attributed to $H_2$ plasma chemistry. The black curve shows the roughly expected power loss directly explainable by the $H\alpha$ photon (e.g. if one would have the power loss of a $H\alpha$ photon ($3\rightarrow2$) plus associated $Ly\alpha$ ($2\rightarrow1$) and $Ly\beta$ photons ($3\rightarrow1$) per emitted $H\alpha$ photon).}
    \label{fig:RadPerDa}
\end{figure}

Examples of these coefficients are shown in figure \ref{fig:RadPerDa} and are compared to the minimum power loss expected from a $H\alpha$ photon due to molecular reactions (when also accounting for associated $Ly\beta$ and $Ly\alpha$ emission). This indicates roughly 20-100 eV power loss per observed $H\alpha$ photon that is due to  $H_2^+$ and $H^-$ contributions to $H\alpha$. This is (especially for $H_2^+$) significantly larger than the minimum expected radiative losses based on $H\alpha$, which indicates that $H_2$ plasma chemistry can also result in $n=2$ excited atoms, without having to go through the $n=3$ state.

\subsubsection{Inferring ion sinks/sources (MAR/MAI) from plasma-molecule interactions}
\label{ch:analysis_molMARMAI}

Similarly to how the radiative losses per $H\alpha$ photon are calculated above, one can also calculate ion sinks/sources (MAR/MAI) per $H\alpha$ photon - equation \ref{eq:MARMAIDaTOT} for $H_2$ and $H^-$. For the MAR/MAI rates we use AMJUEL \cite{Reiter2008,Kotov2007,Sawada1995,Reiter2005} rates H4 7.2.3a - MAR $H^-$; H4 2.2.10 - MAI $H_2$.

\begin{equation}
\begin{aligned}
    \left.\frac{MAR}{H_\alpha^{mol}}\right|_{H^-}&= \frac{MAR_{H^-} (n_e, T_e)}{PEC_{3\rightarrow2}^{H^- + H^+} (n_e, T_e)} \\
    \left.\frac{MAI}{H_\alpha^{mol}}\right|_{H_2} &= \frac{MAI_{H_2} (n_e, T_e)}{PEC_{3\rightarrow2}^{H_2} (n_e, T_e)} 
    \end{aligned}
    \label{eq:MARMAIDaTOT}
\end{equation}

Calculating MAR/MAI ion sinks/sources for $H_2^+$ requires additional steps as not only the \emph{destruction} of $H_2^+$ matters, which can result in excited $H$ atoms thus providing the $B_{n\rightarrow2}^{H_2^+}$ we infer, \emph{but also the creation process of $H_2^+$}. $H_2^+$ can be created either through molecular charge exchange (CX) ($H_2 + H^+ \rightarrow H_2^+ + H$), \emph{which turns a plasma ion into a neutral}, or ionisation of $H_2$ ($e^- + H_2 \rightarrow 2 e^- + H_2^+$) which \emph{does \underline{not} turn a plasma ion into a neutral}. When $H_2^+$ reactions with an electron, there are $3\times2=6$ possible outcomes: 1,2) $e^- + H_2^+ \rightarrow H + H$ (AMJUEL reaction H4 2.2.14) is MAR for molecular CX and MAD for $H_2$ ionisation; 3,4) $e^- + H_2^+ \rightarrow H^+ + H$ (AMJUEL reaction H4 2.2.12) is MAD for molecular CX and MAI for $H_2$ ionisation; 5,6) $e^- + H_2^+ \rightarrow H^+ + H^+$ (AMJUEL reaction H4 2.2.14) is MAI for molecular CX and MAI (x2) for $H_2$ ionisation.

As neither of those $H_2^+$ creation processes result in excited atoms, we cannot extract information on which process is dominant using only the Balmer line spectra. Instead, we need to model the relative strength of the two $H_2^+$ creation process based on $n_e$ and $T_e^E$ using their reaction rates (equation \ref{eq:ffromCX}). This assumes the electron density equals the hydrogen ion density and makes assumptions on the distribution of vibrational states (see section \ref{ch:discussion_isotope}). For $<\sigma v>_{H^+ + H_2 \rightarrow H + H_2^+}$ we use data from \cite{Kukushkin2017} (for deuterium), whereas from $<\sigma v>_{H^+ + H_2 \rightarrow H + H_2^+}$ we use data from AMJUEL H4 2.2.9. 

\begin{equation}
    f_{H_2^+ from CX} = \frac{<\sigma v>_{H^+ + H_2 \rightarrow H + H_2^+}}{<\sigma v>_{H^+ + H_2 \rightarrow H + H_2^+} + <\sigma v>_{e^- + H_2 \rightarrow 2 e^- + H_2^+}}
    \label{eq:ffromCX}
\end{equation}

We use this model and combine it with the possible MAI/MAR outcomes to calculate the MAI/MAR to $H_\alpha$ emission ratios for $H_2^+$ shown in equation \ref{eq:MARMAIDaH2p}. The notation MAR/MAI/MAD for the rates of equation \ref{eq:MARMAIDaH2p} refers to what the process would be if $H_2^+$ is purely created through molecular charge exchange (e.g. $f_{H_2^+ from CX} = 1$). The impact of different reaction rates on $f_{H_2^+ from CX}$ and subsequently the "MAR and MAI to $H\alpha$ emission coefficient ratios" are discussed in section \ref{ch:discussion_isotope}.

\begin{equation}
\begin{aligned}
    \left.\frac{MAR}{H_\alpha^{mol}}\right|_{H_2^+}&= \frac{f_{H_2^+ from CX} (n_e, T_e) MAR (n_e, T_e)}{PEC_{3\rightarrow2}^{H_2^+} (n_e, T_e)} \\
    \left.\frac{MAI}{H_\alpha^{mol}}\right|_{H_2^+}&= \frac{(2 - f_{H_2^+ from CX} (n_e, T_e)) MAI (n_e, T_e) + (1-f_{H_2^+ from CX}) MAD (n_e, T_e)}{PEC_{3\rightarrow2}^{H_2^+} (n_e, T_e)}
    \end{aligned}
    \label{eq:MARMAIDaH2p}
\end{equation}

\begin{figure}[htb]
    \centering
    \includegraphics[width=0.8\linewidth]{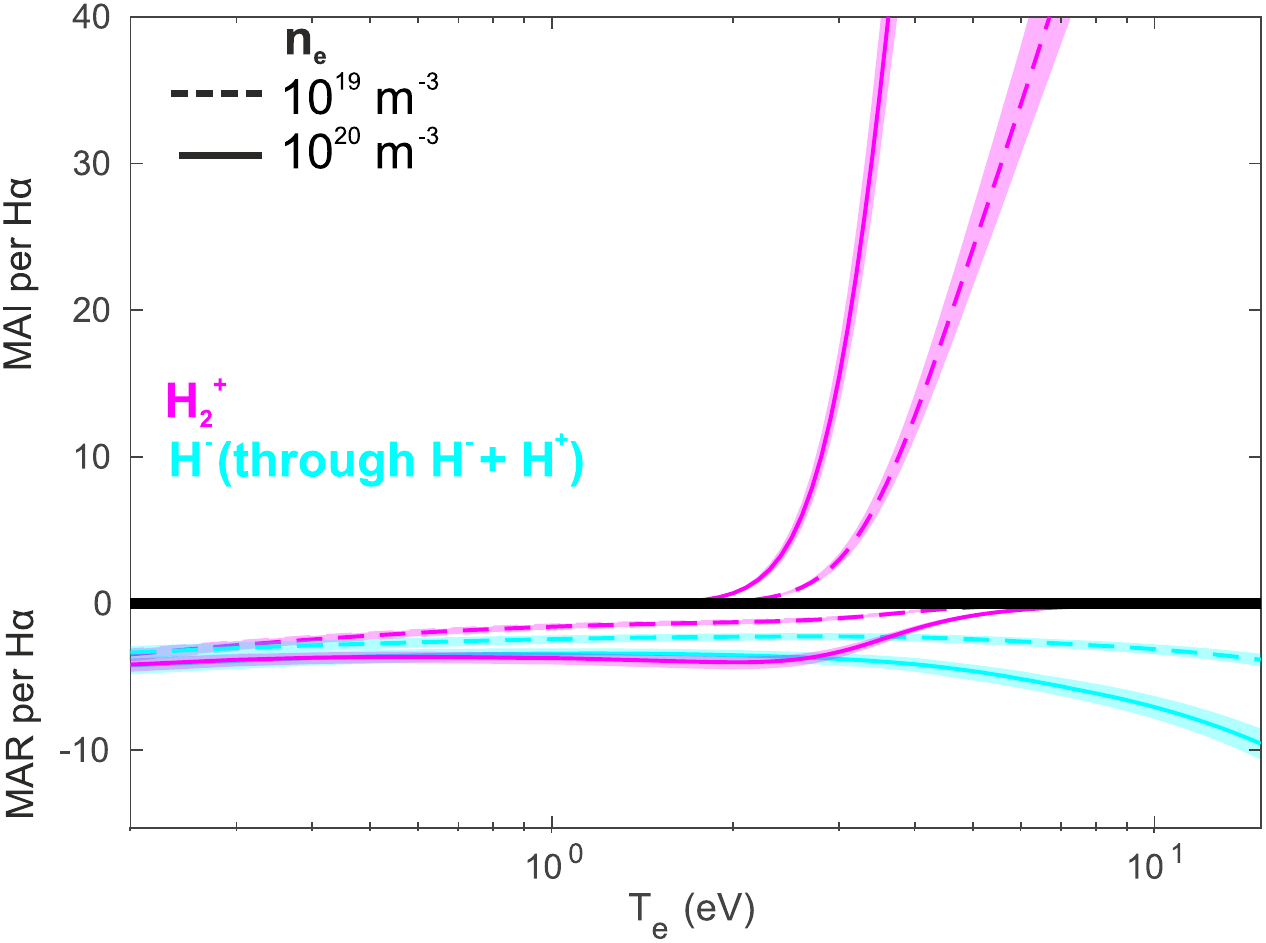}
    \caption{MAR (negative) and MAI (positive) per emitted $H\alpha$ photon for $H_2^+, H^-$ at different electron densities. The black curve represents zero. The uncertainties are provided by the default assumption of an uncertainty of 25 \% on all used molecular reaction rates and emission coefficient in addition to an assumed $H_2$ temperature range ([0.37 - 10] eV using a log-uniform distribution) - see section \ref{ch:discussion_isotope}. The indicated uncertainties are 68\% confidence margins}
    \label{fig:MARMAIperDa}
\end{figure}

Figure \ref{fig:MARMAIperDa}, which shows the calculated MAR/MAI per $H\alpha$ photon, indicates that at detachment relevant temperatures ($T_e < 3$ eV) $H_2^+$ and $H^-$ have (within experimental uncertainty) similar MAR per $H\alpha$ photon ratios (4-7). MAI starts to dominate over MAR for $H_2^+$ at $T_e > 3$ eV. The MAI per $H\alpha$ ratio is particularly sensitive to $T_e$ for $T_e > 3$ eV due to the dependence of $f_{H_2^+ from CX}$ on $T_e$. Considering that the inferred $T_e$ will have an uncertainty, this likely leads to large uncertainties in the MAI estimations.  %In all cases, $n_e$ can have a significant impact on the MAR/MAI per $D\alpha$ rate and tends to increase it.

\section{Verification using synthetic diagnostic techniques}
\label{ch:analysis_solps}

There are numerous ways in which the analysis approach uses a simplified emission model, which may not accurately reflect reality. For instance, the analysis approach simplifies the emission along the line of sight as a dual slab model (\ref{ch:analysis_BalmerEmissModel} and section \ref{ch:analysis_atomic}) with the same electron density and two different electron temperatures. In reality the plasma profiles along the line of sight vary and the various emission processes can occur at different positions along the line of sight \cite{Verhaegh2019a,Perek2020}. Additionally, the analysis assumes $Z_{eff} = 1$, which is not necessarily true. Those limitations are not necessarily problematic as the aim of the analysis is not to retrieve the emission profile along the line of sight but to extract the various line-integrated ion sources/sinks and power losses in the divertor.

The performance of the analysis to extract ion sources/sinks and power losses must be tested. One way of doing this is to verify its outcomes against a 'known' case, which can be achieved by using plasma-edge simulations to simulate the spectra a spectrometer would see synthetically. This can then be analysed in the same way as experimental data and those outputs can be compared 'directly' outputs from the simulations. 

In this work we apply this synthetic testing approach to SOLPS-ITER simulations of TCV and MAST-U plasmas. This involves both a $D_2$ gas puff scan (TCV, MAST-U) and a $N_2$ gas puff scan (MAST-U). The methods used for this have been developed in \cite{Verhaegh2019a,Verhaegh2018} and account accurately for the various spectrometer uncertainties.

To simulate the Balmer line brightnesses attributed to $H_2$ chemistry involving $H_2, H_2^+, H^-, H_3^+$, Yacora (on the Web) collisional radiative modelling results \cite{Wunderlich2020,Wuenderlich2016} are used in conjunction with the simulated electron temperature, electron density, molecule ($H_2$) density as well as the ion ($H^+$) temperature. The temperature of $H^-$ is assumed to be equal to the $H_2$ temperature plus a random number between 0 - 2.2 eV as $H^-$ arises from reactions between the plasma and $H_2$ and a part of the Franck-Cordon energy binding $H_2$ is released to $H^-$. ADAS is used for the electron excitation impact (of $H$) and electron-ion recombination (of $H^+$) Balmer line emission contributions \cite{Summers2006,OMullane}. 

The densities for $H_2^+$, $H^-$, $H_3^+$ must be known to accurately model the Balmer line emission from excited hydrogen atoms after those ions react. Such species are, by default, not ('fully') treated in SOLPS-ITER. Generally, only $H_2^+$ is included. However, it is designated as a 'test specie' in Eirene where it remains static (e.g. there is no transport) after being created. Additionally, there is some discussion on the isotope dependency of the rates leading to and/or breaking up $H_2^+, H^-$ \cite{Kukushkin2016}; which is further discussed in section \ref{ch:discussion_isotope}. %As a consequence, the $H_2^+$ densities for the TCV SOLPS simulations from SOLPS-ITER are insufficient to  %and so cannot be considered accurate. 

We overcome the above limitations of the information from SOLPS-ITER corresponding to $H_2^+, H_3^+, H^-$ to $H_2$ by using a balance (which neglects transport) between the creation and destruction rates of these species from $H_2$ to 'post-process' the $H_2^+, H_3^+, H^-$ densities after obtaining the SOLPS-ITER results \cite{Verhaegh2019a}. For the $H_2^+$ rates we employ the same rates as discussed in section \ref{ch:analysis_molMARMAI} (using the reported $H_2$ temperatures from the simulation). It is important to warn the reader that these ratios are still being debated in literature and may have large uncertainties, see section \ref{ch:discussion_isotope}. Therefore, significant deviations can occur between the post-processed results, the direct SOLPS-ITER outputs and the experimental results when it comes to the $H_2^+$ (and $H^-, H_3^+$) densities. 

Although the goal of this analysis is to retrieve line-integrated parameters from line-integrated spectroscopy, it would be beneficial to have estimates of the various parameters also along the lines-of-sight. Given the complexities of modelling the various molecular densities along the line of sight, this could be achieved by applying the shown techniques to 2D filtered camera images of the Balmer line emission in the divertor (see section \ref{ch:discus_applicability}).  %How and whether to include such species and reactions more accurately in SOLPS-ITER is still debated \cite{Kukushkin2016,Kotov2007} - see section \ref{ch:discussion_isotope}. In this case, $\frac{H_2^+}{H_2} = \frac{SCD_{H_2 \rightarrow H_2^+}}{ACD_{H_2^+ \rightarrow H_2}}$ where $SCD_{H_2 \rightarrow H_2^+}$ is the sum of the creation ('ionisation') rates of $H_2^+$ from $H_2$ and $ACD_{H_2^+ \rightarrow H_2}$ is the sum of the destruction ('recombination') rates of $H_2^+$, for instance. 

%This approach is identical to taking the tabulated ratios from AMJUEL \cite{} (rates H12 2.0a ($H_2^+/H_2$), H11 7.0a ($H^-/H_2$) and H11 4.0a ($H_3^+/H_2$)). 
 % synthetic diagnostic results, the experimental data and SOLPS.

\subsection{Description of results from modeling the Balmer line emission on the SOLPS grid}

\begin{figure}[htb]
    \centering
    \includegraphics[width=1\linewidth]{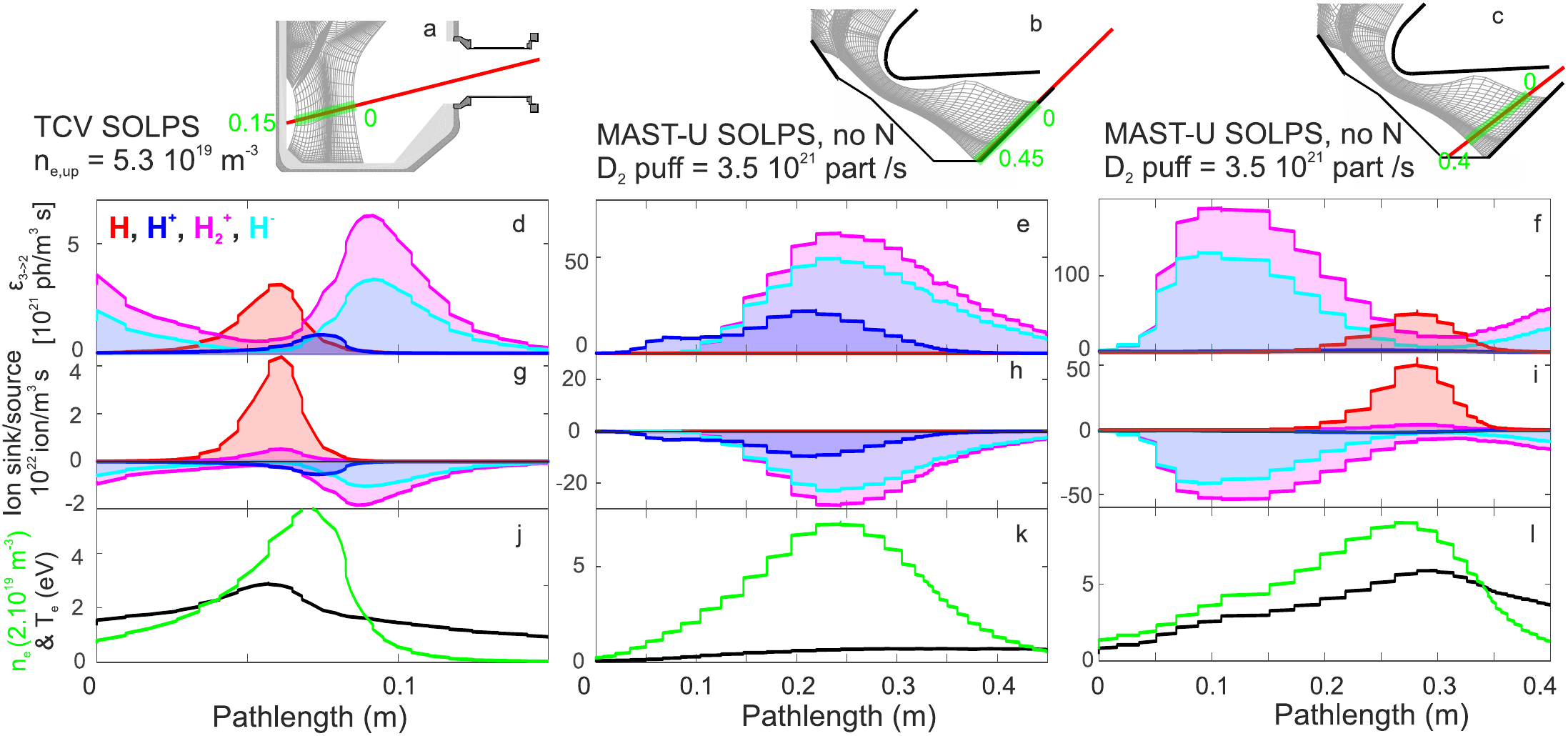}
    \caption{Profiles along an indicated line of sight of the various emission processes, particle source/sink processes as well as the electron temperature and density. This is shown for one TCV case and for two different line of sights for the same MAST-U case. The used line of sights, as well as the divertor geometry and the region where the lines of sight intersect the SOLPS grids are shown.}
    \label{fig:SOLPS_EmissProfs}
\end{figure}

Now that we have explained how we model the Balmer line emission on the SOLPS grid, we will later use this to perform synthetic testing on the analysis using simulations from both TCV as well as MAST-U. First, however, we show in figure \ref{fig:SOLPS_EmissProfs} d,e,f three example emission profiles along a line of sight for both a TCV and MAST-U simulation, together with the respective reaction profiles (figure \ref{fig:SOLPS_EmissProfs} g,h,i), electron temperature and electron density profiles (figure \ref{fig:SOLPS_EmissProfs} j,k,l). From this we see, indeed, that there is a spatial separation between the various emission profiles in all three cases. Furthermore, in figure \ref{fig:SOLPS_EmissProfs} we observe that MAI and MAR from $H_2^+$ occur at different locations spatially. \emph{We can thus conclude that the actual emission and reaction profiles along the line of sight are far more complicated in the test case than is assumed in the analysis chain.}

Secondly, we discuss how the Balmer line emission associated with $H_2$ plasma chemistry, under the assumptions/limits described, changes the synthetic brightnesses (compared to only accounting for electron-impact excitation and electron-ion recombination) and how this compares to experimental observations. The simulations used \cite{Fil2017} have been compared previously against the accompanying experiment in \cite{Verhaegh2019} from a view point of atomic interactions. This provides us with \emph{qualitative} arguments as to how representative the analysed synthetic diagnostic results are of the experiment. %An illustration of the analysis technique on this discharge will be presented in section \ref{ch:results}, while a more detailed analysis will be shown in a future paper \cite{}. 

\begin{enumerate}
    \item The synthetic diagnostic brightnesses are in quantitative agreement with the experiment if only electron-ion recombination and electron-impact excitation is considered for the medium-n Balmer lines. The total synthetic $H\alpha$ brightness (related to atomic interactions and $H_2$ plasma chemistry) is in rough agreement with the total measured $H\alpha$ brightness.
    \item However, the simulated results indicate a significant fraction of the $n=5$ Balmer line emission is due to plasma-molecule interactions (mostly due to $H_2^+$). This lowers the simulated $n=6 / n=5$ Balmer line ratio from its atomic estimate ($\sim 0.5$) to $\sim 0.4$ near the target; while the experimental measurement is closer to $0.5$ near the target. 
    
    As explained in \cite{Verhaegh2019a}, such changes in the Balmer line ratio are expected to have a relatively strong influence on the inferred excitation Balmer line emission. The larger the modification of the medium-n Balmer line ratio by $H_2$ plasma-chemistry related processes, the more complex and uncertain it is to fully disentangle the 'atomic only' line ratio required for estimating accurately the excitation emission contribution.
\end{enumerate}

 \emph{Therefore, the application of the analysis is more complex (and has higher uncertainties) in the synthetic diagnostic case than in the experiment}.
    
\begin{figure}[htb]
    \centering
    \includegraphics[width=1\linewidth]{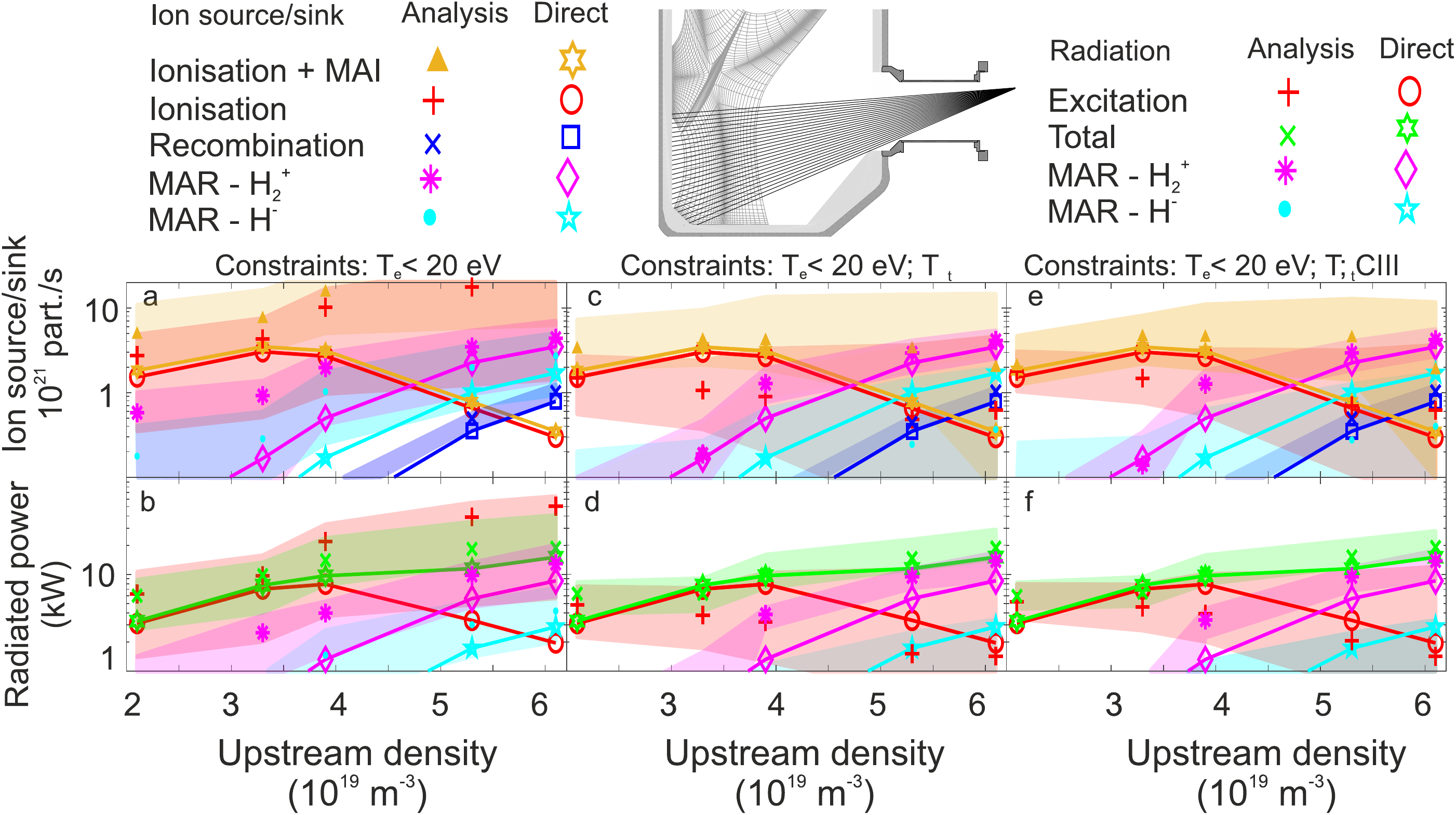}
    \caption{Comparisons between 'Direct' output of SOLPS-ITER modelling of TCV density scan to detachment and the same quantities derived ('Analysis' - with uncertainty margins) from synthetic diagnostic measurements of the same SOLPS-ITER cases: a) Inferred particle balance, including atomic ionisation, electron-ion recombination recombination, MAR from $H_2^+$ and $H^-$. b) Inferred radiative loss channels from atomic (mostly line) emission, including atomic excitation, radiation due to excited atoms from reactions involving $H_2^+$ and $H^-$ c/d) Power and particle balance comparison between 'Direct' outputs and outputs from the 'Analysis' with the added constraint of target temperature. e/f) Power and particle balance comparison between 'Direct' outputs and outputs from 'Analysis' with an added constraint based on the CIII emission front as well as the target temperature.}
    \label{fig:SynDiag_TCV}
\end{figure}

\subsection{Synthetic testing on TCV SOLPS simulations}

Now that we have discussed some of the results from simulating the Balmer line emission associated with $H_2$ plasma chemistry as well as plasma-atom interactions, we show the synthetic testing results using SOLPS simulations for TCV. Figure \ref{fig:SynDiag_TCV} shows a comparison between various processes obtained 'Direct'(ly) from simulations of a TCV density scan and the same quantities evaluated ('Analysis') through synthetic measurements. Each column of plots corresponds to different sets of constraints that are applied. The technicalities of these constraints are described in more detail in appendix \ref{ch:analysis_solps_Tt}. For all the cases, an upper electron temperature limit constraint of 20 eV is applied, which is characteristic for these TCV simulations.

Figure \ref{fig:SynDiag_TCV} a,b shows that the synthetically inferred MAR/EIR ion sinks as well as the radiative power loss associated with $H_2^+$ and $H^-$ are in quantitative agreement with the direct SOLPS output if no constraints are employed. There is, however, a strong difference in the ionisation source as well as the radiation associated with electron-impact excitation after the detachment onset (around an upstream density of $3.5 \times 10^{19} m^{-3}$). This difference after the detachment onset is caused by an overestimate of the atomic excitation emission caused by underestimating the (atomic only) line ratio $n=6 / n=5$ near the end of the discharge($\sim 0.45$ instead of $\sim 0.5$). The analysis technique shown can thus be used to obtain adequate estimates on electron-ion recombination, MAR and power losses arising from plasma-molecule interactions. However, during detachment, the ionisation as well as MAI ion source inferences can become unreliable if no constraints are employed. %Although the analysis does correct to some degree for this, the contribution of plasma-molecule interactions to the medium-n Balmer line emission has too much of an effect to completely correct for it. 

%When comparing the direct inferences of the various power loss channels, we see, again, a quantitative agreement for the power loss channels arising from $H_2^+$, $H^-$ and electron-ion recombination with a strong disagreement for the power loss channels arising from atomic excitation. Again, this is caused by the (relatively) strong overestimation of the atomic excitation emission channel.

The periods of poor inference of ion sources can be improved by including additional constraints. As explained in \cite{Verhaegh2019a}, the overestimation of excitation emission is a known complication in cases where the excitation emission is relatively small. Since an overestimation of the excitation emission manifests in an overestimation of the excitation temperature, one can improve the analysis by enforcing temperature constraints \cite{Verhaegh2019a}. We include two temperature constraints: a) a constraint at the 'target' (lines of sight near the target) on the electron-impact excitation emission derived target temperature ($T_e^E$) based on other target temperature estimates (for the synthetic case a $\pm 1$ eV (68 \% confidence interval) uncertainty is assumed); b) a temperature constraint based on the observation of the CIII emission front: \emph{below} that front $T_e^E>8$ eV is given a lower probability while \emph{above} that front $T_e^E<4$ eV is given a lower probability.

Adding only the target temperature constraint (c and d of Figure \ref{fig:SynDiag_TCV}) leads to a strong improvement of the quantitative agreement of the inferred/directly obtained excitation estimates until even in the detached phase. This can be further improved by adding additional constraints based on the CIII front location. For the synthetic test in figure \ref{fig:SynDiag_TCV}e,f, the impact of the CIII front constraints is marginal ($\sim 5\%$ change in maximum likelihood estimates). However, for other cases (such as the experimental case shown in section \ref{ch:results}) the impact of the CIII front location constraint can be of a similar magnitude than the impact of the target temperature constraint. These additional constraints also reduce the level of uncertainty in the various estimates. The uncertainties would likely improve further with more detailed profile (e.g. along the divertor leg) temperature estimate constraints.

Even with constraints, the MAI estimates have a significant uncertainty during detachment. This is related to the strong $T_e$ dependence of the MAI/$H\alpha$ ratio (figure \ref{fig:MARMAIperDa}), which is related to the change-over from $H_2^+$ being created from molecular charge exchange to it being created from $H_2$ ionisation (see section \ref{ch:analysis_molMARMAI}). This implies that the MAI estimates are sensitive to inaccuracies in the $T_e$ estimate, which also implies that they are relatively more sensitive to chordal integration effects. The uncertainties in MAI and atomic ionisation are however anti-correlated, and the total uncertainty is reduced when MAI and atomic ionisation is summed (as is done in figure \ref{fig:SynDiag_TCV}).  

 In the remainder of this work, both temperature constraints from the estimated target temperature as well as the CIII front location are employed for the ionisation and (atomic) radiation estimates when the full analysis (figure \ref{fig:schem_analysis}) is applied, unless stated otherwise. 
 
 We observe the electron density is \emph{different} for the EIR, electron-impact excitation, $H_2^+$ and $H^-$ emission regions (figure \ref{fig:SOLPS_EmissProfs} d,g,j). As the analysis assumes the same electron density for all interaction regions based on Stark broadening $n_e$ estimates of the $n=7$ Balmer line (which is mostly dominated by EIR and is obtained from the synthetic spectrometer in this case), the analysis will overestimate the characteristic electron density for the plasma-molecule interaction processes. Despite this overestimate, when $T_e$ constraints are employed, the inferred parameters (given their uncertainties) agree with those obtained directly from the simulation.

 \subsection{Further synthetic testing through 'code experiments' on TCV SOLPS simulations}
 \label{ch:analysis_indepth_syndiag}
 
\begin{figure}[htb]
    \centering
    \includegraphics[width=1\linewidth]{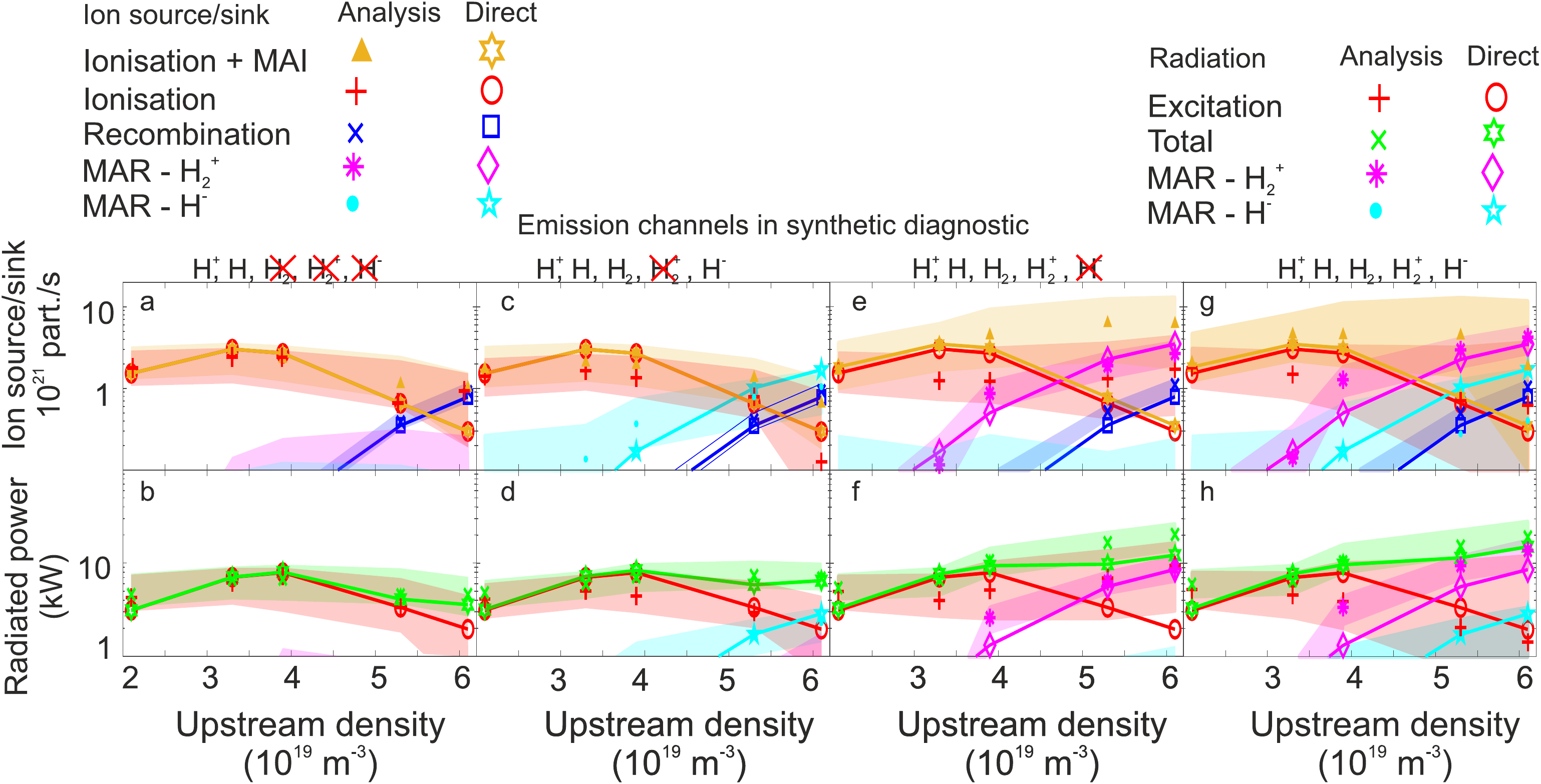}
    \caption{Power and particle balance, similar to figure \ref{fig:SynDiag_TCV}, where certain emission channels have been disabled in the synthetic diagnostic to investigate its influence on the analysis outputs}
    \label{fig:SynDiag_TCV_Testing}
\end{figure}

 We can perform further synthetic testing on the simulations shown in the previous section through 'code experiments' by removing certain emission channels from the input of the synthetic brightnesses, after which the full analysis is used to analyse the 'modified' synthetic brightnesses. This is an important part of testing the robustness of the analysis scheme as it enables us to see how well the analysis copes with excluding processes which are not present. This is investigated by: 
\begin{enumerate}
    \item Removing all molecular emission channels (figure \ref{fig:SynDiag_TCV_Testing} a,b).
    \item Removing the $H_2^+$ emission channel (figure \ref{fig:SynDiag_TCV_Testing} c,d).
    \item Removing the $H^-$ emission channel (figure \ref{fig:SynDiag_TCV_Testing} e,f).
\end{enumerate}

These cases are shown in figure \ref{fig:SynDiag_TCV_Testing}, together with a copy of the analysis in which all emission channels are included, previously shown in figure \ref{fig:SynDiag_TCV}.

Figure \ref{fig:SynDiag_TCV_Testing} generally shows a quantitative agreement between the various particle sinks/sources and power sinks estimated from the analysis and those obtained directly from the code, when one considers the uncertainty of the analysis estimates (68\% confidence levels are shown). One exception to this is the MAI estimate in figure \ref{fig:SynDiag_TCV_Testing}e,f, where $H^-$ was not accounted for. This is related with the large uncertainties of MAI discussed previously. We observe that the upper uncertainty level of MAR from $H_2^+$ and/or $H^-$ are negligible (although not zero) when they have been omitted in the synthetic diagnostic brightness during detachment. This test shows the analysis can correctly point out the lack/presence of MAR and separate MAR from $H_2^+$ and $H^-$ - as long as their impacts are 'significant'. %For instance, we observe that at an upstream density of $4 x 10^{19} m^{-3}$ that MAR from $H_2^+$ and $H^-$ are detected in equal amounts even if all MAR is due to $H^_$. The reason for this is that the additional $H\beta$ emission from $H^-$ is not sufficient to be comfortably detected and thus the $H\beta/H\alpha$ ratio due to $H_2^+$ and $H^-$ has a large uncertainty.  $H\alpha$ emission from $H^-$ at this point is just becoming sufficient to be detected - given the various uncertainties; while additional $H\beta$ emission from $H^-$ is below the detection threshold. Additionally, the analysis seems to be able to distinguish $H_2^+$ and $H^-$ related processes correctly - provided their impact is above a certain detection threshold. This detection threshold for MAR is significantly higher than for electron-ion recombination. 

We also observe in figure \ref{fig:SynDiag_TCV_Testing} that the quality of the excitation-dependent inferences as well as MAI deteriorates as more emission channels are present in the input synthetic brightnesses. As the contribution of molecules to the $n=5,6$ Balmer line increases, the quality of the excitation inferences decreases. This illustrates the necessity of including the various temperature constraints introduced in the previous section.

 \subsection{Synthetic testing on MAST-U SOLPS simulations}
 
\begin{figure}[htb]
    \centering
    \includegraphics[width=1\linewidth]{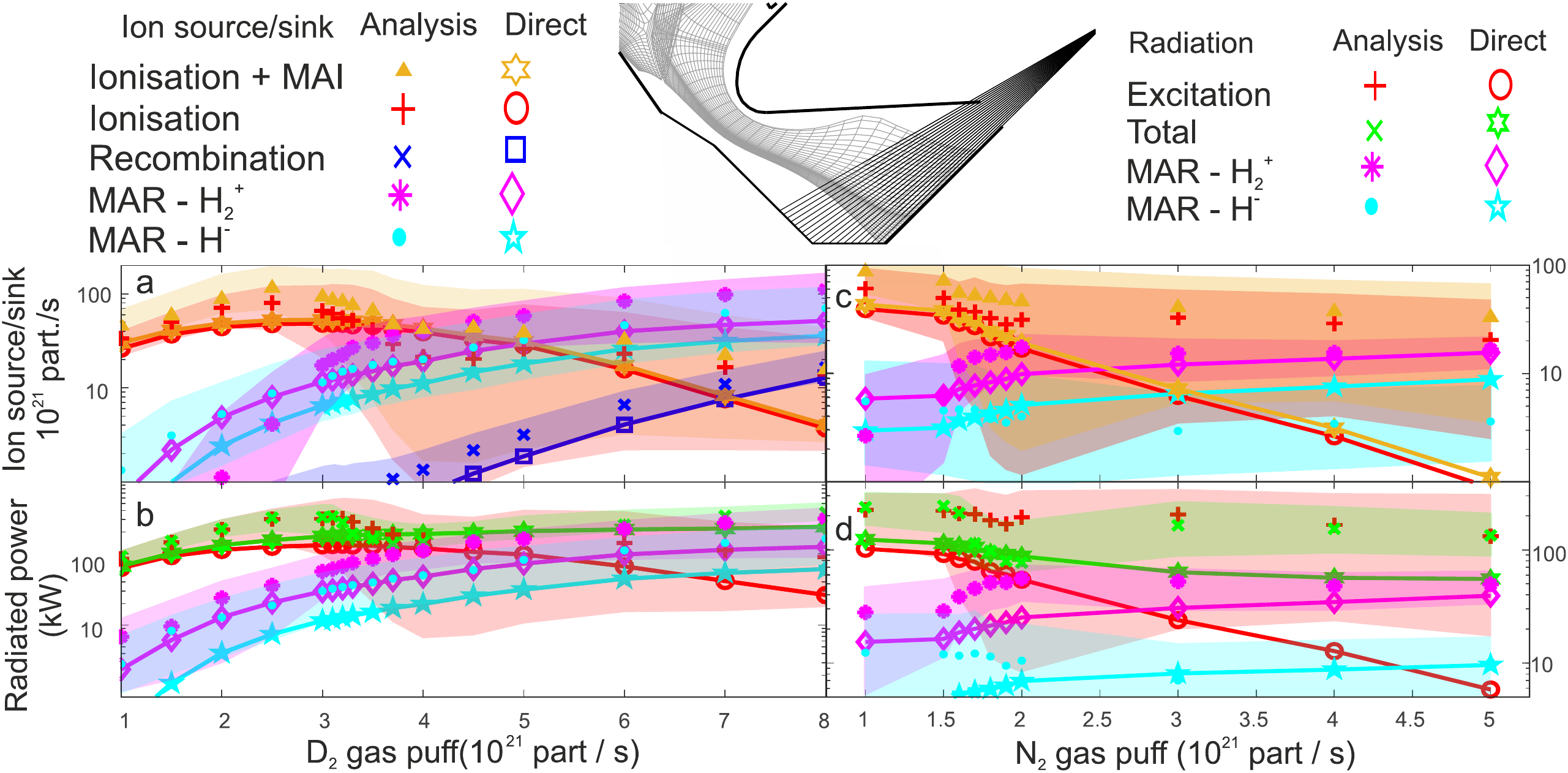}
    \caption{Power and particle balance from estimates from synthetic diagnostic analysis and directly obtained from SOLPS MAST-U simulations \cite{Myatra}, similar to figure \ref{fig:SynDiag_TCV}. The results of both a density ramp (with intrinsic carbon impurities) and that of a fixed $D_2$ puff ($2 \cdot 10^{21}$ part/s) with a $N$ seeding ramp (with intrinsic carbon impurities) are shown.}
    \label{fig:SynDiag_MASTU}
\end{figure}

We have applied the similar synthetic testing procedure shown throughout this section to MAST-U SOLPS simulations \cite{Myatra} of a core density ramp as shown in figure \ref{fig:SynDiag_MASTU}a, b. In addition, we have applied our synthetic testing procedure to a $N_2$ seeded scan (with intrinsic carbon impurities) (figure \ref{fig:SynDiag_MASTU}c, d) to have more capabilities of testing our analysis as the plasma fields are different between the fuelling and seeding scans \cite{Myatra}. In this case, we have not used the temperature exclusion constraint based on the CIII front introduced previously as, given the magnetic geometry of the MAST-U Super-X divertor, the CIII front cannot be comfortably tracked using line integrated spectroscopy and instead requires camera diagnostics, such as \cite{Ravensbergen2020, Perek2019submitted}.
 
We observe that, generally, there is an agreement within uncertainty (68\% confidence intervals are shown) between the parameters inferred from the synthetic diagnostic and those obtained directly. The strongest exception to this are excitation related estimates (e.g. ionisation + MAI - orange, ionisation - red and excitation radiation - red (which is also a part of the total radiation - green)) at the highest $N$ puff rates ($geq 4 \cdot 10^{21}$ part/s), which are significantly overestimated. These overestimates occur because the excitation-related estimates drop below the levels which can be comfortably detected. Those 'detection threshold' levels are higher for the nitrogen seeded case than the core density ramp case because of electron-ion recombination is fully negligible in these cases. This also explains the large uncertainty of the ionisation estimates shown.  

In general we see an improved quantitative agreement (especially for MAI) for the MAST-U synthetic testing than the TCV synthetic testing. This is likely attributed to the closed divertor/higher electron densities in MAST-U, resulting in shorter mean free paths. Shorter neutral mean free paths would result in a more strongly localised ionisation region as the neutrals cannot penetrate through the entire divertor leg (as is the case on TCV \cite{Verhaegh2019}). As three-body Electron-Ion Recombination (EIR) takes over radiative EIR \cite{Verhaegh2017,Terry1998} at higher electron densities, the total EIR rate becomes a stronger function of electron density at higher electron densities, resulting in a more localised and stronger EIR region near the target at MAST-U (considering the density is increasing from the x-point towards the target). Both these results lead to a stronger spatial separation between the ionizing and recombining regions along the divertor leg. 

\subsection{Summary of synthetic testing}

The analysis chain has been tested synthetically using both TCV and MAST-U SOLPS simulations in both seeded and non-seeded conditions in combination with synthetic spectroscopy diagnostics to simulate what a spectrometer would observe, which is then analysed in an identical way as the experiment. The emission and reaction profiles along the lines of sight in the analysis are significantly more complicated than the simplified dual-slab model assumed in the analysis chain.

Although the various emission processes occur at different positions along the line of sight, the line-integrated estimates obtained by the analysis during synthetic testing are generally, considering their uncertainties, in agreement with those obtained by directly integrating the profiles along the line of sight. In addition, we observe that if certain emission processes are removed from the \emph{input} of the analysis, the analysis correctly points out that their contribution is negligible. 

This testing suggests the analysis is fairly robust for chordal-integration effects. This is particularly true for the estimates on radiative losses related to excited atoms from $H_2$ plasma chemistry, MAR and electron-ion recombination. Ionisation and MAI estimates require additional temperature constraints for higher accuracy.

\section{Illustration of the analysis using experimental data from TCV}
\label{ch:results}

%Before applying the molecular analysis techniques developed in this work, we first briefly discuss the experimental set-up and discuss some of the experimental observations Balmer line observations in comparisons against power source/sink estimates \emph{obtained by neglecting plasma-molecule interactions} \cite{Verhaegh2019,Verhaegh2019a}. 

Although the performance of an analysis can be analysed in detail through synthetic testing, it is beneficial to test an analysis using experimental data. This is particularly true for the analysis used here as there are many uncertainties in simulating the $H_2^+$ and $H^-$ densities required for simulating the Balmer line emissivities associated with $H_2$ plasma chemistry in SOLPS-ITER simulations, as discussed in section \ref{ch:analysis_solps}. 

\begin{figure}[H]
    \centering
    \includegraphics[width=0.5\linewidth]{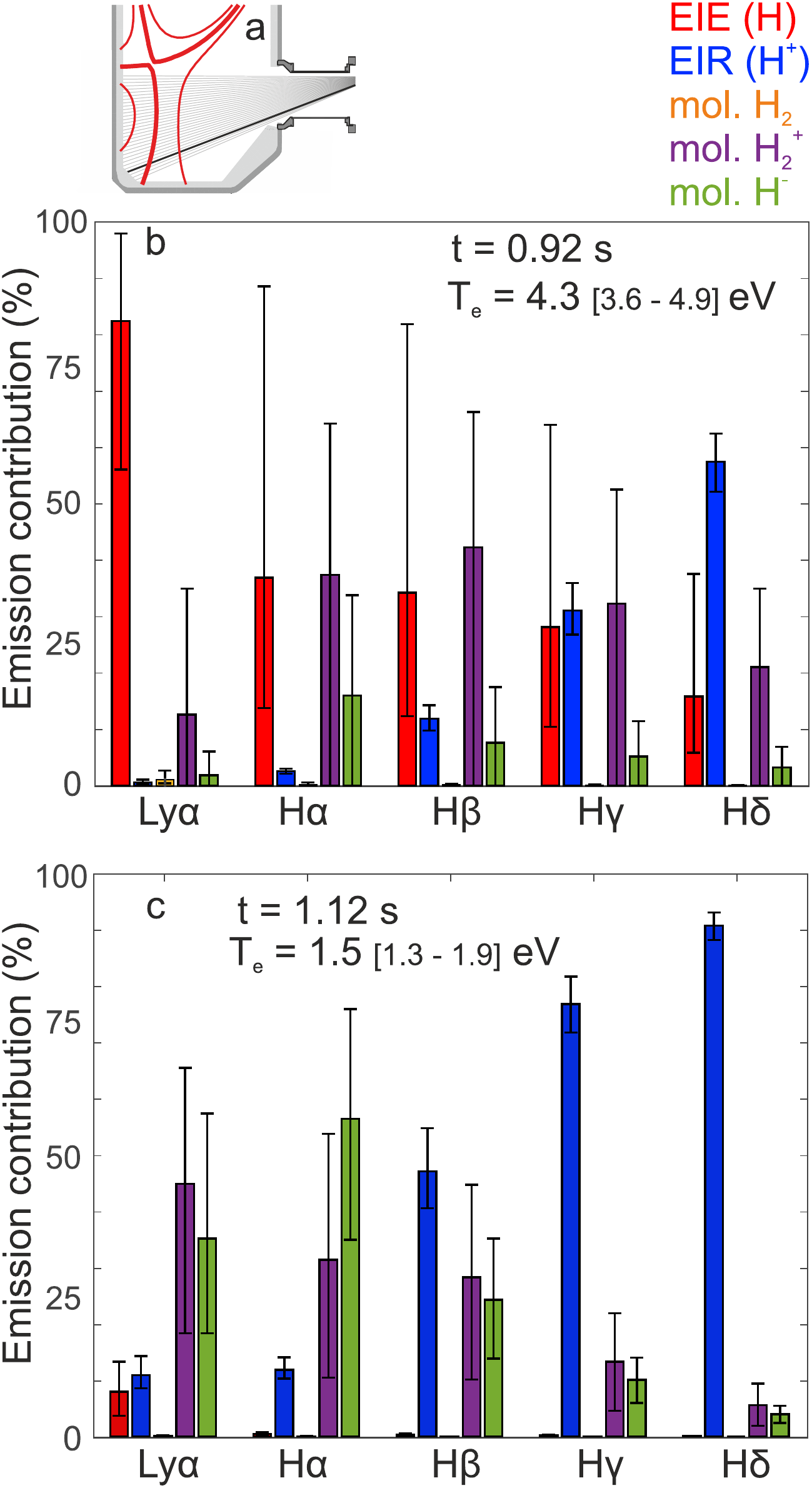}
    \caption{A schematic illustration of the divertor geometry and line of sight is shown (a). A bar-chart of the contributions (\%) of various processes is shown (electron-impact exctiation 'EIE (H)', electron-ion recombination 'EIR ($H^+$)', plasma-molecule interaction ('mol.') with $H_2$, $H_2^+$ and $H^-$ for various hydrogenic series lines at two different times for a chord close to the target together with indicated estimated electron temperature ranges (b,c). }
    \label{fig:SepaEmiss}
\end{figure}

We illustrate an example of the \emph{self-consistent} results of the full BaSPMI analysis to separate the hydrogen line brightnesses into its various atomic (excitation / recombination) and molecular ($H_2$, $H_2^+$, $H^-$) contributions. For this we use a conventional divertor L-mode reversed field (unfavourable for H-mode) density ramp discharge with a plasma current of 340 kA. The divertor physics of this discharge has been discussed previously in \cite{Verhaegh2019,Verhaegh2020a}. The emission spectra is diagnosed using the TCV Divertor Spectroscopy System (DSS) diagnostic  \cite{Verhaegh2017,Verhaegh2019a}. The divertor geometry with the lines of sight coverage for this diagnostic can be seen in figure \ref{fig:SepaEmiss}, adapted from \cite{Verhaegh2019,Verhaegh2019a}. Diagnostic repeat discharges are used in order to obtain sufficient diagnostic coverage. The reproducibility of this has been demonstrated in \cite{Verhaegh2019}. Three different temperature constraints (for $T_e^E$) have been employed: 1) the upper temperature limit is 25 eV; 2) temperature constraint based on the CIII 465 nm emission line front which is measured throughout the discharge using line-of-sight spectroscopy (see details in \ref{ch:analysis_solps_Tt}); 3) a target temperature constraint based on the estimated target temperature by power balance ($T_t^{PB}$), which was shown and compared against various target temperature estimates (measured and modelled) in Figure 10 of \cite{Verhaegh2019} yielding a good agreement between the various temperature estimates. %It should be noted that the discharge analysed in this section is a \emph{deuterium} discharge. However, collisional radiative models such as ADAS \cite{Summers2006,OMullane} and Yacora Online \cite{Wuenderlich2016} are applicable to \emph{hydrogen}. This will be further discussed in section \ref{ch:discussion_isotope}. 

The results of the emission contributions are shown in figure \ref{fig:SepaEmiss} for one line of sight at two different times as a bar-chart. This is shown for a single line of sight for both the measured Balmer lines ($H\alpha$,$H\beta$,$H\gamma$,$H\delta$) used in the analysis as well as an extrapolated analysis estimate of the $Ly\alpha$ ($B_{2\rightarrow1}$) line, whose totals and individual contributions has been obtained through 'extrapolating' the experimental data of the molecular contributions of $H\alpha$ and the atomic contributions of the medium-n Balmer line $n$ using equation \ref{eq:LyAUp} based on combining equations \ref{eq:MolContamin} and \ref{eq:DaUp}. 

\begin{equation}
\begin{split}
    B_{2\rightarrow1} &= \underbrace{\frac{PEC_{2\rightarrow1}^{H_2} (n_e, T_e^E)}{{PEC_{3\rightarrow2}^{H_2} (n_e, T_e^E)}} \times B_{3\rightarrow2}^{H_2}}_{B_{2\rightarrow1}^{H_2}} + \underbrace{\frac{PEC_{2\rightarrow1}^{H_2^+} (n_e, T_e^E)}{{PEC_{3\rightarrow2}^{H_2^+} (n_e, T_e^E)}} \times B_{3\rightarrow2}^{H_2^+}}_{B_{2\rightarrow1}^{H_2^+}} + \underbrace{\frac{PEC_{2\rightarrow1}^{H^-} (n_e, T_e^E)}{{PEC_{3\rightarrow2}^{H^-} (n_e, T_e^E)}} \times B_{3\rightarrow2}^{H^-}}_{B_{2\rightarrow1}^{H^-}} + \\ &\underbrace{\frac{PEC_{2\rightarrow1}^{exc} (n_e, T_e^E)}{{PEC_{n\rightarrow2}^{exc} (n_e, T_e^E)}} \times B_{n\rightarrow2}^{atom, exc}}_{B_{2\rightarrow1}^{exc}} + \underbrace{\frac{PEC_{2\rightarrow1}^{rec} (n_e, T_e^R)}{{PEC_{n\rightarrow2}^{rec} (n_e, T_e^R)}} \times B_{n\rightarrow2}^{atom,rec}}_{B_{2\rightarrow1}^{rec}}
    \end{split}
    \label{eq:LyAUp}
\end{equation}

The illustration of the technique in figure \ref{fig:SepaEmiss} indicates that depending on the plasma conditions (in this case dictated by the timestep in the discharge): 

\begin{itemize} 
\item Plasma-molecule interactions can contribute considerably to hydrogenic line emission. It can \emph{dominate the $H\alpha$, $H\beta$ emission} and it can have a significant impact on \emph{$Ly\alpha$ emission} as well as \emph{medium-n Balmer line emission} ($H\gamma, H\delta$). This has important implication for hydrogenic radiation losses as well as the interpretation of Balmer line divertor spectroscopy measurements.
\item A large range of different emission processes can be significant simultaneously; e.g. both electron-ion recombination, plasma-molecule interactions from $H_2^+$ and $H^-$ appear to be significant for $H\beta$ at $t=1.12$ s. This shows the importance of separating the various emission channels.  
\item The emission processes can change strongly between each hydrogenic transition. We observe that the sensitivity to plasma-molecule interactions diminishes with increasing $n$ of the hydrogenic transition while the sensitivity to electron-ion recombination increases \cite{Verhaegh2019a}. Plasma-molecule interactions involving $H^-$ seem to excite the $n=3$ populational state (e.g. $H\alpha$ emission) in particular.
\item We observe that the uncertainties in the electron impact excitation (of $H$) (EIE) and the emission contribution from $H_2^+$ are substantial. A closer inspection shows that these uncertainties are \emph{anti-correlated}: low value estimates of the EIE contributions in the statistical samples correspond to high values of the $H_2^+$ contributions (and visa versa). The EIE contribution is strongly correlated with the excitation-inferred temperature. This illustrates why the various temperature constraints introduced in \ref{ch:analysis_solps_Tt} are important: without such constraints it is uncertain to distinguish, given the measured data and its uncertainties, electron impact excitation (of $H$) and emission from excited atoms after plasma reacts with $H_2^+$.
\end{itemize}

%To summarise our findings:
%\begin{enumerate}
%    \item Plasma-molecule interactions can contribute strongly to especially $H\alpha, H\beta$ but also significantly to $H\gamma$
%    \item Depending on the state of detachment and position, four different emission channels can have significant contributions to the Balmer line emission simultaneously: atomic excitation, EIR, plasma-molecule interaction ($H_2^+$ and $H^-$). This shows the importance of separating the emission into its different emission channels.
%    \item The shown results may indicate that plasma-molecule interactions with $H^-$ occur near the cold target region. This is, however, under the various assumptions made in the analysis and requires further investigation (see section \ref{ch:discussion_isotope}).
%    \item Plasma-molecule interactions can contribute significantly to $H\gamma$ between the recombination and ionisation region ($>40\%$), modifying the line ratio between $H\delta / H\gamma$ and altering the estimations of their emission fractions compared to an atomic only analysis \cite{Verhaegh2019a}. Excitation emission (and thus ionisation) may thus be overestimated in the deepest detached phases by the technique in \cite{Verhaegh2019a}, as shown in figure \ref{fig:HaMeasurement}.
%\end{enumerate}

\section{Discussion}
\label{ch:discussion}

%In this work we first established an analysis technique which can be used to quantitatively infer the influence of plasma-molecule interactions (with $H_2^+$ and $H^-$) on 1) particle balance (Molecular Activated Ionisation - MAI \& Molecular Activated Recombination - MAR); 2) the hydrogenic line emisson including hydrogenic radiation from excited atoms after plasma-molecule interactions. We have synthetically tested this analysis chain using both TCV and MAST-U SOLPS simulations. Experimental results from TCV highlight the measured $H\alpha$ and the estimated atomic contribution of $H\alpha$ bifurcate at the detachment onset which is attributed due to plasma-molecule interaction. 

\subsection{Estimating the Balmer line emission associated with $H_2$}
\label{ch:fH2_Treatment}

In section \ref{ch:analysis_DaMolSepa} we discussed methods to separate the Balmer line emission attributed to $H_2$ chemistry in its various components (related to $H_2, H_2^+, H^-$). We started that procedure with assuming an a priori $g_{H_2} \approx \Delta L n_{H_2}$ which is a function of $T_e^E$. That allows us to estimate the brightness associated with $H_2$: $B_{n\rightarrow2}^{H_2} = g_{H_2} (T_e^E) n_e PEC_{H_2} (n_e, T_e^E)$. In this section we highlight how we obtain this a priori function and we discuss its implications. This shows that the expected $B_{n\rightarrow2}^{H_2}$ is insignificant and can be neglected in most divertor conditions.

%Previous studies in limiter devices (where the molecules reach a much hotter plasma than in the divertor) have showed that $H_2$ dissociation may contribute to the $H\alpha$ emission \cite{Brezinsek2005}. We, however, find that the contribution of $H_2$ to $H\alpha$ in the studied divertor conditions is expected to be very small, especially during detached conditions. Instead the contribution of $H_2$ plasma chemistry to $H\alpha$ is found to be mainly through $H_2^+$ and/or $H^-$ (figure \ref{fig:SepaEmiss}). Those estimates for the $H_2$ contribution are based on obtaining a functional form of the 'effective' $H_2$ density ($n_{H_2}$) times pathlength ($\Delta L$) as function of the (impact excitation of $H$ region) electron temperature $T_e^E$ - $g_{H_2} (T_e^E) \approx \Delta L \times n_{H_2}$. Using that relation, we can then estimate the Balmer line brightness associated with $H_2$ (equation \ref{eq:DaH2}). 

We obtain this functional form $g_{H_2} (T_e^E)$ by combining TCV \cite{Fil2017} and MAST-U \cite{Myatra} SOLPS-ITER simulations in combination with synthetic spectroscopy diagnostics \cite{Verhaegh2019a,Verhaegh2018} (see section \ref{ch:analysis_solps}). To obtain an estimate for $g_{H_2} \approx \Delta L n_{H_2}$ (equation \ref{eq:fH2}) we take the synthetic brightness associated with $H_2$ - $B_{n\rightarrow2}^{H_2}$ (which is obtained by integrating the emissivity associated with $H_2$ along the line of sight) and divide this by $n_e PEC_{3\rightarrow2}^{H_2}$ estimated using the electron-impact excitation-emission weighted electron temperature $T_e^E$ and the Stark broadening inferred electron density $n_e$ for that chord using the synthetic diagnostic \cite{Verhaegh2019a}). We have chosen this formulation because using $g_{H_2}$ with those same electron densities/temperatures in a plasma-slab model would  bring us back to - by definition - the synthetically obtained $B_{n\rightarrow2}^{H_2}$. 

\begin{equation}
    g_{H_2} \equiv \frac{B_{3\rightarrow2}^{H_2}}{n_e PEC_{3\rightarrow2}^{H_2} (n_e, T_e^E)} \approx \Delta L \times n_{H_2}
\label{eq:fH2}
\end{equation} 

\begin{figure}[htb]
    \centering
    \includegraphics[width=1\linewidth]{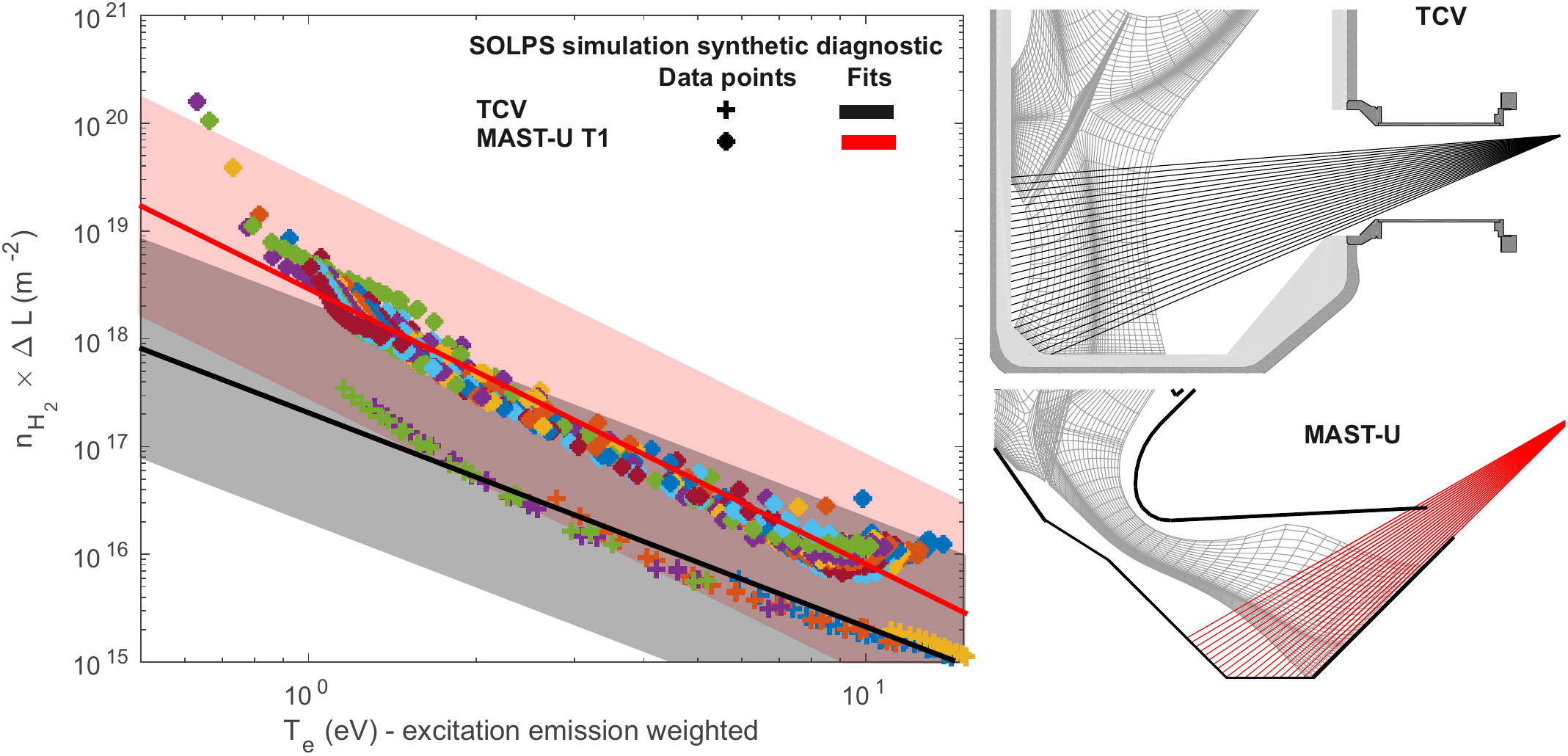}
    \caption{Relation between the excitation Balmer line emission weighted temperature $T_e^E$ and $g_{H_2} (T_e^E) \equiv \frac{B_{3\rightarrow2}^{H_2}}{n_e PEC_{3\rightarrow2}^{H_2} (n_e, T_e^E)} \approx n_{H_2} \times \Delta L$ (where $n_e$ is the synthetic Stark density). Each colour corresponds to a different simulation. Fits through each of the data sets are shown. The TCV data set consists out of 5 simulations \cite{Fil2017} (density scan) (26 lines of sight) while the MAST-U data set consists out of 35 simulations \cite{Myatra} (density scan and $N_2$ seeded) with 20 lines of sight. The corresponding SOLPS grid cells and spectroscopy lines of sight for MAST-U and TCV are also shown.}
    \label{fig:fH2_fit_TCV_MASTU}
\end{figure}

We then take all the spectroscopy chords in the synthetic diagnostic for all SOLPS simulations and show the obtained $g_{H_2}$ in figure \ref{fig:fH2_fit_TCV_MASTU} as function of the estimated $T_e^E$. We find there is a strong relation between $g_{H_2}$ and $T_e^E$ for both TCV and MAST-U simulations. This is a remarkable result as the points in figure \ref{fig:fH2_fit_TCV_MASTU} all come from different simulations and different chords (thus different plasma positions) of the synthetic diagnostic. In essence, this indicates that having information about the kind of device (e.g. TCV vs MAST-U), the electron excitation temperature and the electron density is sufficient for providing rough estimates on the Balmer line brightness attributed to $H_2$: $B_{n\rightarrow2}^{H_2} = g_{H_2} (T_e^E) n_e PEC_{n\rightarrow2}^{H_2} (n_e, T_e^E)$.

We use the $g_{H_2}$ values obtained from the simulations to estimate the relation between $g_{H_2}$ and $T_e^E$ using a fit (linear in log-log space) to which we ascribe an of a factor 100 uncertainty (from a factor 0.1 to 10 - log-uniformly distributed) when it is used to estimate $B_{n\rightarrow2}^{H_2}$.

\begin{figure}[htb]
    \centering
    \includegraphics[width=0.5\linewidth]{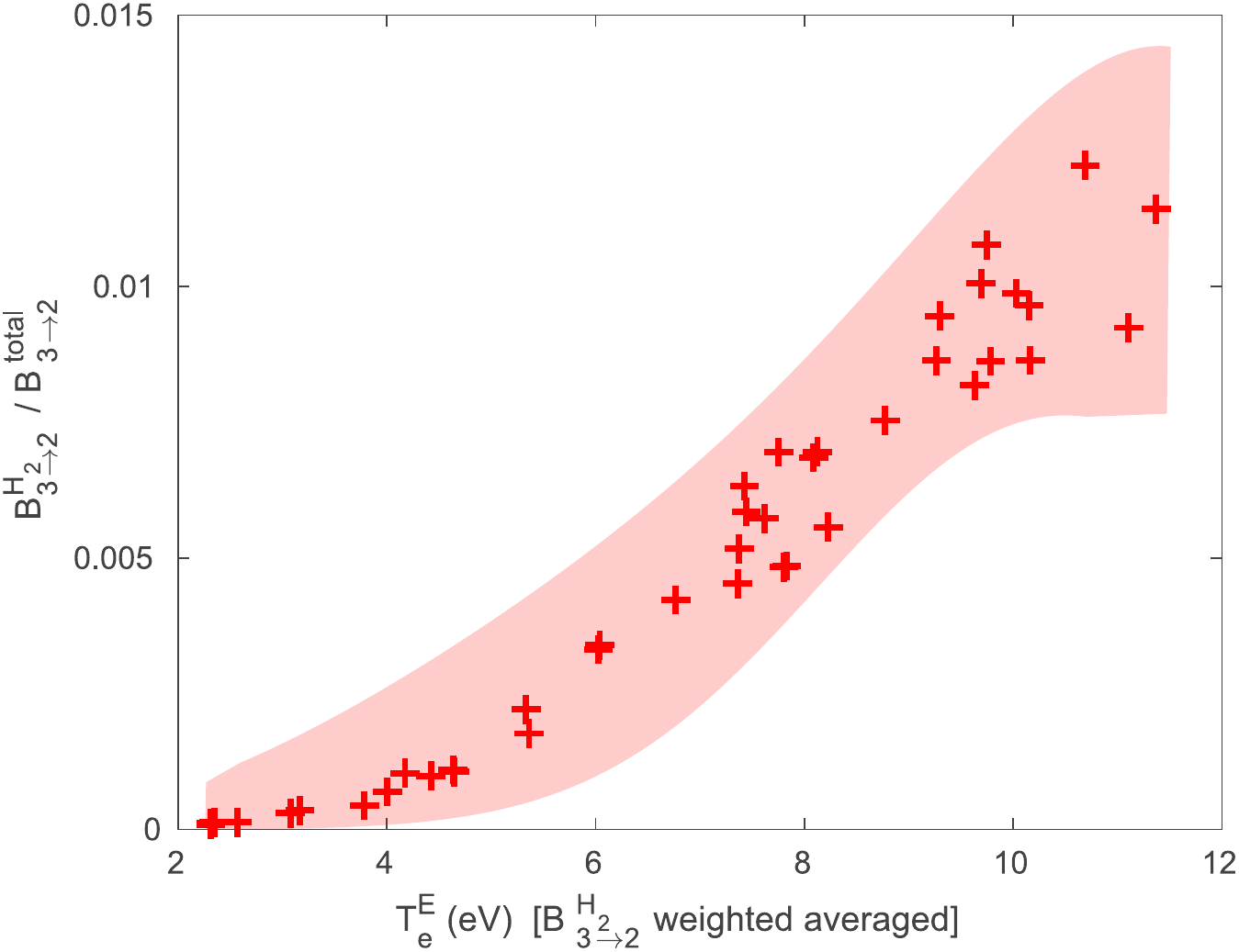}
    \caption{Estimated $H\alpha$ emission fraction attributed to $H_2$ (summed along the viewing fan) for the experimental discharge analysed in \cite{Verhaegh2020a} as function of the characteristic $T_e^E$ for that viewing fan (weighted averaged by $B_{3\rightarrow2}^{H_2}$)}
    \label{fig:EmissFracH2}
\end{figure}

The inferred fraction of $H\alpha$ attributed to $H_2$ (e.g. $B_{3\rightarrow2}^{H_2} / B_{3\rightarrow2}^{total}$ along the total viewing fan) for the experimental discharge analysed in \cite{Verhaegh2020a} is shown in figure \ref{fig:EmissFracH2} as function of the 'characteristic' excitation electron temperature (weighted (by $B_{3\rightarrow2}^{H_2}$) average $T_e^E$ along the viewing fan). We observe that the \emph{relative} contribution of $H_2$ to $H\alpha$ is highest at high temperatures. At relatively low temperatures (such as the cases shown in section \ref{ch:results})  \ref{ch:results}), $\frac{B_{n\rightarrow2}^{H_2}}{B_{n\rightarrow2}} < 10^{-4}$ (for the result indicated in figure \ref{fig:SepaEmiss}). Therefore, even if $g_{H_2} (T_e^E)$ obtained from SOLPS-ITER is strongly underestimated, it would be unlikely that this would influence the obtained solutions.

This result is somewhat in contrast to results from previous studies in limiter devices \cite{Brezinsek2005}, which have shown that $H_2$ dissociation may contribute to $H\alpha$ emission. In such conditions, there is a relatively higher molecular density at the location of the hot temperature plasma, whereas in a divertor we obtain high molecular densities at low electron temperatures (see figure \ref{fig:fH2_fit_TCV_MASTU}). Despite this anti-correlation between $n_{H_2} \Delta L$ and $T_e$, the emission fraction of $H\alpha$ attributed to $H_2$ increases with $T_e$ (figure \ref{fig:EmissFracH2}) because $PEC_{3\rightarrow2}^{H_2} (n_e, T_e)$ is strongly correlated with $T_e$ for $T_e < 10$ eV. Furthermore, as was also mentioned in \cite{Brezinsek2005}, a significant amount of this $H\alpha$ emission in those limiter devices studies could also have arisen from dissociative recombination of $H_2^+$.

\subsection{Additional $H\alpha$ emission contributions not related to plasma-atom interaction and $H_2$ chemistry}
\label{ch:OtherDa_Emiss}

In carbon machines, such as TCV and MAST-U, reactions with hydrocarbons could lead to excited $n=3,4$ atoms leading to additional $H\alpha$,  $H\beta$ emission. Additionally, opacity of $Ly\beta$ and $Ly\gamma$ can also lead to additional excited $n=3,4$ atoms (and thus $H\alpha$ and $H\beta$ emission). We discuss these two processes here and estimate their importance for TCV.

To obtain an upper limit estimate for the possible atomic emission of hydrocarbons, we assume that \emph{all} neutral carbon from validated SOLPS simulations for TCV \cite{Fil2017,Fil2019submitted} exists in the form of hydrocarbons. \footnote{Since neutral carbon recombination has been deactivated for most of these simulations (\cite{Fil2017} - the SOLPS-ITER default at the time), the sum of the neutral and $C^+$ densities is utilised as an upper estimate of the neutral carbon density.} To map these hypothetical hydrocarbon densities to the $H\alpha$ emission we utilise reaction cross-sections from \cite{Shirai2002} for $C H_4$ (the cross-sections for $H\alpha$ emission from \cite{Shirai2002} are similar for the full range of hydrocarbons presented: $C H_4$, $C_2 H_2$, $C_2 H_6$, $C_2 H_4$). For this extreme case, the estimated $H\alpha$ emission from $n=3$ excited atoms after hydrocarbon reactions is more than $10^4$ smaller than the total $H\alpha$ emission. It is thus unlikely that hydrocarbon chains contribute significantly to the $H\alpha$ emission.

%Opacity of $Ly\beta$ can lead to significant increases of the $H\alpha$ emission. Additionally, the plasma-molecule interactions discussed alter the $H\alpha$ emission profile spatially: while $Ly\alpha$ is brightest at the ionisation region; plasma-molecule interactions near the target (involving $H_2^+, H^-$) shift the $H\alpha$ (and thus $Ly\beta$) emission region towards the target below the ionisation region, where the neutral density is higher and opacity is more likely.

With respect to the effect of opacity, the neutral density along the spectroscopic line of sight remains smaller than $10^{18} m^{-2}$ in SOLPS simulations for TCV \cite{Fil2017,Verhaegh2019a}. At this level, not much opacity is expected \cite{Terry1998,Behringer2000}, which is indeed confirmed by post-processing the SOLPS simulations using ray-tracing techniques. For the cases shown in figure \ref{fig:SepaEmiss}, the impact of photon opacity on $H\alpha$ is estimated to be around 2\%. However, opacity can be much more significant on devices with higher neutral densities than TCV. This would impact our analysis as $Ly\beta$ opacity can raise the $H\alpha$ brightness and the analysis would have to be modified to account for this (section \ref{ch:discus_applicability}). %These indicate very minor modifications to the population escape factors due to opacity, indicating ultimately modifications to the 2D profile of the $H\alpha$ emissivity of up to 3\%. Therefore, opacity is not expected to significantly alter the $H\alpha$ emission on TCV.

\subsection{$H\alpha$ as a monitor for MAR and atomic line radiation associated with $H_2$ plasma chemistry}
\label{ch:Dalpha_mon}

%REVISE BASED ON PREVIOUS CHANGES

\begin{figure}[htb]
    \centering
    \includegraphics[width=0.7\linewidth]{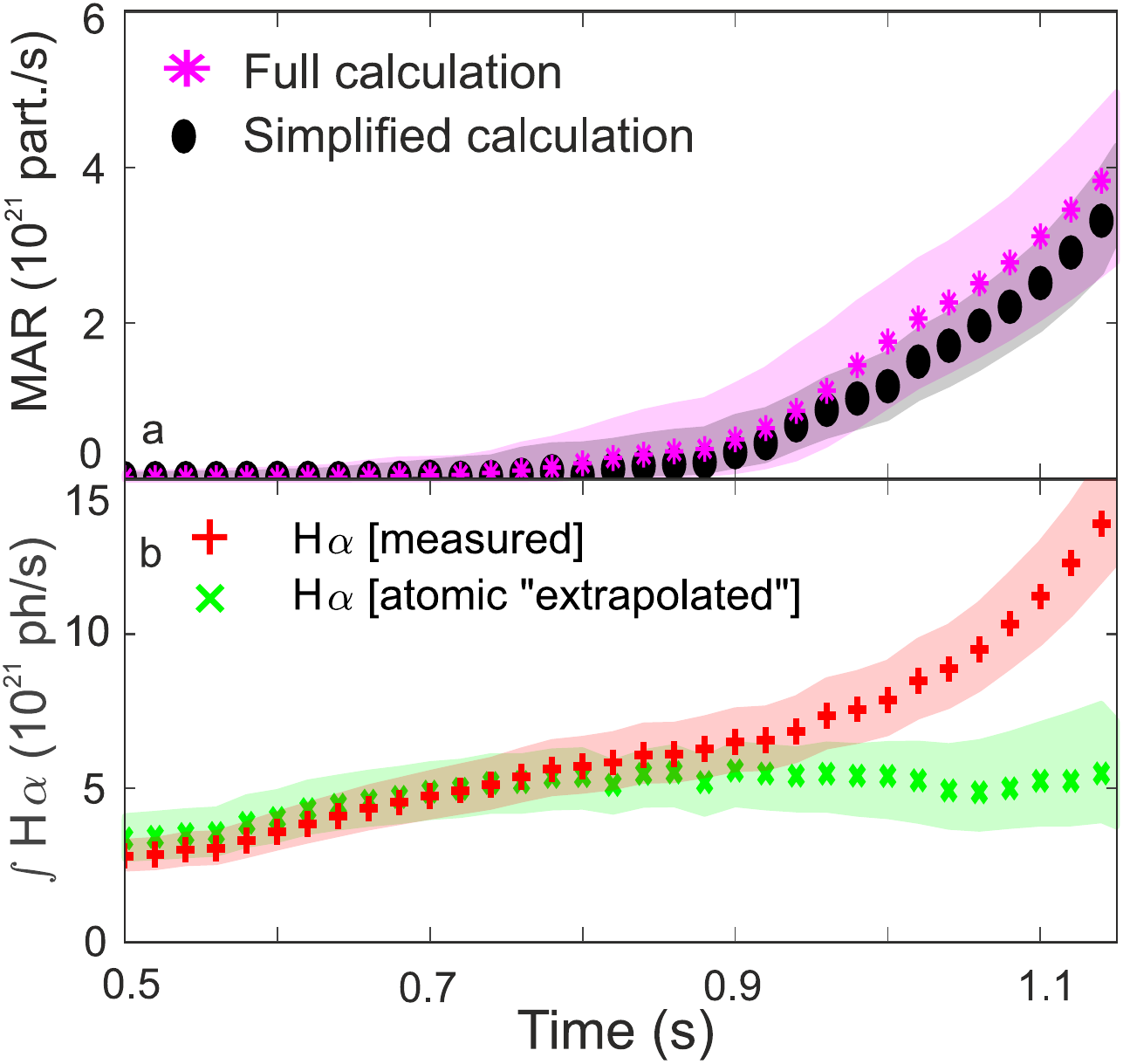}
    \caption{Comparison of analysis techniques based on TCV experimental results for \# 56567 which have been discussed, from a divertor physics point of view, in \cite{Verhaegh2020a} from which the full BaSPMI MAR result and the measured and extrapolated $H\alpha$ have been adopted. a) Comparison of the 'simplified' divertor MAR calculation (based on multiplying the difference between the atomic extrapolated $H\alpha$ and the measured $H\alpha$ (figure b) with the 'MAR/$H\alpha$ photon' ratio for $H_2^+$) against the full BaSPMI result.}
    \label{fig:MAR_Simplified}
\end{figure}

The increase of $H\alpha$ during detachment, or more specifically the 'anti-correlation between $H\alpha$ and the ion target current' during detachment is observed on several devices \cite{Stangeby2000,Lomanowski2020,Hollmann2006}. The results derived and described in this paper were applied to a detachment discharge in TCV \cite{Verhaegh2020a}. This shows the extrapolated $H\alpha$ atomic estimate matches the measured $H\alpha$ until the detachment onset after which the measured $H\alpha$ keeps on increasing while the atomic estimate of $H\alpha$ saturates. We can conclude two things from this result, which is repeated in figure \ref{fig:MAR_Simplified}b. 

First, we have shown in \cite{Verhaegh2020a} that the increase of $H\alpha$ during detachment cannot be explained through electron-ion recombination on TCV, but is explained through plasma-molecule interactions with $H_2^+$ and $H^-$ (figure \ref{fig:MAR_Simplified}b). Although electron-ion recombination can be higher on higher divertor density machines, it is likely that the increase of $H\alpha$ in such machines is also, at least partially, due to plasma-molecule interactions with $H_2^+$ and $H^-$. In fact, contributions of $H_2^+$, and/or $H^-$ were also suspected in JET \cite{Lomanowski2020} and DIII-D \cite{Hollmann2006} based on the experimentally measured trend and magnitude of $H\alpha$ emission.

Secondly, we have shown in \cite{Verhaegh2020a} that comparing $H\alpha$ measurements against its atomic extrapolation based on the medium-n Balmer lines is a quick and useful monitor for plasma-molecule interactions involving $H_2^+$ and/or $H^-$ during detachment, which can be achieved with only the atomic part of the BaSPMI technique. This is also illustrated in figure \ref{fig:MAR_Simplified} where we observe that the MAR ion sink starts to appear when there is a bifurcation between the atomic extrapolation of $H\alpha$ and the measured $H\alpha$.

This quick monitor can also be used as a quantitative estimate for radiative power loss and MAR from plasma-molecule interactions, which we denote as a "'simplified' MAR calculation". This is done by taking the difference between the measured $H\alpha$ and the atomic extrapolation for each plasma chord, which is multiplied  with the 'MAR per $H\alpha$ ratio' for $H_2^+$ (obtained from figure \ref{fig:MARMAIperDa}), obtained using the inferred $n_e, T_e^E$ from the atomic analysis. The result is a MAR estimate for each chord (e.g. ions per metre squared per second) which is integrated to provide the total MAR sink in the viewing fan (e.g. ions per second). A similar MAR ion sink is obtained if the divertor-integrated $H\alpha$ attributed to plasma-molecule interactions is multiplied with the 'MAR per $H\alpha$ ratio' for $H_2^+$ using an assumed $n_e = 10^{20} m^{-3}$ and $T_e = 1 eV$.

The above approach only uses the atomic portion of BaSPMI and neglects the impact of $H_2$ plasma chemistry on the medium-n Balmer lines. Furthermore, this assumes that Balmer line emission attributed to $H_2$ plasma chemistry arises from interactions of the plasma with $H_2^+$. The result of this for the total divertor MAR ion sink is shown in figure \ref{fig:MAR_Simplified} a, where it is compared against the full BaSPMI analysis (obtained from \cite{Verhaegh2020a}). A similar agreement between the simplified and full analysis is obtained for the radiative power loss estimates associated with $H_2$ plasma chemistry. 

These estimates appear to be fairly accurate despite the lack of quantitatively separating $H\alpha$ emission from plasma-molecule interactions with $H_2^+$ and $H^-$ and despite accounting for the influence of $H_2$ plasma chemistry on the medium-n Balmer lines. The reason for this is likely that, within experimental uncertainties, the 'MAR per $H\alpha$ ratios' for $H_2^+$ and $H^-$ are similar (figure \ref{fig:MARMAIperDa}). This also indicates that, at least for estimating the MAR rate, the calculation is insensitive to chordal integration effects - which is supported with the results from section \ref{ch:analysis_solps}.

However, the full analysis chain would be required to estimate the 'molecular' contributions to the medium-n Balmer line emission, which can be important for ionisation estimates in detached plasmas \cite{Verhaegh2020a}.

Therefore, monitoring $H\alpha$ and comparing it to its atomic estimate is sufficient to: 
\begin{enumerate}
    \item Show that plasma-atom interactions involving $H_2^+$ (and possibly $H^-$) may occur (in environments with low/negligible $Ly\beta$ opacity).
    \item Estimate what their influence on the plasma is in terms of particle and power losses.
\end{enumerate} 

Afterwards, one could consider running the full analysis presented to: 
\begin{enumerate}
    \item Propagate this information to all Balmer lines to get a self-consistent picture which separates each hydrogenic line into its individual contributions, similar to figure \ref{fig:SepaEmiss}). 
    \item Delineate the plasma-molecule contributions from $H_2^+$ and $H^-$.
\end{enumerate} %Of course, the situation is not this simple in the case where significant opacity of $Ly\beta$ occurs (since in this case the $H\alpha$ emission can increase due to opacity). 
 
%However, we have shown that the $H\alpha$ signal can be used for more as 'just' a monitor for plasma-molecule interactions with $H_2^+$ and $H^-$: using the mismatch between the measured $H\alpha$ signal and the atomic estimate we were able to predict radiative power losses (due to plasma-molecule interactions with $H_2^+$ and $H^-$) using figure \ref{fig:RadPerDa} as well as MAR rates using figure \ref{fig:MARMAIperDa} in quantitative agreement with the full calculation (figure \ref{fig:PowerPartBal}). The observation of $H\alpha$ in comparison of that of an 'atomic' estimate could thus lead to quantitative predictions of MAR and radiative loss rates.

\subsection{Balmer line contributions from plasma-molecule interactions and Fulcher band investigations}
\label{ch:discus_fulcher}

%REVISE SECTION

%It should be emphasized that the analysis technique does not employ a model or reaction rates for the various 'creation' and 'destruction' reactions of $H_2^+$ or $H^-$ to determine their respective densities or emission contributions. Instead, it splits the various emission pathways quantitatively into its separate atomic ($H^0, H^+$) and 'molecular' contributions ($H_2, H_2^+, H^-$) using the \emph{measured Balmer line brightnesses}. This is in contrast to how MAR rates are estimated from Fulcher band measurements in literature \cite{Hollmann2006,Fantz2002,Fantz2001}, where such reaction rates are used to model $H_2^+/H_2$ ratios. Extracting the $H_2$ density then from the Fulcher band brightness results in an indirect measure of the $H_2^+$ density. This distinction is important because the validity of such reaction rates; their relative importance and their isotope dependencies are still debated in literature \cite{Kukushkin2017}. They could also differ between carbon and metallic walls (which influences the vibrational state of molecules reflected from the wall, influencing the creation/destruction mechani\textbf{}sms behind $H_2^+, H^-$) \cite{Wischmeier2005,Wischmeier2004,Cadez2011,Miyamoto2003}. As such, this analysis chain can be a used as a tool to answer such questions by inferring the presence of $H_2^+$ and $H^-$ and its influence on the plasma.

Previous research on investigating plasma-molecule interactions in the divertor spectroscopically generally focused on monitoring the molecular band emission, such as the Fulcher band which comes from electronically excited molecules \cite{Fantz2002,Fantz2001,Hollmann2006} after plasma-molecule \emph{collisions}. %Such measurements provide important information on \emph{collisions} between the plasma and $H_2$, which \emph{excite the molecules rovibronically}. Such collisions \emph{are different} from \emph{reactions} between the plasma and $H_2^+$ and $H^-$, which result in MAR, MAI and \emph{excited atoms} resulting in atomic line radiation/radiative power loss. %This is further illustrated schematically in figure \ref{fig:FulcherBalmerSchem}.

Although plasma-molecule \emph{collisions} are different from plasma-molecule \emph{reactions}, MAR rate estimates from \emph{reactions} with $H_2^+$ and $H^-$ have been estimated previously using measurements of the Fulcher band. Those measurements provide information on $H_2$ and its vibrational distribution, which is combined with $n_e, T_e$ estimates and a model or simulations to extrapolate the $H_2$ density to the $H_2^+$ density and its resulting MAR rate \cite{Fantz2002,Fantz2001,Hollmann2006}.% which requires several assumptions (as discussed in \ref{ch:analysis_solps}. % have been combined with $n_e, T_e$ estimates to a) model the $H_2^+ / H_2$ ratio by making assumptions on the (lack of) transport of $H_2^+$ and its 'ionisation' (e.g. $H_2$ to $H_2^+$) and 'recombination' (e.g. $H_2^+$ to $H_2$) rates - e.g. given those assumptions one approximates the $H_2^+$ density from the $H_2$ density; b) multiply such $H_2^+$ densities with the MAR/MAI rate of $H_2^+$ \cite{Fantz2002,Hollmann2006}. 

%\begin{figure}[H]
%    \centering
%    \includegraphics[width=0.4\linewidth]{FulcherBalmerSchematic.pdf}
%    \caption{Schematic overview of the processes between the plasma and the molecular cloud resulting in Fulcher emission through excitation of the molecular cloud through \emph{collisions} and Balmer emission through plasma-molecule \emph{reactions} resulting in excited atoms.}
%    \label{fig:FulcherBalmerSchem}
%\end{figure}

This differs from the approach in this work which aims to extract the Balmer line emission arising directly from the excited atoms after plasma-molecule interactions with $H_2^+, H^-$. Therefore, it does not require assuming that the location of the $H_2$ electronic excitation (e.g. Fulcher band emission) is the same as the location of the MAR reactions along the lines of sight. That assumption could be problematic as electronic excitation of $H_2$ requires fairly high electron temperatures ($T_e > 3-4$ eV), whereas MAR can occur at lower temperatures ($T_e = [1.5 - 4]$ eV). Our measurements in \cite{Verhaegh2020a} indicate, for instance, that the peak Balmer line emission from excited atoms after \emph{reactions} between the plasma and $H_2^+$ (and/or $H^-$) occurs at a different position than the region with the brightest Fulcher band emission \cite{Verhaegh2020a}. 

Extrapolating MAR rates out of a Fulcher band analysis requires using a model to predict the creation and destruction rates of $H_2^+$ and $H^-$, which may have large uncertainties and isotope dependencies (see section \ref{ch:discussion_isotope}). The BaSPMI analysis, however, does not \footnote{There is a very weak dependence on such rates for estimating the "MAR/MAI per $H\alpha$ photon" ratio as this changes depending on whether $H_2^+$ was created through molecular charge exchange or $H_2$ ionisation - see section \ref{ch:discussion_isotope}} rely on such assumptions as it monitors \emph{the destruction of $H_2^+$ and/or $H^-$ into excited atoms} directly, rather than using rates and models to model the $H_2^+$ and/or $H^-$ densities based on estimations of the $H_2$ density. %are strongest near the target, whereas the Fulcher band emission is dim (e.g. hard to detect experimentally) near the target and is brighter below the ionisation region. %, than the Fulcher band emission location. The Fulcher band is observed to emit brightest near the ionisation region, while Balmer line emission from $H_2^+$ (and/or $H^-$) remains peaked near the target.

BaSPMI can be used as an alternative tool to the Fulcher spectra to investigate more closely how such interactions with $H_2^+, H^-$ influence the plasma and provides an indirect tool to investigate the conditions which promote $H_2^+, H^-$ creation. It could for instance be used to study differences in MAR from $H_2^+$ and/or $H^-$ between carbon and metallic walls (which influences the vibrational state of molecules reflected from the wall, influencing the creation/destruction mechanisms behind $H_2^+, H^-$ \cite{Wischmeier2005,Wischmeier2004,Cadez2011,Miyamoto2003}). As BaSPMI uses Balmer line measurements, it could be a tool which is more straightforward to employ as Balmer line measurements are more routinely employed on tokamaks. They are often easier to diagnose than the Fulcher band, given the high spectral resolution and high sensitivity often required for molecular band studies. Given that BaSMPI uses Balmer line measurements, which can be measured using 2D multi-spectral imaging diagnostics \cite{Perek2019submitted}, its analysis could in principle be extended to a 2D analysis. For this, however, electron density estimates would be required, which could be obtained using Helium line spectroscopy \cite{Lisgo2009}, coherence imaging techniques (for Stark broadening) or Bayesian analysis techniques \cite{Bowman2020}.

BaSPMI is, however, influenced by opacity and requires high quality collisional radiative model results to provide information on how the various $H_2$ plasma chemistry processes lead to excited hydrogen atoms and resulting atomic line emission \cite{Wunderlich2020,Wuenderlich2016}. Furthermore, as $H_2$ ionisation occurs in a similar $T_e$ window as Fulcher band emission, a Fulcher band analysis may provide more accurate MAI (from $H_2^+$) estimates than BaSPMI.

\subsection{Reliance on molecular data: isotope effects and impacts from vibrational states}
\label{ch:discussion_isotope}

The Yacora (on the Web) \cite{Wuenderlich2016,Wunderlich2020} collisional radiative model and AMJUEL database \cite{Reiter2008} does not (yet) provide explicit parallel information for $H$ and $D$ or $T$ related processes. Rather the preponderance of rates for $H$ and various assumptions must be made in their application to $D$ (and $T$). This is an important caveat of this analysis. If  collisional radiative results become available for $D$ (and $T$) in the future, they could be used instead in the outlines analysis approach. %It should be noted, however, that no assumptions are made on the isotope effect on the creation/destruction mechanisms of $H_2^+, H^-$ (since no assumptions are made on the creation/destruction rates of $H_2^+, H^-$ at all); which is currently debated \cite{}.

In this discussion, it is important to distinguish between two categories of atomic/molecular data. First, there are the actual \emph{creation and destruction rates} of $H_2^+$ and/or $H^-$. Those rates are important for modelling the $H_2^+$ and/or $H^-$ densities based on the $H_2$ density. Secondly, there are the photon emission coefficients which provide estimates on the distribution of the excited states of hydrogen atoms \emph{after $H_2^+$ and/or $H^-$ undergoes a reaction resulting in hydrogen neutrals}. Particularly, some of the rates of the first category are discussed in literature to have potentially strong isotope dependencies \cite{Kukushkin2016,Krishnakumar2011} as well as significant dependencies on the vibrational distribution of $H_2$. 
 %Additionally could provide clarity on the isotope dependencies of the creation of $H_2^+$ as well.% for the same reason, although particularly the MAI ion source estimates depend (weakly) on ratio between the charge exchange and total $H_2^+$ creation rates (see section \ref{ch:analysis_molMARMAI}). 

Our analysis almost fully depends on only rates from the second category. However, we use rates from the first category for estimating the "MAR/MAI per $H\alpha$ photon" ratio for $H_2^+$ (equation \ref{eq:MARMAIDaH2p}) as we must distinguish between $H_2^+$ creation through molecular charge exchange and $H_2$ ionisation (equation \ref{eq:ffromCX}). Distinguishing between these two different $H_2^+$ creation mechanisms depends on the $H_2^+$ molecular charge exchange rate (equation \ref{eq:ffromCX}), which in particular is expected to be both isotope and $H_2$ vibrational level dependent \cite{Kukushkin2016}. 

\begin{figure}[htb]
    \centering
    \includegraphics[width=0.8\linewidth]{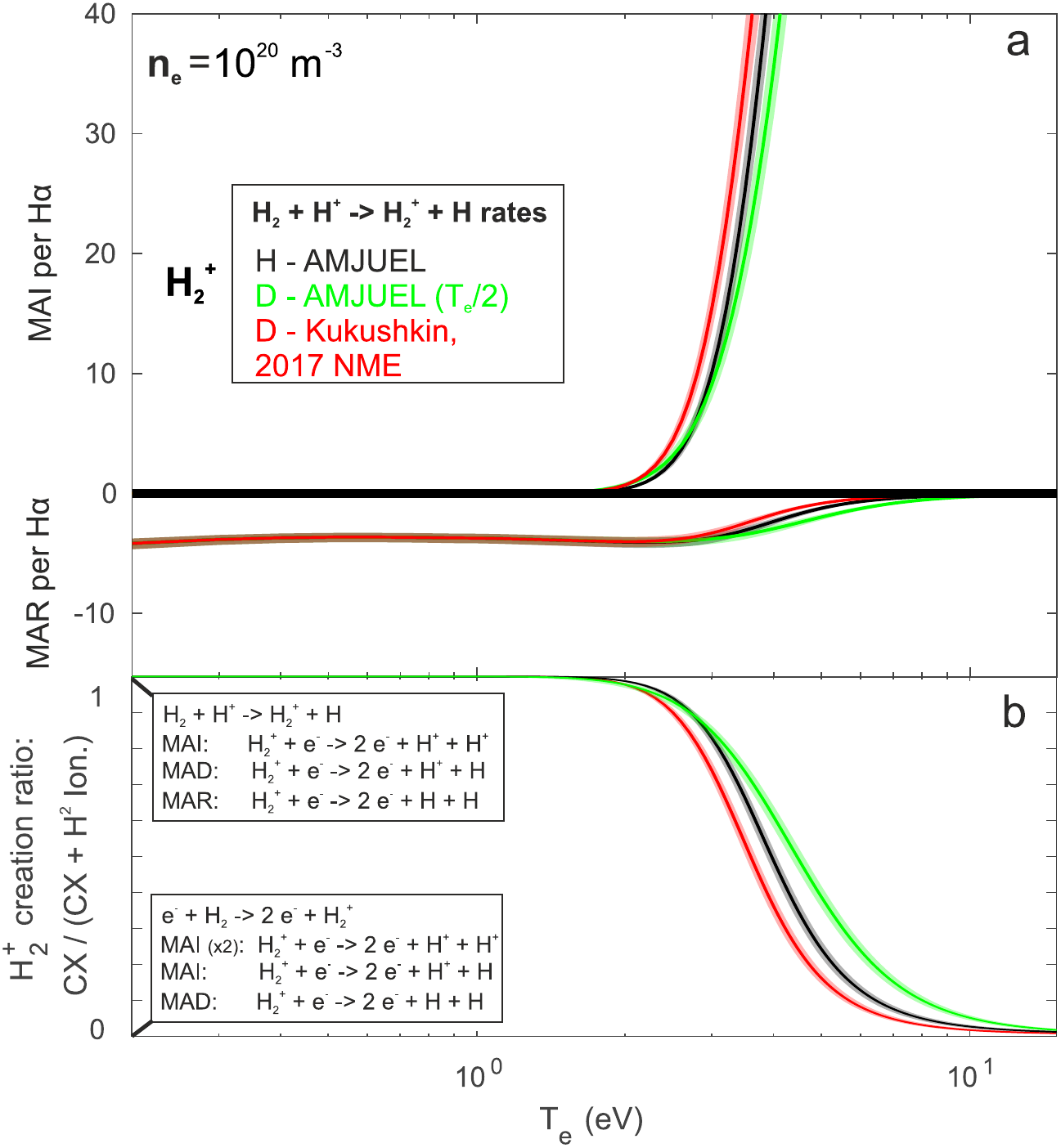}
    \caption{A more detailed version of figure \ref{fig:MARMAIperDa}. a) "MAR/MAI per $H\alpha$" ratios for various molecular charge exchange reaction rates for the creation of $H_2^+$ as function of $T_e$. These depend on $f_{H_2^+\text{ from CX}}$ (e.g. ratio between the CX $H_2^+$ creation rate and the total $H_2^+$ creation rate (CX + $H_2$ ionisation)), which is shown in figure b, through equation \ref{eq:MARMAIDaH2p}.}
    \label{fig:MARMAI_per_Halpha_complete}
\end{figure}

From a detailed analysis we, however, find that using different models for the molecular charge exchange rate only has a negligible impact on the "MAR/MAI per $H\alpha$ photon" ratio, despite the molecular charge exchange rate itself changing by an order of magnitude in detachment-relevant conditions ($T_e = [1 - 3] eV$) between the different models used. We perform this analysis by calculating the fraction of $H_2^+$ created by molecular charge exchange $f_{H_2^+\text{ from CX}}=\frac{<\sigma v>_{H_2 + H^+ \rightarrow H_2^+ + H}}{<\sigma v>_{H_2 + H^+ \rightarrow H_2^+ + H} + <\sigma v>_{e^- + H_2 \rightarrow 2 e^- + H_2^+}}$ - equation \ref{eq:ffromCX} (figure \ref{fig:MARMAI_per_Halpha_complete} b) and its impact on the "MAI/MAR rate per $H\alpha$ photon ratios" for $H_2^+$ (equation \ref{eq:MARMAIDaH2p}), shown in figure \ref{fig:MARMAI_per_Halpha_complete} a. The result for three different molecular charge exchange rates are shown: 1 - the default rates from AMJUEL for hydrogen; 2 - the default rates from AMJUEL where the rates are shifted by dividing the electron temperature by two to model the deuterium rate \footnote{This is the default in Eirene \cite{Reiter2008}.}; 3 - an alternative rate for deuterium investigated in \cite{Kukushkin2016}. The vibrational distribution is modelled using an assumed $H_2$ temperature, which has been varied in the Monte Carlo uncertainty processing throughout the entire validity regime [0.37 - 10 eV] of the data. We find in figure \ref{fig:MARMAI_per_Halpha_complete} that the impact of the various rates on the calculated "MAI/MAR rate per $H\alpha$ photon ratios" for $H_2^+$ is small. Therefore, our analysis seems to be robust against these uncertainties.

The reason for this is that our analysis only depends on $f_{H_2^+\text{ from CX}}$. Modelling the $H_2^+/H_2$ ratio instead based on a no-transport model, however, would depend on the \emph{relative ratio between the sum of the $H_2^+$ creation and destruction mechanisms} - e.g. proportional to $<\sigma v>_{H_2 + H^+ \rightarrow H_2^+ + H}+ <\sigma v>_{e^- + H_2 \rightarrow 2 e^- + H_2^+}$. The latter ratio would change by an order of magnitude if the molecular charge exchange rate is dominant and changes by an order of magnitude. $f_{H_2^+\text{ from CX}}$, however would be insensitive to such changes in the molecular charge exchange rate (as long as this rate is significantly larger than the $H_2$ ionisation rate). 

%is divided by two First, the reaction rates used are intended for hydrogen; while the discharges here simulated are deuterium. These creation/destruction rates can have strong dependencies on the isotope; but their isotope dependencies are still heavily debated \cite{Kukushkin2016}. Secondly, these rates depend on the vibrational distribution of $H_2$, which is not normally ('fully') accounted for in SOLPS-ITER. Instead, the vibrational temperature is set according to an assumed temperature (0.1 eV) \cite{Reiter2008,Reiter2005,Kotov2007}.

There is not only an expected isotope dependence on the $H_2^+$ creation rate, but also on the $H^-$ creation rate. Experimental evidence indicates the cross-sections for creating $D^-$ at low vibrational levels is much less likely than creating $H^-$ \cite{Krishnakumar2011}. However, to allow for the largest degree of flexibility, we have opted to allow for the possibility of reactions with $H^-$ resulting in Balmer line emission in our analysis. As our analysis does not depend on the reaction rates for creating $H^-$, it could potentially provide clarity on the presence of $H^-$ - but that requires further investigation.

The discussion in this section also applies to the application of this analysis for different isotope mixtures such as D\&T or H\&D. If the impact of such isotope mixtures on the molecular PECs as well as on the MAR/MAI per $H\alpha$ photon ratios is limited ($<50\%$), it should be possible to apply this analysis. Possibly high resolution spectroscopy to separate the H, D, T Balmer lines could help separate reactions per isotope combination in such conditions. 

\subsection{The applicability of these techniques to different devices and its implications}
\label{ch:discus_applicability}
In this work we have applied an analysis technique to separate the Balmer line emission from its various atomic and molecular channels; after which the power losses due to each individual channel as well as the ion sources and sinks can be estimated. Its workings have been demonstrated analysing synthetic diagnostic results obtained from SOLPS simulations of both TCV and MAST-U. Emission characteristics very likely differ, however, between TCV, MAST-U and higher power and/or density tokamaks such as ASDEX-Upgrade and JET. This raises the question how generally applicable our presented analysis techniques are. Below we address this from the point of view of plasma conditions, viewing geometry and diagnostic capabilities. %To obtain this, post-processing has been applied in order to 'model' the $H_2^+$ and $H^-$ densities in the divertor. Post-processing as been applied instead of the values directly obtained by SOLPS since the integration of those reactions in SOLPS is not complete and is still under discussion. In addition to this, sanity checking has been employed to make sure that a) the analysis technique correctly infers the lack of plasma-molecule interactions in the emission channels when the plasma-molecule interaction emission channels are excluded in the input brightnesses; b) the analysis technique correctly infers that most/all plasma-molecule interaction arises from $H_2^+$ if the $H^- + H^+$ emission channel is excluded from the input brightnesses. 

\paragraph{Applicability related to plasma conditions} Generally, the applicability of this analysis technique has been discussed in depth in \cite{Verhaegh2019a} where only atomic processes are considered for the Balmer line emission. It was reasoned that the atomic analysis of the Balmer lines should be generally applicable in both attached and detached conditions. We can make various quantitative estimates of the roles of plasma-molecule interactions on MAR and radiative losses, based on comparing the atomic contribution of $H\alpha$ with the total measured $H\alpha$, which only depends on the atomic analysis part of the analysis and should be generally applicable to other devices.

Although there are no strict $n_e$, $T_e$ limits of this analysis, 'soft' limits are estimated at $n_e = [10^{19} - 5 \cdot 10^{20}] m^{-3}$ (based on Stark broadening inferences and photon opacity - more information will follow) and $T_e = [0.2 - 50] eV$ (based on the availability of collisional-radiative model data \cite{OMullane,Wunderlich2020}.  

For the most part, based on the synthetic testing result, we would expect also the full analysis chain to be fairly well applicable also to other devices, given some caveats. 

Since the total $H\alpha$ emission in the divertor associated with plasma-molecule interactions (photons/s) is sufficient for quantitatively estimating MAR ion sinks and radiative power losses associated with $H_2$ chemistry, those two analysis estimates are robust against chordal integration effects, which is in agreement with the synthetic testing results. The separation between emission from $H_2^+$ and $H^-$ could be more sensitive to chordal integration effects, given its relatively strong dependence on the electron density (figure \ref{fig:RatHmHp_RatH2p}). Chordal integration effects could occur as the Stark-broadened $n_e$ estimate from the higher-n Balmer lines (which are more sensitive to EIR) could be different from the electron density at the $H_2^+$ and $H^-$ emission location (see for example figure \ref{fig:SOLPS_EmissProfs}d,j); although this was not found to lead to a discrepancy in the synthetic testing results. MAI estimates from the BaSPMI are more sensitive to chordal integral effects, given the strong temperature dependence between the MAI per $H\alpha$ ratio (figure \ref{fig:MARMAI_per_Halpha_complete}). This is also shown in the synthetic testing results in section \ref{ch:analysis_solps}.

This analysis relies on the lower-n Balmer lines and as such is susceptible to opacity. Photon opacity occurs at high neutral densities which are often correlated with high electron densities. In devices where $Ly\beta$ opacity is significant, which can be monitored using VUV spectroscopy based on the measured $Ly\beta / H\alpha$ ratio, such as JET \cite{Lomanowski2020} and C-Mod \cite{Terry1998}, modifications to this analysis have to be employed to separate the $H\alpha$ increase due to molecular processes and due to opacity. 

\paragraph{Applicability in terms of viewing geometry}

One caveat to the general applicability of the atomic Balmer line analysis (and thus also BaSMPI), is the placement of the lines of sight \cite{Verhaegh2019}. If the lines of sight are placed in such a way that they go through both a significant ionisation and bright recombination emission region, the electron-impact excitation contribution to the total Balmer line emission could be lower than a few percent. In that case, there is insufficient information about electron-impact excitation in the signal resulting in large uncertainties in the ionisation estimates. This can occur if there is a large shift between the respective electron densities and electron temperature profiles along the line of sight \cite{Reimold2017}. 

It should be noted, however, that a closed and higher density divertor may also facilitate a more natural separation of the various emission regions as the characteristics mean-free-paths become smaller. This, for instance, is why the synthetic testing results are more consistent for MAST-U than TCV (see section \ref{ch:analysis_solps}).

\paragraph{Applicability in terms of diagnostic capabilities}

As illustrated in this work, inferring information from plasma-molecule interactions simultaneously with the ionisation source complicates extracting the ionisation rate in strongly detached conditions unless temperature 'constraints' are employed. Although the temperature constraints employed here may only be applicable to specific situations or specific devices, other constraints could be employed, for instance based on divertor Thomson scattering or impurity line spectroscopy. Essentially, what is required is a way of estimating whether the inferred electron-impact excitation temperature ($T_e^E$) for a single (or multiple) chord(s) is 'likely' or 'unlikely'. $T_e^E$ will correspond to the characteristic temperature of the high temperature region along the line of sight. Those temperature constraints enable obtaining ionisation estimates even when the electron-impact excitation (of $H$) component of the Balmer line emission is fairly small.

The full BaSPMI analysis puts requirements on the divertor spectroscopy system. It requires inferred electron densities (from Stark broadening for line-of-sight spectroscopy - $n_e > 10^{19} m^{-3}$ \cite{Verhaegh2018}) as well as high quality absolute brightness of two medium-n Balmer lines in addition to $H\alpha$ and $H\beta$. Therefore, BaSPMI requires a flexible divertor spectroscopy system which can be used to measure 4-5 Balmer lines. Given the large differences between the brightnesses of the various Balmer lines, these measurements may likely have to be restricted to measuring 1-2 Balmer lines at the same time. In that case, either repeat discharges or multiple spectrometers would be required to measure the 4-5 required Balmer lines. Neutral density filters may need to be employed to attenuate the emission of the particularly bright Balmer lines ($H\alpha$, $H\beta$). 

%The necessity of adding additional constraints shows that the 'full analysis' is less 'robust' than the 'atomic analysis'. This is further highlighted in appendix \ref{ch:analysis_indepth_syndiag}, where the analysis result for a synthetic case where no emission due to plasma-molecules is present is shown. The analysis result does not directly show there is no emission due to plasma-molecule interactions, instead it indicates that the emission estimate from plasma-molecule interactions is likely small but has a high upper uncertainty. Those overestimates of the plasma-molecule interactions corresponds to underestimates of the excitation emission estimates. Therefore, it would be recommended only to run the full analysis if there are indications that various plasma-molecule interactions are contributing to the Balmer line emission by checking if there is any mismatch between the measured and atomic contribution of $H\alpha$ emission.

 %As explained previously, the opacity for $Ly\alpha$ and $Ly\beta$ line integrated measurements may have to be decoupled in the analysis, however, as both are differently influenced by plasma-molecule interactions; as the emission of those two is at different regions of the line of sight.

Ultimately, the entire analysis technique can be improved through the inclusion of multiple diagnostics in a consistent statistical framework such as in \cite{Bowman2020}. Such a technique would use '2D spectroscopy' using filtered camera imaging \cite{Perek2019submitted}. This could be further improved by complementing the Balmer line measurements with impurity lines, such as He-I lines, providing more information on electron temperature and electron density (see discussion in section \ref{ch:discus_fulcher}. Using toroidally-view filtered camera imaging spectroscopy would also enable a more precise localisation of all the different processes involved both along and across the field lines. This (partially) resolves the difficulty of 'line integration' effects and facilitates the separation of the various processes - since they are already spatially separated \cite{Perek2020}. Additionally, such a 2D variety of the analysis could enable estimating 2D maps of the $H_2^+$ and $H^-$ densities, which is not feasible otherwise. 

%The necessity of including such constraints is, in fact, due to the same complications as described previously in \cite{Verhaegh2019a}: if a line of sight goes through two emission regions with two different processes and one is much brighter than the other; than the relative uncertainty on the inference of the other process is increased. In other words, if the relative contribution of excitation to the Balmer line emission is very small; its relative uncertainty is increased. The inclusion of plasma-molecule interactions has further reduced the relative contribution of excitation to the Balmer line emission. Therefore, adding the various temperature constraints, also facilitates higher accuracy ionisation estimates in cases where other emission pathways are much brighter than the excitation emission either due to increased volumetric recombination / plasma-molecule interactions or due to the placements of the lines of sight. 

\section{Summary}
\label{ch:conclusion} 

%It is often observed that the $H\alpha$ emission rises strongly during detachment. We have shown quantitatively that such an increase is not consistent with plasma-atom interactions on TCV and is indicative of plasma-molecule interactions involving $H_2^+$ (and possibly $H^-$), leading to excited atoms which emit atomic line radiation as well as Molecular Activated Recombination (MAR).

Both plasma-atom and $H_2$ plasma chemistry (involving $H_2^+$ and/or $H^-$) can result in excited atoms leading to hydrogen atomic line emission. We have developed a new quantitative analysis technique, Balmer Spectroscopy Plasma Molecule Interaction - BaSPMI, to separate the emission of all Balmer lines into their electron-impact excitation (of $H$), electron-ion recombination (of $H^+$) and $H_2$ plasma chemistry related (involving $H_2, H_2^+, H^-$) contributions. This is facilitated using the consistency between the medium-n ($H\gamma, H\delta$) Balmer lines, which are less sensitive to plasma-molecule interactions, and lower-n Balmer lines ($H\alpha, H\beta$). The individual emission contributions are then used to:
\begin{itemize}
    \item Estimating the particle sources/sinks through plasma-atom (ionisation, recombination) and plasma-molecule (Molecular Activated Recombination / Ionisation - MAR/MAI) interactions.
    \item Estimating the radiative loss from \emph{excited atoms} arising from plasma-atom and plasma-molecule interactions.
\end{itemize} 

%Experimental results from TCV indicate a bifurcation between the measured $H\alpha$ and the atomic estimate of $H\alpha$ at the detachment onset stage. The expected $H_2$ concentrations are expected to have a negligible impact on the measured $H\alpha$. Reactions between the plasma and $H_2^+$ (and possibly $H^-$), however, are expected to result in this bifurcation. Our experimental illustration shows that such reactions can lead to significant modifications to the entire hydrogenic line series (and thus also hydrogenic line radiation). 

This analysis technique is validated by analysing synthetic spectra obtained from a synthetic divertor spectrometer using SOLPS simulations of both TCV and MAST-U. The Balmer line emissivity profiles along each line of sight showed strong spatial variations depending on the type of plasma-atom/molecule interaction. Despite this, however, the analysis result was in fair agreement (e.g. within uncertainty) with the direct outputs from SOLPS-ITER. The analysis was further tested by artificially removing certain plasma-atom/molecule interaction processes from the synthetic brightness. In this additional testing, the analysis correctly pointed out the lack of the removed processes.% the Balmer line emission profiles of the various plasma-atom/molecule interactions along the line of sight This analysis result is compared against the direct outputs from SOLPS indicating an agreement within uncertainty. Furthermore, artificially removing certain emission contributions in the synthetic brightnesses 

The analysis makes several assumptions which have been discussed in detail. It has been shown that these have only minor impacts on the analysis result. In particular, we have shown that the MAR ion sink in the plasma can be readily estimated by comparing the expected atomic contribution of $H\alpha$ (based on only the analysis of a medium-n ($n=5,6,7$) Balmer line pair) to the measured $H\alpha$. Those simplified MAR estimates are in quantitative agreement to the MAR estimates of the full BaSPMI result. The full BaSPMI analysis is, however, required to separate the contributions of $H_2^+$ and $H^-$ to the Balmer line emission as well as to estimate the impact of $H_2$ plasma chemistry on the medium-n Balmer lines. 

An experimental illustration of the analysis on TCV has been presented, indicating that plasma-molecule interactions can significantly contribute to the Balmer line emission. This has important implications for the diagnosis of tokamak divertors using hydrogen atomic line spectroscopy. We believe that this analysis technique should be generally applicable (in conditions where there is no significant photon opacity) to other tokamak devices to address those implications.  %he MAR ion sink estimates are shown to be robustly estimated as the full BaSPMI result is in agreement with a simplified MAR ion sink estimate

%Our analysis - Balmer Spectroscopy Plasma Molecule Interactions - BaSPMI - shows that plasma-molecule interactions can have a significant impact on the Balmer line spectra. BaSPMI provides a path forwards to investigate those interactions and separate their atomic line contributions from that due to plasma-atom interactions.  %(MAR, larger than electron-ion recombination) and significant hydrogenic line radiation (up to more than 50 \% of all hydrogenic radiative losses) from excited atoms arising from reactions between the plasma and $H_2^+$ (and possibly $H^-$). %Comparisons between the measured $H\alpha$ emissions to the atomic estimate of the $H\alpha$ emission (based on analysis of the medium-n Balmer lines) can facilitate quick estiamtes of the MAR and hydrogenic radiative loss rates. 

\section{Acknowledgements}

This work has received support from EPSRC Grant EP/T012250/1 and has been carried out within the framework of the EUROfusion Consortium and has received funding from the Euratom research and training programme 2014-2018 and 2019-2020 under grant agreement No 633053. This work was supported in part by the Swiss National Science Foundation. The views and opinions expressed herein do not necessarily reflect those of the European Commission. 

\appendix

\section{Balmer line emission model description for plasma-molecule interactions}
\label{ch:analysis_BalmerEmissModel}

Balmer line emission attributed to $H_2$ plasma chemistry can arise from interactions with $H_2, H_2^+, H_3^+$ and $H^-$ (figure \ref{fig:emiss_pathways}). In addition, Balmer line emission associated with $H^-$ can arise from either reactions starting with $H^- + H_2^+$ or $H^- + H^+$. Using a slab model for the plasma, we can describe the Balmer line brightness associated with $H_2$ chemistry ($B_{n\rightarrow2}^{H_2, H_2^+, H^-}$ - $photons/m^2 s$) using equation \ref{eq:BalMol}. Such plasma-slab models assume that all processes occur at the same location physically and implications of this have been discussed in detail for atomic reactions in literature \cite{Verhaegh2019,Verhaegh2019a,Verhaegh2018,Verhaegh2017}. 

The PEC coefficients in equation \ref{eq:BalMol}, obtained through Yacora (on the Web) \cite{Wuenderlich2016,Wunderlich2020}, are functions of the electron density, electron temperature, as well as the temperatures of the molecular species ($H_2, H_2^+, H_3^+, H^-$). Those latter temperature dependencies have, however, been found to be insignificant ($\ll 1$ \%) for most pathways (except $H^-$ \footnote{The additional temperature dependencies for $H^-$ are similar for all transitions: $PEC_{n\rightarrow2}^{H^- + H^+} \approx f(T_{H^+}, T_{H^-}) \times g(n, n_e, T_e)$ and therefore only impact the "MAR/$H\alpha$ emission coefficient" ratios employed in section \ref{ch:analysis_molMARMAI} in the analysis. A random temperature between 0.5-3 eV is assumed for the $H^-$ temperature as it can get some of the Franck-Cordon energy of the $H_2$ bond (2.2 eV) when $H_2$ dissociatively attaches with an electron to form $H^-$ ($e^- + H_2 \rightarrow H^- + H$). A random value between 0.8 to 1.5 times $T_e^E$ is assumed for the $H^+$ temperature, as estimated from SOLPS-ITER simulations \cite{Fil2017,Fil2019submitted,Wensing2019}.}) and thus a 1 eV temperature for $H_2, H_2^+, H_3^+$ has been assumed.

\begin{equation}
\begin{split}
    B_{n\rightarrow2}^{H_2, H_2^+, H^-} = &\Delta L n_e n_{H_2} PEC_{n\rightarrow2}^{H_2} (n_e, T_e) +  \\ &\Delta L n_e n_{H_2^+} PEC_{n\rightarrow2}^{H_2^+} (n_e, T_e) +  \\ &\Delta L n_e n_{H_3^+} PEC_{n\rightarrow2}^{H_3^+} (n_e, T_e) + \\ &\Delta L n_{H^+} n_{H^-} PEC_{n\rightarrow2}^{H^- + H^+} (n_e, T_e, T_H^+, T_{H^-}) + \\ &\Delta L n_{H_2^+} n_{H^-} PEC_{n\rightarrow2}^{H^- + H_2^+} (n_e, T_e, T_{H_2^+}, T_{H^-})
    \label{eq:BalMol}
    \end{split}
\end{equation}

To further simplify equation \ref{eq:BalMol}, we ignore the emission contribution from $H_3^+$ (which we estimate to be negligible based on post-processing of SOLPS simulations - section \ref{ch:analysis_solps}) and we assume that all emission from $H^-$ occurs from $H^-$ interacting with $H^+$ (rather than $H_2^+$) as the $H^+$ density is far larger than the $H_2^+$ density while their PECs are similar at the region where we would expect emission from such processes to occur. With those simplifications, we now obtain equation \ref{eq:BalMolSimpl} for $B_{n\rightarrow2}^{H_2, H_2^+, H^-}$.

\begin{equation}
\begin{split}
    B_{n\rightarrow2}^{H_2, H_2^+, H^-}\ &\approx \underbrace{\Delta L n_e n_{H_2} PEC_{n\rightarrow2}^{H_2} (n_e, T_e)}_{B_{n\rightarrow2}^{H_2}} + 
    \underbrace{\Delta L n_e n_{H_2^+} PEC_{n\rightarrow2}^{H_2^+} (n_e, T_e)}_{B_{n\rightarrow2}^{H_2^+}} + \\ &\underbrace{\Delta L n_e n_{H^-} PEC_{n\rightarrow2}^{H^- + H^+} (n_e, T_e)}_{B_{n\rightarrow2}^{H^-}} 
    \label{eq:BalMolSimpl}
    \end{split}
\end{equation}

\section{Detailed information on the iterative scheme and convergence}
\label{ch:analysis_iterative}

The analysis scheme uses an Euler iterative scheme in order to obtain self-consistent results between the various atomic and molecular contributions of the Balmer lines. The convergence of this \emph{relative change} in the estimated molecular contribution to the medium-n Balmer line is tracked per each iteration until it is 'converged'. The convergence criteria for this are listed below and are applied to the \emph{statistical output sample} (which is determined from all the various input distributions) for this relative change: %This statistical output sample covers all the various expected uncertainties in the input values as well as the atomic/molecular coefficients. The output criteria for the relative change in the output sample for the \emph{relative change} in the estimated molecular contribution to the medium-n Balmer line are:

\begin{enumerate}
\item At least 16 \% of the output sample should have a \emph{negative} change in the estimated molecular contribution (to make sure the analysis result is not 'drifting' towards a positive change).
\item At least 16 \% of the output sample should have a \emph{positive} change in the estimated molecular contribution (to make sure the analysis result is not 'drifting' towards a positive change).
\item The median of the change of the output sample should be between -0.2 and +0.2 \% (assuming the median is a proxy for the maximum likelihood, this makes sure that the analysis estimates are converged).
\item 68 \% of the output sample should have a relative absolute change below 2 \% (assuming the equal-tailed 68 \% quantile \cite{Cowles2013} is a proxy for the highest density interval \cite{Cowles2013} confidence intervals, this makes sure that the estimated uncertainties are converged).
\end{enumerate}

These convergence criteria have to be obeyed for at least 4 iterations simultaneously. These settings have been made after verifying that the results and their uncertainty have converged before reaching these criteria while keeping the number of iterations required acceptable (usually between 7-20).

\section{Improving the analysis through temperature constraints}
\label{ch:analysis_solps_Tt}

We introduce here two possible temperature constraints which can improve the analysis output estimates: one based on the excitation temperature near the target and one based on the observation of the CIII front. The goal of these 'constraints' is to provide some 'probability' for having a certain temperature at a certain location of the divertor. Other temperature constraints could be employed in a similar fashion. Before introducing our constraints, first we will explain how they are employed in the analysis technically.

Each Monte Carlo output sample point contains an estimate for the excitation-derived temperature $T_t^E$. Given these constraints, we can compute the probability of that sample point being true (for this we assume an asymmetric Gaussian probability distribution for $T_t^E$). The samples and their probabilities are then mapped to a probability density functions (PDFs) using a \emph{weighted} Gaussian Kernel density estimator (as opposed to an adaptive \emph{non-weighted} one when no contraints are employed \cite{Botev2010}). From the PDF estimates, the maximum likelihood and shortest interval corresponding to 68 \% uncertainty can be extracted, representing the estimated outputs and its uncertainty in a similar way as done in \cite{Verhaegh2019a}. 

This way of implementing constraints also changes how the integrated values should be obtained. Since the uncertainties are assumed to be systematic, the uncertainties applied to each chord per sample are the same - there is thus a correlation between the uncertainties of different chords when calculating integrated values (such as the total ionisation source). This could interfere with the way the constraints are built up. For instance, if all the analysis outputs would, hypothetically, be isothermal, then the maximum likelihood values of the temperature profile along the divertor leg would, after applying the constraints, not be isothermal (since a probability per point on the poloidal profile is ascribed). However, the integrated ionisation values would be determined all from isothermal solutions (since in this case a probability per poloidal profile is ascribed rather than a probability per point on the poloidal profile). Given these technicalities, we therefore determine the maximum likelihood of the poloidal profiles with their 68 \% confidence intervals of ionisation, recombination, etc. and integrate these profiles (and their upper/lower estimates) to get the estimates for the integrated (ionisation source, recombination sink, etc. parameters); which is more consistent with applying the constraints per point on the poloidal profile.

One drawback of employed constraints is that it strongly reduces the 'effective' Monte Carlo sample size of the simulation (since many sample points are given low probabilities and are thus 'effectively excluded'). Therefore, the analysis would require a larger number of Monte Carlo samples and thus more computational time when such constraints are employed. Furthermore, the requirement of using a weighted Kernel density estimator makes the choice for a suitable Kernel density estimator more restricted.

Employing temperature constraints in the analysis is only necessary for electron-impact excitation (of $H$) derived quantities in detached conditions. Adding the constraints to the other quantities, however, changes the maximum likelihoods insignificantly, although it does reduce their uncertainties.

\subsection{Target temperature constraint}

Assuming that we have a estimate for a range of possible target temperatures, we can use this to constrain the analysis. In this, we assume that this target temperature estimate is similar to the excitation emission weighted temperature of the nearest chord at the target ($T_t^E$). For synthetic testing we obtain this estimate directly from the SOLPS output (assuming an uncertainty of $\pm 1$ eV), while for the experimental analysis the target temperature has been estimated using power balance using the result from \cite{Verhaegh2019}. 

\subsection{CIII temperature 'exclusion' constraints}

An additional temperature constraint can be employed along the viewing chord fan; rather than a single point at the target. The front of the CIII (465 nm) emission line is an emission line frequently used in the qualitative characterisation of edge physics experiments in carbon devices, especially at TCV \cite{Theiler2017,Ravensbergen2020,Harrison2017} where it is used as a 'proxy' for the 'cold front' taking off the target \cite{Harrison2017} during detachment experiments. Depending on transport, the expected temperature of such a 'front (1/e fall-off point)' (assuming the carbon concentration does not change dramatically over the field line) is 4-8 eV. Below the CIII front the electron temperature will likely not be hotter than 8 eV. Likewise, above the front, the temperature will likely not be colder than 4 eV: the CIII emission front thus provides us with information to spatially 'exclude' (e.g. lower the likeliness of) certain temperatures. We can thus constrain the temperature samples further by adding a probability function which represents this argument - equation \ref{eq:ProbCIII}. 

In here $z$ represents the $z$ position of the line of sight intersecting the divertor leg, $z_f$ represents the CIII front location estimate and $T_{f,l}, T_{f,h}$ corresponds to the lowest/highest-temperature estimate of the front respectively. In this case, $z_f$ is determined analogously to \cite{Theiler2017,Harrison2017} as the 1/e fall-off-length of the CIII emission profile, which is determined by line of sight spectroscopy. The probablity used for each line of sight shown in equation \ref{eq:ProbCIII} represents an analytical depiction of the multiplication of two block-functions making two clauses likely: below CIII front $z_f$ and below temperature $T_{f,h} =8$ eV \& above CIII front $z_f$ and above temperature $T_{f,l} = 4$ eV. The  fall-off length of the functions are set to $kz = 2$ cm and $kT=1.5$ eV respectively. The solutions are largely insensitive to relatively modest changes of these fall-off parameters and temperature points.

\begin{equation}
    P(T_e) = \frac{1}{1 + \exp{-( \frac{z - z_f}{kz}})} \frac{1}{1 + \exp{-( \frac{T_e - T_{f,l}}{kT}})} + \bigg[1 - \frac{1}{1 + \exp{- (\frac{z - z_f}{kz}})}\bigg] \bigg[ 1 - \frac{1}{1 + \exp{-( \frac{T_e - T_{f,h}}{kT}})}\bigg]
    \label{eq:ProbCIII}
\end{equation}

\bibliographystyle{iopart-num}
\bibliography{all_bib.bib}

\providecommand{\newblock}{}
\begin{thebibliography}{10}
\expandafter\ifx\csname url\endcsname\relax
  \def\url#1{{\tt #1}}\fi
\expandafter\ifx\csname urlprefix\endcsname\relax\def\urlprefix{URL }\fi
\providecommand{\eprint}[2][]{\url{#2}}

\bibitem{Pitts2013}
Pitts R~A, Carpentier S, Escourbiac F, Hirai T, Komarov V, Lisgo S, Kukushkin
  A~S, Loarte A, Merola M, Naik A~S, Mitteau R, Sugihara M, Bazylev B and
  Stangeby P~C 2013 {\em Journal of Nuclear Materials\/} {\bf 438} S48--S56
  ISSN 0022-3115

\bibitem{Loarte2007}
Loarte A, Lipschultz B, Kukushkin A~S, Matthews G~F, Stangeby P~C, Asakura N,
  Counsell G~F, Federici G, Kallenbach A, Krieger K, Mahdavi A, Philipps V,
  Reiter D, Roth J, Strachan J, Whyte D, Doerner R, Eich T, Fundamenski W,
  Herrmann A, Fenstermacher M, Ghendrih P, Groth M, Kirschner A, Konoshima S,
  LaBombard B, Lang P, Leonard A~W, Monier-Garbet P, Neu R, Pacher H, Pegourie
  B, Pitts R~A, Takamura S, Terry J, Tsitrone E and Phy I~S~o~L~D 2007 {\em
  Nuclear Fusion\/} {\bf 47} S203--S263 ISSN 0029-5515

\bibitem{Pitcher1997}
Pitcher C~S and Stangeby P~C 1997 {\em Plasma Physics and Controlled Fusion\/}
  {\bf 39} 779--930 ISSN 0741-3335 1361-6587

\bibitem{Stangeby2018}
Stangeby P~C 2018 {\em Plasma Physics and Controlled Fusion\/} {\bf 60} 044022
  ISSN 0741-3335

\bibitem{Verhaegh2019}
Verhaegh K, Lipschultz B, Duval B, Février O, Fil A, Theiler C, Wensing M,
  Bowman C, Gahle D, Harrison J, Labit B, Marini C, Maurizio R, de~Oliveira H,
  Reimerdes H, Sheikh U, Tsui C, Vianello N and Vijvers" W 2019 {\em Nuclear
  Fusion\/} {\bf 59}

\bibitem{Krasheninnikov2017}
Krasheninnikov S~I and Kukushkin A~S 2017 {\em Journal of Plasma Physics\/}
  {\bf 83} 155830501 ISSN 0022-3778

\bibitem{Terry1999}
Terry J~L, Lipschultz B, Bonnin X, Boswell C, Krasheninnikov S~I, Pigarov A~Y,
  LaBombard B, Pappas D~A and Scott H~A 1999 {\em Journal of Nuclear
  Materials\/} {\bf 266-269} 30--36 ISSN 0022-3115
  \urlprefix\url{http://www.sciencedirect.com/science/article/pii/S0022311598008125}

\bibitem{Lipschultz1998}
Lipschultz B, Terry J~L, Boswell C, Hubbard A, LaBombard B and Pappas D~A 1998
  {\em Physical Review Letters\/} {\bf 81} 1007--1010 ISSN 0031-9007

\bibitem{Verhaegh2017}
Verhaegh K, Lipschultz B, Duval B~P, Harrison R, Reimerdes H, Theiler C, Labit
  B, Maurizio R, Marini C, Nespoli F, Sheikh U, Tsui C~K, Vianello N, Vijvers
  W~A~J and Team T~T~~E~M 2017 {\em Nuclear Materials and Energy\/} {\bf 12}
  1112--1117 ISSN 2352-1791

\bibitem{Verhaegh2019a}
Verhaegh K, Lipschultz B, Duval B, Fil A, Wensing M, Bowman C and Gahle D 2019
  {\em Plasma Phys. Control. Fusion\/} {\bf 61}

\bibitem{Lomanowski2020}
Lomanowski B, Groth M, Coffey I~H, Karhunen J, Maggi C~F, Meigs A, Menmuir S
  and O'Mullane M 2020 {\em Plasma Physics and Controlled Fusion\/}

\bibitem{Krasheninnikov1997}
Krasheninnikov S, Pigarov A~Y, Knoll D, LaBombard B, Lipschultz B, Sigmar D,
  Soboleva T, Terry J and Wising F 1997 {\em Physics of Plasmas\/} {\bf 4}
  1638--1646 ISSN 1070-664X

\bibitem{Fantz2002}
Fantz U 2002 {\em Contributions to Plasma Physics\/} {\bf 42} 675--684 ISSN
  0863-1042

\bibitem{Fantz2001}
Fantz U, Reiter D, Heger B and Coster D 2001 {\em Journal of Nuclear
  Materials\/} {\bf 290} 367--373 ISSN 0022-3115

\bibitem{Sakamoto2017}
Sakamoto M, Terakado A, Nojiri K, Ezumi N, Nakashima Y, Sawada K, Ichimura K,
  Fukumoto M, Oki K, Shimizu K, Ohno N, Masuzaki S, Togo S, Kohagura J and
  Yoshikawa M 2017 {\em Nuclear Materials and Energy\/} {\bf 12} 1004--1009
  ISSN 2352-1791

\bibitem{Hollmann2006}
Hollmann E~M, Brezinsek S, Brooks N~H, Groth M, McLean A~G, Pigarov A~Y and
  Rudakov D~L 2006 {\em Plasma Physics and Controlled Fusion\/} {\bf 48} 1165
  ISSN 0741-3335

\bibitem{Groth2019}
Groth M, Hollmann E, Jaervinen A, Leonard A, McLean A, Samuell C, Reiter D,
  Allen S, Boerner P, Brezinsek S, Bykov I, Corrigan G, Fenstermacher M,
  Harting D, Lasnier C, Lomanowski B, Makowski M, Shafer M, Wang H, Watkins J,
  Wiesen S and Wilcox R 2019 {\em Nuclear Materials and Energy\/} {\bf 19}
  211--217

\bibitem{Kukushkin2017}
Kukushkin A~S, Krasheninnikov S~I, Pshenov A~A and Reiter D 2017 {\em Nuclear
  Materials and Energy\/} {\bf 12} 984--988 ISSN 2352-1791

\bibitem{Wischmeier2004}
Wischmeier M, Pitts R~A, Alfier A, Andrebe Y, Behn R, Coster D, Horacek J,
  Nielsen P, Pasqualotto R, Reiter D and Zabolotsky A 2004 {\em Contributions
  to Plasma Physics\/} {\bf 44} 268--273

\bibitem{Terry1998}
Terry J~L, Lipschultz B, Pigarov A~Y, Krasheninnikov S~I, LaBombard B, Lumma D,
  Ohkawa H, Pappas D and Umansky M 1998 {\em Physics of Plasmas\/} {\bf 5}
  1759--1766 ISSN 1070-664x

\bibitem{Terakado2019}
Terakado A, Sakamoto M, Ezumi N, Nojiri K, Mikami T, Kinoshita Y, Togo S,
  Iijima T, Sawada K, Kado S and Nakashima Y 2019 {\em Nuclear Materials and
  Energy\/} {\bf 20} 100679 ISSN 2352-1791

\bibitem{Wuenderlich2016}
W\"{u}nderlich D and Fantz U 2016 {\em Atoms\/} {\bf 4} ISSN 2218-2004

\bibitem{Wunderlich2020}
W\"{u}nderlich D, Giacomin M, Ritz R and Fantz U 2020 {\em Journal of
  Quantitative Spectroscopy and Radiative Transfer\/} {\bf 240} 106695 ISSN
  0022-4073

\bibitem{Verhaegh2020a}
Verhaegh K, Lipschultz B, Harrison J~R, Duval B~P, Bowman C, Fil A, Gahle D~S,
  Moulton D, Myatra O, Perek A, Theiler C and Wensing M 2021 {\em Nuclear
  Materials and Energy\/}
  \urlprefix\url{https://doi.org/10.1016/j.nme.2021.100922}

\bibitem{Verhaegh2018}
Verhaegh K 2018 {\em Spectroscopic Investigations of detachment on TCV\/}
  Thesis University of York
  \urlprefix\url{http://etheses.whiterose.ac.uk/22523/}

\bibitem{Summers2006}
Summers H~P, Dickson W~J, O'Mullane M~G, Badnell N~R, Whiteford A~D, Brooks
  D~H, Lang J, Loch S~D and Griffin D~C 2006 {\em Plasma Physics and Controlled
  Fusion\/} {\bf 48} 263--293 ISSN 0741-3335 1361-6587

\bibitem{OMullane}
O’Mullane M 2013 Adas: Generalised collisonal radiative data for hydrogen
  Tech. rep. ADAS \urlprefix\url{http://www.adas.ac.uk}

\bibitem{Hinkley1969}
Hinkley D~V 1969 {\em Biometrika\/} {\bf 56} 635--639

\bibitem{Kotov2007}
Kotov V, Reiter D and Kukushkin A~S 2007 Numerical study of the iter divertor
  plasma with the b2-eirene code package Journal article Forschungszentrum
  J\"{u}lich

\bibitem{Reiter2008}
Reiter D {\em et~al.\/} 2008 The eirene code user manual Report
  Forschungszentrum Jülich GmbH
  \urlprefix\url{http://www.eirene.de/manuals/eirene.pdf}

\bibitem{Reiter2005}
Reiter D, Baelmans M and Börner P 2005 {\em Fusion Science and Technology\/}
  {\bf 47} 172--186 ISSN 1536-1055
  \urlprefix\url{https://doi.org/10.13182/FST47-172}

\bibitem{Stangeby2017}
Stangeby P~C and Chaofeng S 2017 {\em Nuclear Fusion\/} {\bf 57} 056007 ISSN
  0029-5515

\bibitem{McLean2019}
McLean A 2019 Understanding plasma divertor detachment in fusion power reactors
  Tech. rep. Lawrence Livermore National Lab.(LLNL), Livermore, CA (United
  States)

\bibitem{Sawada1995}
Sawada K and Fujimoto T 1995 {\em Journal of applied physics\/} {\bf 78}
  2913--2924

\bibitem{Perek2020}
Perek A, Vijvers W~A~J, Andrebe Y, Classen I~G~J, Duval B~P, Galperti C,
  Harrison J~R, Linehan B~L, Ravensbergen T, Verhaegh K and de~Baar M~R 2019
  {\em Review of Scientific Instruments\/} {\bf 90} 123514

\bibitem{Kukushkin2016}
Kukushkin A and Pacher H 2016 {\em Contributions to Plasma Physics\/} {\bf 56}
  711--716 ISSN 1521-3986

\bibitem{Fil2017}
Fil A~M~D, Dudson B~D, Lipschultz B, Moulton D, Verhaegh K~H~A, Fevrier O and
  Wensing M 2017 {\em Contributions to plasma physics\/} {\bf 58} ISSN
  0863-1042

\bibitem{Myatra}
Myatra O, Moulton D, Fil A, Dudson B and Lipschultz B 2018 Taming the flame:
  Detachment access and control in mast-u super-x {\em Plasma Surface
  Interactions\/}

\bibitem{Ravensbergen2020}
Ravensbergen T, van Berkel M, Silburn S~A, Harrison J~R, Perek A, Verhaegh K,
  Vijvers W~A~J, Theiler C, Kirk A and de~Baar M 2020 {\em Nuclear Fusion\/}
  \urlprefix\url{https://doi.org/10.1088\%2F1741-4326\%2Fab8183}

\bibitem{Perek2019submitted}
Perek A, Linehan B, Wensing M, Verhaegh K, Classen I~G~J, Duval B~P,
  F\'{e}vrier O, Reimerdes H, Theiler C, Wijkamp T and de~Baar M 2021 {\em
  Nuclear Materials and Energy\/}

\bibitem{Brezinsek2005}
Brezinsek S, Sergienko G, Pospieszczyk A, Mertens P, Samm U and Greenland P~T
  2005 {\em Plasma Physics and Controlled Fusion\/} {\bf 47} 615--634
  \urlprefix\url{https://doi.org/10.1088\%2F0741-3335\%2F47\%2F4\%2F003}

\bibitem{Fil2019submitted}
Fil A, Lipschultz B, Moulton D, Dudson B~D, F{\'{e}}vrier O, Myatra O, Theiler
  C, Verhaegh K, Wensing M and and 2020 {\em Plasma Physics and Controlled
  Fusion\/} {\bf 62} 035008

\bibitem{Shirai2002}
Shirai T, Tabata T, Tawara H and Itikawa Y 2002 {\em Atomic Data and Nuclear
  Data Tables\/} {\bf 80} 147 -- 204 ISSN 0092-640X

\bibitem{Behringer2000}
Behringer K and Fantz U 2000 {\em New Journal of Physics\/} {\bf 2} 23--23

\bibitem{Stangeby2000}
Stangeby P 2000 {\em The Plasma Boundary of Magnetic Fusion Devices. Series:
  Series in Plasma Physics, ISBN: 978-0-7503-0559-4. Taylor \& Francis, Edited
  by Peter Stangeby, vol. 7\/} {\bf 7}

\bibitem{Wischmeier2005}
Wischmeier M 2005 {\em Simulating divertor detachment in the TCV and JET
  tokamaks\/} Thesis EPFL

\bibitem{Cadez2011}
Cadez I, Markelj S and Milosavljevic A~R 2011 {\em Nuclear Engineering and
  Design\/} {\bf 241} 1267 -- 1271 ISSN 0029-5493 international Conference on
  Nuclear Energy for New Europe 2009
  \urlprefix\url{http://www.sciencedirect.com/science/article/pii/S0029549310002827}

\bibitem{Miyamoto2003}
Miyamoto K, Hatayama A, Ishii Y, Miyamoto T and Fukano A 2003 {\em Journal of
  Nuclear Materials\/} {\bf 313-316} 1036 -- 1040 ISSN 0022-3115 plasma-Surface
  Interactions in Controlled Fusion Devices 15

\bibitem{Lisgo2009}
Lisgo S, Borner P, Counsell G~F, Dowling J, Kirk A, Scannell R, O'Mullane M,
  Reiter D and Team M 2009 {\em Journal of Nuclear Materials\/} {\bf 390-91}
  1078--1080 ISSN 0022-3115 \urlprefix\url{<Go to ISI>://WOS:000267747300249}

\bibitem{Bowman2020}
Bowman C, Harrison J~R, Lipschultz B, Orchard S, Gibson K~J, Carr M, Verhaegh K
  and Myatra O 2020 {\em Plasma Physics and Controlled Fusion\/} {\bf 62}
  045014

\bibitem{Krishnakumar2011}
Krishnakumar E, Denifl S, \ifmmode \check{C}\else
  \v{C}\fi{}ade\ifmmode~\check{z}\else \v{z}\fi{} I, Markelj S and Mason N~J
  2011 {\em Phys. Rev. Lett.\/} {\bf 106}(24) 243201
  \urlprefix\url{https://link.aps.org/doi/10.1103/PhysRevLett.106.243201}

\bibitem{Reimold2017}
Reimold F, Wischmeier M, Potzel S, Guimarais L, Reiter D, Bernert M, Dunne M,
  Lunt T, Team A~U and Mst1Team E 2017 {\em Nuclear Materials and Energy\/}
  {\bf 12} 193--199 ISSN 2352-1791 \urlprefix\url{<Go to
  ISI>://WOS:000417293300026}

\bibitem{Wensing2019}
Wensing M, Duval B, Fevrier O, Fil A, Galassi D, Havlickova E, Perek A,
  Reimerdes H, Theiler C, Verhaegh K and Wischmeier M 2019 {\em Plasma Phys.
  Control. Fusion\/} {\bf 61}

\bibitem{Cowles2013}
Cowles M~K 2013 {\em Applied Bayesian statistics: with R and OpenBUGS
  examples\/} vol~98 (Springer Science \& Business Media) ISBN 1461456967

\bibitem{Botev2010}
Botev Z~I, Grotowski J~F and Kroese D~P 2010 {\em The Annals of Statistics\/}
  {\bf 38} 2916--2957 ISSN 0090-5364

\bibitem{Theiler2017}
Theiler C, Lipschultz B, Harrison J, Labit B, Reimerdes H, Tsui C, Vijvers
  W~A~J, Boedo J~A, Duval B~P, Elmore S, Innocente P, Kruezi U, Lunt T,
  Maurizio R, Nespoli F, Sheikh U, Thornton A~J, van Limpt S~H~M, Verhaegh K,
  Vianello N, Team T and Team E~M 2017 {\em Nuclear Fusion\/} {\bf 57} 072008
  ISSN 0029-5515

\bibitem{Harrison2017}
Harrison J~R, Vijvers W~A~J, Theiler C, Duval B~P, Elmore S, Labit B,
  Lipschultz B, van Limpt S~H~M, Lisgo S~W, Tsui C~K, Reimerdes H, Sheikh U,
  Verhaegh K~H~A and Wischmeier M 2017 {\em Nuclear Materials and Energy\/}
  {\bf 12} 1071--1076 ISSN 23521791

\end{thebibliography}






% \end{thebibliography}

\end{document}